\documentclass[11pt, a4paper]{article}

\usepackage[margin = 1in]{geometry}
\usepackage{palatino}

\usepackage{amsmath}
\usepackage{amssymb}
\usepackage{amsfonts}
\usepackage{amsthm} 
\usepackage{graphicx}
\usepackage{comment}
\usepackage{thm-restate}
\usepackage{hyperref}
\usepackage{here}
\usepackage{cite}
\usepackage{enumerate}
\usepackage{tabularx}
\usepackage{environ}
\usepackage{url}
\usepackage[normalem]{ulem}
\usepackage{color}
\usepackage[T1]{fontenc}

\newtheorem{theorem}{Theorem}
\newtheorem{lemma}[theorem]{Lemma}

\newtheorem{claim}[theorem]{Claim}
\newtheorem{observation}[theorem]{Observation}

\newtheorem{remark}[theorem]{Remark}

\usepackage[linesnumbered,noend,ruled,vlined]{algorithm2e}
\SetKwInput{KwInput}{Input}     
\SetKwInput{KwOutput}{Output}   
\SetKw{KwTo}{to}

\usepackage[framemethod=tikz]{mdframed}

\newenvironment{wrapper}[1]
{
	\smallskip
	\begin{center}
		\begin{minipage}{\linewidth}
			\begin{mdframed}[hidealllines=true, backgroundcolor=gray!20, leftmargin=0cm,innerleftmargin=0.35cm,innerrightmargin=0.35cm,innertopmargin=0.375cm,innerbottommargin=0.375cm,roundcorner=10pt]
                \centering
				#1}
			{\end{mdframed}
		\end{minipage}
	\end{center}
	\smallskip
}

\global\long\def\M{\mathcal{M}}%
\global\long\def\I{\mathcal{I}}%
\global\long\def\tO{\widetilde{O}}%
\global\long\def\eps{\varepsilon}%
\global\long\def\epsilon{\varepsilon}%
\newcommand{\cQ}{\mathcal{Q}}
\global\long\def\tO{\tilde{O}}%
\global\long\def\eps{\varepsilon}%
\newcommand{\cP}{\mathcal{P}}
\newcommand{\cR}{\mathcal{R}}

\newcommand{\Span}{\mathrm{span}}
\newcommand{\poly}{\mathrm{poly}}
\newcommand{\Exp}{\mathrm{exp}}
\newcommand{\sig}{\mathrm{sig}}

\ifdefined\ShowComment
\newcommand{\tatsuya}[1]{{\bf \color{red} tatsuya: #1}}
\newcommand{\later}[1]{{\color{red} [to be revised] #1}} 
\else
\newcommand{\tatsuya}[1]{}
\newcommand{\later}[1]{}
\fi

\title{Faster graphic matroid parity: Breaking the $nm$ time complexity barrier
}

\author{Tatsuya Terao\thanks{
Research Institute for Mathematical Sciences, Kyoto University.
E-mail: ttatsuya@kurims.kyoto-u.ac.jp
}
}

\date{}

\begin{document}

\maketitle

\begin{abstract}
In the graphic matroid parity problem, we are given an undirected multigraph $G=(V,E)$ with $n$ vertices, together with a partition of $E$ into $m$ pairs of distinct edges, and the goal is to find a maximum-cardinality collection of pairs whose union is a forest. This problem generalizes both graph matching and graphic matroid intersection. Graph matching admits the celebrated $O(\sqrt{n}m)$-time algorithm of Micali--Vazirani~[FOCS'80], and graphic matroid intersection admits $o(nm)$-time algorithms due to Gabow--Stallmann~[ICALP'85] and Gabow--Xu~[FOCS'89]. In contrast, for graphic matroid parity, Gabow--Stallmann~[FOCS'84 \& Combinatorica'86] gave an $O(n^2 m)$-time algorithm, and its $\tilde{O}(nm)$-time implementation by Gabow--Stallmann~[ICALP'85] had long remained the best known bound, except in very dense regimes where $m=\omega(n^3)$, in which case the algebraic $\tilde{O}(n^4)$-time algorithm of Cheung--Lau--Leung~[SODA'11 \& TALG'14] is faster.

We present a randomized $\tilde{O}(n^{3/2} \sqrt{m} + \sqrt{n} m)$-time algorithm for graphic matroid parity. This is the first algorithm to break the $O(nm)$-time complexity barrier for this problem outside the very dense regime. Moreover, it improves over the previous bounds except in very sparse regimes where $m=\tilde{O}(n)$ and in very dense regimes where $m = \tilde{\Omega}(n^{3.5})$. In particular, when $m=\Theta(n^2)$, our algorithm runs in $\tilde{O}(n^{2.5})$ time, which nearly matches the $\tilde{O}(n^{2.5})$-time bounds of Micali--Vazirani for graph matching and of Gabow--Xu for graphic matroid intersection in this regime.
\end{abstract}


\newpage

\setcounter{page}{1}

\section{Introduction} \label{sec:intro}

The matroid parity problem~\cite{lawler1976combinatorial} (also known as the matroid matching problem) is a common generalization of graph matching and matroid intersection problems.
For general matroids, it requires an exponential number of independence oracle queries~\cite{jensen1982complexity, lovasz1980matroid}.
In contrast, for linear matroids, Lov{\'a}sz established a min-max theorem for the problem and presented a polynomial-time algorithm~\cite{lovasz1978matroidmatching, lovasz1980matroid, lovasz1980selecting-independent-lines}.

In this work, we give an efficient algorithm for the important special case in which the matroid is graphic, known as the graphic matroid parity problem, or the spanning tree parity problem~\cite{garey2002computers}.
In this problem, we are given an undirected multigraph $G=(V,E)$ with
$n$ vertices, together with a partition of $E$ into $m$ pairs of
distinct edges.  
We refer to each such pair as a \emph{line}.
The goal is to find a maximum-cardinality collection of lines whose union is a forest.
A set $M\subseteq E$ is called a \emph{(graphic) matroid matching} if $M$ is the union of a collection of lines and forms a forest.
We note that, unlike the graph matching problem, where a simple graph has at most $n^2$ edges, the relevant objects here are lines, that is, pairs of edges. Thus, $m$ can be as large as $n^4$ after removing duplicate pair types.

Graphic matroid parity generalizes both graph matching\footnote{General graph matching (on a graph $G$) can be reduced to a graphic matroid parity problem (on a multigraph $H$) as follows.
For each vertex $v$ in $G$, create two vertices $v_0, v_1$ in $H$. 
For each edge $(v, w)$ in $G$, add a line consisting of two edges $(v_0, v_1)$ and $(w_0, w_1)$ in $H$.} and graphic matroid intersection.
More concretely, both graph matching and graphic matroid intersection instances with $n$ vertices and $m$ edges can be reduced to graphic matroid parity instances with $O(n)$ vertices and $O(m)$ lines.

Developing faster algorithms for graphic matroid parity is a natural direction
in light of the rich literature on faster algorithms for matroid problems on
graphic matroids, including graphic matroid intersection \cite{gabow1985efficient, gabow1989efficient}, packing spanning trees (matroid union for graphic matroids)~\cite{roskind1985note, gabow1988forests, gabow1991matroid, blikstad2023fast, quanrud2024faster, arkhipov2026faster}, and several other matroid problems~\cite{gabow1998algorithms, ene_et_al:LIPIcs.ICALP.2019.54,
henzinger_et_al:LIPIcs.ICALP.2023.74, khanna2025optimal}.
Furthermore, the graphic matroid parity problem has applications in a variety of algorithmic settings, including approximation algorithms for the maximum planar subgraph problem~\cite{cualinescu1998better, chalermsook_et_al:LIPIcs.STACS.2019.19,  calinescu_et_al:LIPIcs.ESA.2024.38}, the minimum Steiner tree problem~\cite{promel1997rnc, berman2006applications}, and various other problems~\cite{calinescu2010min, cualinescu20161, cualinescu2024improved}.
It also yields a polynomial-time algorithm for finding maximum-genus embeddings of graphs~\cite{furst1988finding}. Graphic matroid parity has further been used to obtain faster deterministic FPT algorithms for feedback vertex set~\cite{cao2015feedback, kociumaka2014faster, iwata2021improved} (see \cite[Theorem 6]{kociumaka2014faster} and \cite[Section 12.2.2]{cygan2015parameterized}).

Gabow--Stallmann~\cite{gabow1986augmenting} gave an $O(n^2m)$-time algorithm for graphic matroid parity based on the augmenting-path method. Using range-searching data structures, they later improved this to $\tO(nm)$ time\footnote{The $\tO$ notation omits polylog factors.}~\cite{gabow1985efficient}.
Cheung--Lau--Leung~\cite{cheung2014algebraic} gave a randomized algebraic algorithm for graphic matroid parity that runs in $\tO(n^4)$ time, and in $\tO(n^3)$ time when each edge pair shares a common vertex.
The case in which each edge pair shares a common vertex has applications in several approximation algorithms.

For both graph matching and graphic matroid intersection, breaking the
$O(nm)$-time complexity barrier marked an important milestone, and faster algorithms
are now known.

For bipartite graph matching, the celebrated Hopcroft--Karp~\cite{hopcroft1973n} algorithm achieved a running time of $O(\sqrt{n}m)$, improving over the $O(nm)$ bound obtained by a direct application of the Ford--Fulkerson~\cite{ford1956maximal} maximum-flow algorithm.
Recent breakthroughs in faster algorithms for bipartite matching and flow~\cite{madry2013navigating, lee2019solving, cohen2017negative, liu2020faster, axiotis2020circulation, van2020bipartite, van2023deterministic, chuzhoy2024faster, chuzhoy2024maximum, bernstein2024maximum, chen2025maximum, bernstein2025combinatorial} have led to an $m^{1+o(1)}$-time algorithm~\cite{van2023deterministic, chen2025maximum} and a combinatorial $\tilde{O}(n^2)$-time algorithm~\cite{bernstein2025combinatorial} for bipartite matching.

For general graph matching, Edmonds~\cite{edmonds1965paths} first gave a polynomial-time algorithm.
Gabow~\cite{gabow1976efficient} later gave an $O(nm\cdot \alpha(n))$-time implementation of Edmonds' algorithm.
Using the disjoint set union data structure of \cite{gabow1983linear}, this can be improved to $O(nm)$ time.
After the improvement of Even--Kariv~\cite{even1975n2}, the celebrated Micali--Vazirani algorithm~\cite{micali1980v} achieved an $O(\sqrt{n}m)$ running time.
The original paper of Micali--Vazirani did not contain a complete correctness proof, and the first complete and correct proof was given recently by Vazirani~\cite{vazirani2024theory}.
Alternative $O(\sqrt{n}m)$-time algorithms were given by Gabow--Tarjan~\cite{gabow1991faster}, Goldberg--Karzanov~\cite{goldberg2004maximum}, and Gabow~\cite{gabow2017weighted}.
Very recently, Izumi--Kitamura--Yamaguchi~\cite{izumi2025forgetting} proposed a new structural framework for shortest augmenting paths, which yields an $\tO(\sqrt{n}m)$-time algorithm with a simpler correctness proof.
More recently, a preprint released by OpenAI~\cite{OAI:Almost-Linear-Time-Maximum-Cardinality-Matching-in-Sparse-General-Graphs-September-24-2026} claims to give a randomized $m^{1+o(1)}$-time algorithm, and Li--Li--Zhang~\cite{li2026maximum} gave an $n^{2+o(1)}$-time algorithm.

For the matroid intersection problem, Edmonds \cite{edmonds1970submodular, edmonds1979matroid} first presented a polynomial-time algorithm.
For graphic matroid intersection, Edmonds' algorithm can be implemented in $\tO(nm)$ time using appropriate data structures.
Using an idea similar to the Hopcroft--Karp algorithm, Gabow--Stallmann~\cite{gabow1985efficient} presented an $\tO(nm^{2/3} + \sqrt{n}m)$-time algorithm for graphic matroid intersection.
Gabow--Xu~\cite{gabow1989efficient} later improved this to $\tO(\sqrt{n} m)$, matching, up to polylogarithmic factors, the running time of the Hopcroft--Karp algorithm for bipartite matching.
(We note that bipartite matching is a simple special case of graphic matroid intersection.)
More recently, the dynamic-oracle matroid intersection algorithm of Blikstad--Mukhopadhyay--Nanongkai--Tu~\cite{blikstad2023fast} also implies an $\tO(\sqrt{n} m)$-time algorithm for graphic matroid intersection.

While the $O(nm)$-time complexity barrier has been broken for both graph matching and graphic matroid intersection, graphic matroid parity has resisted such an improvement for nearly four decades, except in very dense regimes where $m = \omega(n^3)$~\cite{cheung2014algebraic}.
This naturally raises the following question.

\vspace{-2mm}

\begin{wrapper} 
    Can the graphic matroid parity problem also be solved in $o(nm)$ time? 
\end{wrapper}

\vspace{-2mm}

We answer this question affirmatively, except in the very sparse regime where $m=n^{1+o(1)}$.
For every fixed constant $\delta>0$, our algorithm runs in $o(nm)$ time whenever $m\ge n^{1+\delta}$.

\begin{restatable}{theorem}{mainexact} \label{thm:graphic_matroid_parity_exact_main}
    There is a randomized algorithm that finds a maximum graphic matroid matching with high probability in $\tilde{O}(n^{3/2} \sqrt{m} + \sqrt{n} m)$ time.
\end{restatable}

\begin{figure}[t]
    \centering
    \includegraphics[width=0.5\linewidth]{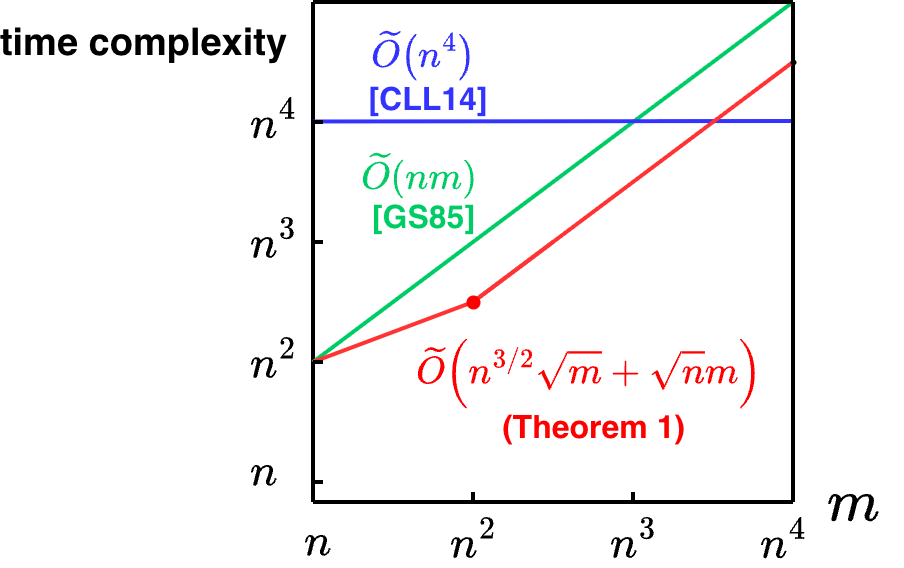}
    \caption{State-of-the-art time complexities for the graphic matroid parity problem.}
    \label{fig:complexity}
\end{figure}

The running time in Theorem~\ref{thm:graphic_matroid_parity_exact_main} is $\tO(n^{3/2}\sqrt{m})$ in the regime where $m\leq n^2$, and $\tO(\sqrt{n}m)$ in the regime where $m>n^{2}$.
Our algorithm in Theorem~\ref{thm:graphic_matroid_parity_exact_main} improves upon the previous algorithms~\cite{gabow1986augmenting, gabow1985efficient, cheung2014algebraic} except in very sparse regime where $m = \tO(n)$ and in very dense regime where $m = \tilde{\Omega}(n^{3.5})$.
For example, when $m = O(n^2)$, Gabow--Stallmann~\cite{gabow1985efficient} takes $\tO(n^3)$ time, whereas our algorithm in Theorem~\ref{thm:graphic_matroid_parity_exact_main} takes only $\tO(n^{2.5})$ time.
Moreover, for $m = O(n^2)$, the algorithms of Micali--Vazirani~\cite{micali1980v} for graph matching and of Gabow--Xu~\cite{gabow1989efficient} for graphic matroid intersection also run in $\tO(n^{2.5})$ time, so our algorithm is nearly as fast as the classical algorithms for the two special cases.
The previous and new bounds over the full range of $m$ are summarized in Figure~\ref{fig:complexity}.

In fact, the algorithm in Theorem~\ref{thm:graphic_matroid_parity_exact_main} is obtained by first developing a faster $(1-\eps)$-approximation algorithm.
Even for obtaining a $(1-\eps)$-approximate solution, Gabow--Stallmann's algorithm~\cite{gabow1985efficient} still takes $\tO(nm)$ time.
Indeed, Gabow--Stallmann's algorithm repeatedly finds a single augmenting path and augments along it.
Finding one augmenting path takes $\tO(m)$ time, and this procedure is repeated $O(n)$ times.
We give a $(1-\eps)$-approximation algorithm running in $o(nm)$ time.
Once a $(1-\eps)$-approximate solution is obtained, only a small number of augmentations remain to obtain an exact solution.
By combining our fast $(1-\eps)$-approximation algorithm with Gabow--Stallmann's augmenting-path algorithm, we obtain the fast exact algorithm in Theorem~\ref{thm:graphic_matroid_parity_exact_main}.

\begin{restatable}{theorem}{mainapprox}  \label{thm:graphic_matroid_parity_approx_main}
    For any $\eps > 0$, there is a randomized algorithm that finds a $(1 - \eps)$-approximate maximum graphic matroid matching with high probability in $\displaystyle \tilde{O}\left( \frac{m + n^2}{\eps} \right)$ time.
\end{restatable}

Our algorithm in Theorem~\ref{thm:graphic_matroid_parity_approx_main} builds on the framework of the recent semi-streaming algorithm for graph matching by Assadi~\cite{assadi2025simple}.
However, we emphasize that obtaining a faster algorithm for graphic matroid parity requires several new ideas.

Since our approximation algorithm builds on the framework of \cite{assadi2025simple}, it can also be implemented in the streaming model.
In the streaming model, the input lines, i.e., pairs of edges, are presented to the algorithm one by one.
The graph matching problem is one of the most extensively studied problems in the graph streaming setting~\cite{feigenbaum2005graph,mcgregor2005finding,eggert2009bipartite,ahn2011laminar,goel2012communication,eggert2012bipartite,konrad2012maximum,ahn2013linear,kapralov2013better,esfandiari2016finding,kale_et_al:LIPIcs.APPROX-RANDOM.2017.15,ahn2018access,konrad2018simple,tirodkar2018deterministic,gamlath2019weighted,assadi2021auction,assadi_et_al:LIPIcs.ICALP.2021.19,assadi2022semi,fischer2022deterministic,huang20231, bernstein2023improved,konrad_et_al:LIPIcs.STACS.2023.41, feldman2024maximum, assadi2025simple,mitrovic_et_al:LIPIcs.ICALP.2025.119, izumi2025forgetting}.
There are also several studies on the matroid intersection problem (and its generalizations) in the streaming setting in the literature~\cite{chekuri2015streaming, chakrabarti2015submodular, feldman2018less, huang_et_al:LIPIcs.APPROX/RANDOM.2020.62, huang2023fpt, garg2023semi, huang2024robust, terao:LIPIcs.WADS.2025.50, dudeja2026weighted, huang2026polynomial,huang2026auction}.
Thus, it is natural to study streaming algorithms for graphic matroid parity, which generalizes both graph matching and graphic matroid intersection.

\begin{restatable}{theorem}{mainstreaming} \label{thm:graphic_matroid_parity_streaming_main}
    For any $\eps > 0$, there is a streaming algorithm that finds a $(1 - \eps)$-approximate maximum graphic matroid matching with high probability using $\displaystyle \tO\left(\frac{n}{\eps^2}\right)$ space and $O(\log(n)/\eps)$ passes.
\end{restatable}

Our streaming algorithm in Theorem~\ref{thm:graphic_matroid_parity_streaming_main} is an extension of the semi-streaming algorithm for graph matching by Assadi~\cite{assadi2025simple}.
However, obtaining the faster sequential algorithm in Theorem~\ref{thm:graphic_matroid_parity_approx_main} requires additional new ideas beyond this extension.

\subsection{Additional related work}

Szigeti~\cite{szigeti2003graphic} gave a relatively simple proof of the min--max theorem specialized to graphic matroid parity (see {\cite[Corollary 43.2c]{schrijver2003combinatorial}} and {\cite[Theorem 1]{szigeti2003graphic}}).\footnote{
Although our algorithm will use a min--max theorem, we do not use this specialized min--max theorem for graphic matroid parity~\cite[Corollary 43.2c]{schrijver2003combinatorial}.
Instead, we rely on the min--max theorem for linear matroid parity.
The reason is that the min--max theorem specialized to graphic matroid parity is not known in an algorithmic form.
This is not an obstacle to obtaining an exponential-time streaming algorithm, since one can enumerate all candidate covers by brute force.
However, it becomes a serious obstacle to obtaining a fast sequential algorithm for graphic matroid parity, where we would need to find a minimum cover efficiently.
For this reason, our algorithm uses the min--max theorem for linear matroid parity, stated as Theorem~\ref{thm:min_max_linear_matroid_parity}.
}

There have been several important advances on the complexity of matroid parity.
Lee--Sviridenko--Vondr\'ak~\cite{lee2013matroid} presented a PTAS for matroid parity for general matroids.
For linear matroid parity, Gabow--Stallmann~\cite{gabow1986augmenting} gave an $O(mn^{\omega})$-time algorithm based on the augmenting-path method.
Orlin--Vande Vate~\cite{orlin1990solving} gave an alternative algorithm that solves linear matroid parity as a sequence of matroid intersection problems.
This approach was later simplified and improved by Orlin~\cite{orlin2008fast}, who gave an $O(mn^3)$-time algorithm.
Cheung--Lau--Leung~\cite{cheung2014algebraic} presented a randomized algebraic $O(mn^{\omega - 1})$-time algorithm for linear matroid parity, extending Harvey's~\cite{harvey2009algebraic} technique for linear matroid intersection.
Very recently, Hamada--Iwata~\cite{hamada2026shortest} presented a deterministic $O(mn^2\log n)$-time algorithm that repeatedly augments the current solution along a shortest augmenting path, improving upon the deterministic bounds of Gabow--Stallmann~\cite{gabow1986augmenting} and Orlin~\cite{orlin2008fast}.
With fast matrix multiplication, the running time can be reduced to $O(mn^2)$.
The breakthrough work of Iwata--Kobayashi~\cite{iwata2021weighted} gave the first polynomial-time algorithm for the weighted version of linear matroid parity.

The search procedure of Hamada--Iwata finds only one shortest augmenting path per search.
To improve upon the $\tO(nm)$ bound for graphic matroid parity along the lines of Micali--Vazirani~\cite{micali1980v}, one would need to find a maximal set of disjoint shortest augmenting paths in a single search of $\tO(m)$ time, which requires both extending the search procedure of Hamada--Iwata so that it finds such a set at once, and designing data structures that handle blossoms and transforms as in Gabow--Stallmann~\cite{gabow1985efficient}.
We are not aware of how to achieve either, and we leave such an approach as an interesting direction for future work.

There have also been many studies on fast matroid intersection algorithms for other important specific classes of matroids~\cite{gabow1979efficient, cunningham1986improved, gabow1989efficient, gabow1996efficient, xu1994fast, harvey2007algebraic, harvey2009algebraic, huang2016exact, terao2026faster}.
Recently, Blikstad--Mukhopadhyay--Nanongkai--Tu \cite{blikstad2023fast} proposed a remarkable new oracle model, called dynamic oracle, which unifies various results on fast matroid intersection algorithms for specific classes of matroids; see \cite[Table 2]{blikstad2023fast}.
There have also been many studies on fast matroid intersection algorithms in the independence (or rank) oracle model, including several recent breakthroughs~\cite{cunningham1986improved, fujishige1995efficient, shigeno1995dual, lee2015faster, chekuri2016fast, huang2016exact, nguyen2019note, chakrabarty2019faster, blikstad2021breaking_STOC, blikstad2021breaking, u2022subquadratic, blikstad2023fast,quanrud:LIPIcs.ICALP.2024.118, blikstad2024efficient, dudeja2026weighted, huang2026auction, terao2026fasterb}.

Gabow--Stallmann~\cite{gabow1985efficient} also gave a fast algorithm to find $k$ edge-disjoint spanning trees on a graph. 
Subsequently, Gabow--Westermann~\cite{gabow1988forests} presented a fast algorithm to find $k$ forests in a graph maximizing the number of edges in their union, which subsumes the $k$ edge-disjoint spanning trees problem.
More recently, several faster algorithms for this problem have been developed~\cite{quanrud2024faster, blikstad2023fast, arkhipov2026faster}.
Note that this problem can be viewed as a matroid intersection of a graphic matroid and a partition matroid.
It can also be viewed as the matroid union problem in which the given matroids are graphic.
There have also been several studies on fast matroid union algorithms in the oracle model~\cite{cunningham1986improved, blikstad2023fast , quanrud2024faster, terao2025faster}.

The framework of Assadi~\cite{assadi2025simple} has already found several applications, including a fast $(1-\eps)$-approximation algorithm for matroid intersection in the independence-oracle model~\cite{quanrud:LIPIcs.ICALP.2024.118}, a streaming algorithm for single-source shortest paths~\cite{assadi2026better}, and a quantum query algorithm for linear programs~\cite{objois2025adaptive}.


\section{Preliminaries} \label{sec:preliminaries}

\subparagraph*{Notation.}
Throughout the paper, $\log$ denotes the logarithm to base $2$.
We write ``with high probability'' (w.h.p.) if the failure probability is at most $1/ \poly(n)$. 
For a finite set $X$, let $|X|$ denote the cardinality of $X$, which is also called the {\em size} of $X$.
For two sets $A$ and $B$, we write $A\oplus B$ for their symmetric difference.

Throughout the paper, we allow graphs to have parallel edges, but not self-loops.
We work in the regime where $m=\Omega(n)$ and $m=O(n^4)$.

For the definitions of matroids, see Appendix~\ref{appendix:matroids}.
In this paper, we use only graphic and linear matroids.

\subparagraph*{Graphic matroids.}
Given an undirected graph $G = (V, E)$, let $\I = \{ F \subseteq E \mid F \text{ is a forest in } $G$ \}$.
Then, $\M = (E, \I)$ is a matroid, called the \emph{graphic matroid} of $G$. 

\subparagraph*{Graphic matroid parity.}
In the graphic matroid parity problem, we are given an undirected multigraph $G=(V,E)$ with $n$ vertices, together with a partition of $E$ into $m$ pairs of distinct edges.  
We refer to each such pair as a \emph{line}.
The goal is to find a maximum-cardinality collection of lines whose union is a forest.
A set $M\subseteq E$ is called a \emph{graphic matroid matching} if $M$ is the union of a collection of lines and forms a forest.
When no confusion can arise, we simply call it a \emph{matroid matching}.

Let $L$ denote the set of all lines.
We denote by $\nu(L)$ the maximum number of lines in a matroid matching in $L$.
For each element $e\in E$, we write $\bar e$ for the mate of $e$, that is, the unique element such that $\{e,\bar e\}\in L$.
A \emph{parity partition} of $E$ is a partition $(E_0,\ldots,E_k)$ of $E$ into unions of lines, that is, $e\in E_i$ if and only if $\bar e\in E_i$ for every $e\in E$ and every $i \in \{0, \ldots, k \}$.

\subparagraph*{Linear matroids.}
A matroid $\M = (E, \I)$ is said to be {\em linearly representable} or a {\em linear matroid}
if there exists a matrix $A$ over some field whose columns are indexed by $E$ such that 
$I \in \I$ if and only if the columns indexed by $I$ are linearly independent. 

It is well known that graphic matroids are linearly representable.
Indeed, it is well known that a graphic matroid can be represented over $\mathbb{F}_2$ by the vertex-edge incidence matrix, where each column corresponds to an edge and has exactly two nonzero entries, both equal to $1$ (see, e.g., \cite[Section~5.1]{oxley2006matroid}).
Under this representation, a set of columns is linearly independent if and only if the corresponding edge set is a forest.

When working with a linear matroid, we fix the representation and identify each element $e\in E$ with the corresponding column vector.
For a set of vectors $S$, we write $\Span(S)$ for the linear subspace generated by $S$.
\textbf{Throughout the paper, $\Span$ always denotes linear span, not matroid closure.}
For any set of vectors $S\subseteq \operatorname{span}(E)$, we define its \emph{rank} by $r(S)=\dim(\operatorname{span}(S))$.

\subparagraph*{Linear Matroid parity.}
In the linear matroid parity problem, we are given an $n \times 2m$ matrix, together with a partition of column set into $m$ pairs of distinct columns. 
We call each such pair a \emph{line}. 
The goal is to find a maximum-cardinality collection of lines whose union is linearly independent.

Graphic matroid parity is a special case of linear matroid parity.
Whenever we use terminology from linear matroid parity, the linear matroid parity instance under consideration is the one obtained from the input graphic matroid parity instance.
That is, we identify each edge of $G$ with its column in the vertex-edge incidence matrix over $\mathbb{F}_2$, and keep the same partition of edges into lines.

\section{Technical overview} \label{sec:technical_overview}

In this section, we give a technical overview of our fast $(1 - \eps)$-approximation algorithm for graphic matroid parity.
Once we have obtained a $(1-\eps)$-approximation algorithm that runs in $\displaystyle \tilde{O}\left(\frac{ m + n^2 }{\eps} \right)$ time, we can turn it into a fast exact algorithm as follows.
After obtaining a $(1 - \eps)$-approximate solution, at most $O(\eps n)$ augmentations are needed to obtain an exact optimal solution.
Since Gabow--Stallmann~\cite{gabow1985efficient} can find one augmenting path in $\tilde{O}(m)$ time, these remaining augmentations take $\tilde{O}(\eps n m)$ time in total.
Thus, the total running time is $\displaystyle \tilde{O}\left(\frac{m + n^2}{\eps} + \eps n m\right)$.
By setting $\displaystyle 1/\eps = \min\left\{\sqrt{\frac{m}{n}},\ \sqrt{n}\right\}$, this becomes $\tO(n^{3/2}\sqrt{m} + \sqrt{n} m)$.

Our fast $(1-\eps)$-approximation algorithm builds on the framework of the recent MWU-based algorithm for graph matching by Assadi~\cite{assadi2025simple}.
There are two main difficulties in turning this framework into a faster algorithm for graphic matroid parity.

The first difficulty is to adapt Assadi's idea from graph matching to graphic matroid parity.
Although this adaptation requires a careful argument, it is not where the main technical novelty lies.
To carry it out, we exploit structural properties of the minimum cover produced by the Gabow--Stallmann algorithm~\cite{gabow1986augmenting,gabow1985efficient}.

The second difficulty is to break the $O(nm)$-time complexity barrier.
The algorithm obtained by an adaptation of Assadi's framework would require $\Theta(nm)$ span-membership tests.
The main technical novelty of our algorithm is to avoid performing these tests
one by one.
To this end, we introduce random fingerprints, which we call \emph{patterns}.
These patterns compactly encode the relevant information, allowing us to
reinterpret the $\Theta(nm)$ span-membership tests as a pattern-matching problem
and process many of them simultaneously.
To make this approach work, we crucially exploit the structural properties of the minimum cover produced by Gabow--Stallmann's algorithm~\cite{gabow1986augmenting,gabow1985efficient}.

\subparagraph*{Organization of the technical overview}
We first review Assadi's MWU framework and view graphic matroid parity as linear matroid parity over $\mathbb{F}_2$, whose min--max theorem supplies the analogue of a vertex cover.
We next describe the structure of the minimum covers produced by Gabow--Stallmann and use it to adapt Assadi's framework.
Finally, we introduce patterns and explain how they avoid the $\Theta(nm)$ span-membership tests required by a direct implementation.

For ease of exposition, the overview mainly explains a $(1-\eps)$-approximation algorithm running in $\displaystyle \tO\left(\frac{nm^{3/4}}{\eps} + \frac{n^2}{\eps^2}\right)$ time, and hence an exact algorithm running in $\tO(n^{4/3} m^{2/3} + nm^{7/8})$ time.
To improve the bounds further, we modify Assadi's framework itself so that it runs faster, and we also speed up the span-membership tests.
Both modifications, which are briefly explained at the end of the overview, yield a $(1-\eps)$-approximation algorithm running in $\displaystyle \tO\left(\frac{m + n^2}{\eps}\right)$ time, and hence an exact algorithm running in $\tO(n^{3/2}\sqrt{m} + \sqrt{n} m)$ time, as stated in Theorems~\ref{thm:graphic_matroid_parity_approx_main} and~\ref{thm:graphic_matroid_parity_exact_main}.

\subparagraph*{Assadi's~\cite{assadi2025simple} MWU-based framework for bipartite matching}
Our graphic matroid parity algorithm builds on Assadi's remarkable framework, so we briefly describe Assadi's algorithm for bipartite matching.
The algorithm maintains an importance value $q_e$ for each edge $e \in E$, with all values initially equal.
The algorithm then repeats the following process for $T = O(\log(n) / \eps)$ iterations: 
In each iteration, it first samples $\tilde{O}(n / \eps)$ edges $E' \subseteq E$ in proportion to $q$.
Then, it computes a maximum matching $M$ and a minimum vertex cover $U$ in the sampled graph $(V, E')$.
Finally, it updates the importance values $q$ by setting $q_e \gets 2q_e$ for every edge $e$ not covered by $U$.
After $T$ iterations, the algorithm outputs the largest matching $M$ found during the process.

\subparagraph*{Viewing graphic matroid parity as linear matroid parity.}
Our algorithm also relies on arguments from Gabow--Stallmann's augmenting-path framework~\cite{gabow1986augmenting} for linear matroid parity.
We therefore briefly describe the relationship between linear matroid parity and graphic matroid parity.
Graphic matroid parity is obtained as the special case of linear matroid parity in which the linear matroid is the $\mathbb{F}_2$-representation of a graphic matroid by its vertex-edge incidence matrix. (See Section~\ref{sec:preliminaries}.)

When we use terminology from linear matroid parity, the linear matroid parity instance under consideration is the one obtained from the input graphic matroid parity instance by the reduction in Section~\ref{sec:preliminaries}.
That is, we identify each edge of $G$ with its column in the vertex-edge incidence matrix over $\mathbb{F}_2$, and keep the same partition of edges into lines.

Gabow--Stallmann~\cite{gabow1986augmenting} gave an augmenting-path algorithm for linear matroid parity.
Applied to the associated linear matroid parity instance over $\mathbb{F}_2$ described above, this framework gives an augmenting-path algorithm for graphic matroid parity. 
Using range-searching data structures, Gabow--Stallmann~\cite{gabow1985efficient} gave a faster implementation of this algorithm. 
Their implementation finds one augmenting path in $\tO(m)$ time and solves graphic matroid parity in $\tO(nm)$ time.

\subparagraph*{Min-max theorem for linear matroid parity}

To extend the framework of Assadi~\cite{assadi2025simple} to graphic matroid parity, we need to identify an object that plays the role of the vertex cover in Assadi's bipartite matching algorithm.
We would like to obtain such an object from a min-max theorem for linear matroid parity.
Lov{\'a}sz~\cite{lovasz1980selecting-independent-lines} proved the following min-max theorem for linear matroid parity, which can be viewed as a generalization of the Tutte--Berge formula for general graph matching.

\begin{restatable}{theorem}{minmax} [See {\cite[Corollary 43.2a]{schrijver2003combinatorial}} and {\cite{gabow1986augmenting}}] \label{thm:min_max_linear_matroid_parity}
    Let $\nu(L)$ denote the maximum number of lines in a matroid matching. 
    For the linear matroid parity, $\nu(L)$ satisfies
    \begin{equation*}
        \nu(L) = \min \left\{ r(A) + \sum_{i = 0}^{k} \left\lfloor \frac{r(A \cup E_i) - r(A)}{2} \right\rfloor \right\}, 
    \end{equation*}
    where the minimum is taken over all subsets $A\subseteq \Span(E)$ and all parity partitions $(E_0,\ldots,E_k)$ of $E$.
\end{restatable}

Here, the minimum is taken over all subsets $A \subseteq \Span(E)$, not only over subsets of the ground set $E$.
We now define covers for linear matroid parity.

\begin{restatable}{definition}{cover} \label{def:matroid_parity_cover}
    A \emph{cover} of a linear matroid parity instance is a tuple $(A;E_0,\ldots,E_k)$, where $A\subseteq \Span(E)$ and $(E_0,\ldots,E_k)$ is a parity partition of $E$.
    The value of a cover $(A;E_0,\ldots,E_k)$ is defined as $\displaystyle r(A) + \sum_{i = 0}^{k} \left\lfloor \frac{r(A \cup E_i) - r(A)}{2} \right\rfloor$.
    A cover is called a \emph{minimum cover} if its value is minimum among all covers.
\end{restatable}

Gabow--Stallmann's augmenting-path algorithm for linear matroid parity~\cite{gabow1986augmenting} constructs an optimality certificate when no augmenting path remains.
In the terminology of Definition~\ref{def:matroid_parity_cover}, this certificate is a minimum cover.\footnote{This is analogous to the standard situation in bipartite matching, where a failed augmenting-path search yields a minimum vertex cover, which certifies the optimality of the current matching.}
Their graphic matroid parity algorithm~\cite{gabow1985efficient} is a specialized
implementation of this linear matroid parity augmenting-path algorithm on the
$\mathbb{F}_2$-representation of the input graph.
Therefore, their graphic matroid parity algorithm~\cite{gabow1985efficient}  also produces a minimum cover for the associated linear matroid parity instance.

\subparagraph*{Structures of a minimum cover by Gabow--Stallmann~\cite{gabow1986augmenting, gabow1985efficient}}

Our adaptation of Assadi's idea~\cite{assadi2025simple} to graphic matroid parity and our approach to breaking the $O(nm)$-time complexity barrier both rely on structural properties of the minimum cover produced by Gabow--Stallmann's algorithm for graphic matroid parity~\cite{gabow1986augmenting, gabow1985efficient}.
We therefore describe the minimum cover obtained from Gabow--Stallmann's algorithm for the associated linear matroid parity instance, and then explain the structural properties of this cover.

In what follows, we view the graphic matroid parity instance as a linear matroid parity instance via the vertex-edge incidence matrix over $\mathbb{F}_2$.
Thus, \textbf{each edge $e\in E$ is also regarded as its corresponding column vector in $\mathbb{F}_2^n$.}
We use the same symbol $e$ for both the edge and this vector.

The algorithm finds a maximum matroid matching $M \subseteq E$, and a parity partition $(E_0, \ldots, E_k)$ of $E$.
For each $i \in \{0, \ldots, k \}$, we set $M_i = E_i \cap M$.
For each $i \in \{0, \ldots, k \}$, the algorithm finds a set $T_i \subseteq E_i$, with $T_0 = \emptyset$ and $T_i \neq \emptyset$ for $i \in \{1, \ldots, k\}$.
For each $i \in \{1, \ldots, k \}$ such that $T_i \subseteq M_i$, we define $\displaystyle \alpha_i = \sum_{\beta\in T_i} \beta$, i.e., the linear combination of the vectors in $T_i$ with all coefficients equal to $1$.
Finally, we set $A = \{ \alpha_i \mid T_i \subseteq M_i \}$.
The above construction has the following structural property. (See Appendix~\ref{sec:gabow_stallmann_min_cover_appendix} for details.)

\begin{restatable}{lemma}{structures}[From {\cite{gabow1986augmenting, gabow1985efficient}}] \label{lem:structual_lemma}
    Let $(A;E_0, \dots, E_k)$ be a minimum cover produced by Gabow--Stallmann's algorithm for graphic matroid parity~\cite{gabow1986augmenting, gabow1985efficient}.
    For every $i \in \{0, \ldots, k \}$ such that $r(E_i \cup A) > r(A) + 1$, the following hold:
    \begin{enumerate}[(i)]
        \item if $T_i = \emptyset$, then $r(E_i \cup A) = r(M_i \cup A) = |M_i \cup A|$,
        \item if $\emptyset \neq T_i \subseteq M_i$, then $r(E_i \cup A) = r(M_i \cup (A \setminus \{ \alpha_i \})) = |M_i \cup (A \setminus \{ \alpha_i \})|$,
        \item if $T_i \not\subseteq M_i$, then, for every $x \in T_i \setminus M_i$, $r(E_i \cup A) = r(M_i \cup A \cup \{ x \}) = |M_i \cup A \cup \{ x \}|$.
    \end{enumerate}
    Here, $T_0=\emptyset$ and $T_i\neq\emptyset$ for every $i\in\{1,\ldots,k\}$, so case~(i) applies exactly to $i=0$.
\end{restatable}

In particular, every relevant space $\Span(E_i \cup A)$ is generated by $A$, the matched edges $M_i$, and at most one additional edge.

Using Lemma~\ref{lem:structual_lemma} together with the linear matroid parity min--max theorem (Theorem~\ref{thm:min_max_linear_matroid_parity}), Gabow--Stallmann proved that, if no augmenting path is found, then the current matroid matching $M$ is maximum.
Equivalently, the tuple $(A;E_0,\ldots,E_k)$ serves as a minimum cover certifying the optimality of $M$.

\subparagraph*{Adapting Assadi's~\cite{assadi2025simple} framework to graphic matroid parity}

We now adapt the framework of Assadi~\cite{assadi2025simple} to graphic matroid parity.
Although this adaptation is nontrivial and requires a careful argument, it is not where the main technical novelty of this paper lies.
The resulting framework is summarized in Algorithm~\ref{alg:streaming_graphic_matroid_parity}, which computes a $(1-\eps)$-approximate matroid matching.
Although this framework extends more generally to linear matroids represented over a constant-size finite field, including binary matroids and hence cographic matroids, we focus on graphic matroids in this paper.

We use the following relaxed notion of a cover.
Unlike a cover in Definition~\ref{def:matroid_parity_cover}, the sets $E_i$ below are not required to form a parity partition.

\begin{restatable}{definition}{goodcover} \label{def:matroid_parity_good_cover}
    Let $(A;E_0,\ldots,E_k)$ be a tuple such that
    $A\subseteq\Span(E)$ and $E_i\subseteq E$ for every
    $i\in\{0,\ldots,k\}$.
    We say that this tuple \emph{covers} a line
    $\ell=\{e,\bar e\}$ if at least one of the following conditions holds:
    \begin{itemize}
        \item $r(A\cup\ell)-r(A)\leq 1$, or
        \item there exists $i\in\{0,\ldots,k\}$ such that
        $r(A\cup E_i\cup\ell)=r(A\cup E_i)$.
    \end{itemize}

    The tuple $(A;E_0,\ldots,E_k)$ is a \emph{pruned cover} of $L$
    if it covers every line $\ell\in L$.
\end{restatable}

Here, we can prove that the minimum value of $\displaystyle r(A)+\sum_{i=0}^{k}\left\lfloor \frac{r(A\cup E_i)-r(A)}{2}\right\rfloor$ over all pruned covers is equal to $\nu(L)$.

Algorithm~\ref{alg:streaming_graphic_matroid_parity} follows the same high-level framework of Assadi's algorithm~\cite{assadi2025simple}.
It maintains an importance value $q_\ell$ for each line $\ell\in L$, with all values initially equal.
In each iteration, the algorithm samples lines according to the current importance values and computes a maximum matroid matching $M^{(t)}$ together with a minimum cover of the sampled instance.
It then removes all sets $E_i$ with $r(A\cup E_i)-r(A)\leq 1$, obtaining a pruned cover.
Finally, it doubles the importance value of every line not covered by this pruned cover.
After $T=O(\log m/\eps)$ iterations, the algorithm outputs the largest matroid matching found during the process.

The proof that Algorithm~\ref{alg:streaming_graphic_matroid_parity} outputs a $(1-\eps)$-approximate solution is almost identical to Assadi's analysis~\cite{assadi2025simple}.
There is only one additional point that requires some care.
We need an $m^{O(n)}$ upper bound on the number of possible covers produced by the sampled instances.
In the bipartite matching case, this is immediate, since the number of vertex covers is at most $2^n$.
For graphic matroid parity, we introduce pruned covers for this purpose.
Using the structural properties of the minimum cover obtained from Gabow--Stallmann's algorithm~\cite{gabow1986augmenting,gabow1985efficient}, together with the pruning operation described above, we show that the number of pruned covers that can be obtained from the minimum covers of sampled instances is at most $m^{O(n)}$.\footnote{More precisely, we bound the number of possible signatures
$\left(\Span(A);\Span(A\cup E_0),\ldots,\Span(A\cup E_k)\right)$, rather than the
number of pruned covers themselves. The idea of bounding the number of spans is also used in Quanrud's
adaptation of Assadi's framework to matroid intersection~\cite{quanrud:LIPIcs.ICALP.2024.118}.}
See Lemma~\ref{lem:candidate_bounds} for the details.
All other parts of the proof follow Assadi's argument without essential changes.

\begin{algorithm}[t]
    \caption{A $(1-\eps)$-approximation algorithm for graphic matroid parity}
    \label{alg:streaming_graphic_matroid_parity}
    Let $q_{\ell}^{(1)} \gets 1$, $Q^{(1)} \gets m$

    \For{$t = 1$ to $T := \frac{4}{\eps} \log m$ }{
        Sample each line $\ell \in L$ independently at random with probability $\displaystyle p_{\ell}^{(t)} = \min \left\{ 1,  \frac{100n \log(n)}{\eps} \cdot \frac{q_{\ell}^{(t)}}{Q^{(t)}} \right\}$. \label{line:sample} \\
        Apply the algorithm of Gabow--Stallmann~\cite{gabow1986augmenting, gabow1985efficient} to compute a maximum matroid matching $M^{(t)} \subseteq E$ and a minimum cover $(A;E_0, \ldots,E_{k'})$ of the sampled instance. \label{line:sample_solve} \\
        Remove every set $E_i$ satisfying $r(A \cup E_i)-r(A) \leq 1$ from the cover, and reindex the remaining sets as $(A;E_0,\ldots,E_k)$. \label{line:removal} \\
        \For{$\ell \in L$ \label{line:forloop}}{
            \If{($r(A \cup \ell ) - r(A) \leq 1$) or (there exists $i \in \{ 0, \ldots, k\}$ such that $r(E_i \cup A \cup \ell ) = r(E_i \cup A)$) \label{line:nm_barrier}}{
                $q_{\ell}^{(t + 1)} \gets q_{\ell}^{(t)}$
            }
            \Else {
                $q_{\ell}^{(t + 1)} \gets q_{\ell}^{(t)} \cdot 2$ \label{line:forloop_end}
            }
        }
        $Q^{(t + 1)} \gets \sum_{\ell \in L} q_{\ell}^{(t + 1)}$
    }

    \Return{a largest matroid matching among $M^{(1)},\ldots,M^{(T)}$}
\end{algorithm}

\subparagraph*{Breaking the $O(nm)$-time complexity barrier via patterns}

In Line~\ref{line:nm_barrier} of Algorithm~\ref{alg:streaming_graphic_matroid_parity}, for each line $\ell\in L$ and each index $i\in\{0,\ldots,k\}$, we need to check whether $r(A\cup E_i \cup \ell)=r(A \cup E_i)$, or equivalently, whether $\ell \subseteq \Span(A \cup E_i)$.
Since the number of sets $E_i$ can be $\Theta(n)$ and $|L| = m$, a direct implementation would require checking $\Theta(nm)$ conditions.
We give a new way to perform these checks implicitly, without explicitly testing all $\Theta(nm)$ conditions.
This is the main technical contribution of our algorithm, and it is where the main novelty lies.

We first explain how to obtain a $(1-\eps)$-approximation algorithm running in $\displaystyle \tO\left(\frac{nm^{3/4}}{\eps} + \frac{n^2}{\eps^2}\right)$ time, which in turn implies an exact algorithm running in $\tilde{O}(n^{4/3} m^{2/3} + nm^{7/8})$ time.
Later in this paragraph, we further improve this running time.

We first bound the running time except for the inner for loop (Lines~\ref{line:forloop}--\ref{line:forloop_end}), which is the main bottleneck and will be discussed below.
In each iteration $t$, the number of sampled lines is $\tO(n/\eps)$ with high probability.
Therefore, by applying the exact algorithm of Gabow--Stallmann~\cite{gabow1985efficient} to the sampled instance, Line~\ref{line:sample_solve} can be implemented in $\tO(n^2/\eps)$ time per iteration, including the computation of a minimum cover.
Since the number of iterations is $T=\tO(1/\eps)$, the total time spent on Line~\ref{line:sample_solve} is $\tO(n^2/\eps^2)$.
Line~\ref{line:sample} can be implemented in $\tO(m)$ time per iteration.
Line~\ref{line:removal} can be implemented in $\tO(m)$ time per iteration, because the minimum cover produced by Gabow--Stallmann's algorithm contains enough information to determine, for each part $E_i$, whether
$r(A\cup E_i)-r(A)\leq 1$.
Since $m=O(n^4)$, we have $ \tO(m/\eps) \leq \tO(nm^{3/4}/\eps)$.
Therefore, it remains to prove that the inner for loop (Lines~\ref{line:forloop}--\ref{line:forloop_end}) can be implemented in  $\tO(n^2 + nm^{3/4})$ time per iteration.

To evaluate the condition in Line~\ref{line:nm_barrier}, for each line $\ell\in L$ and each index $i\in\{0,\ldots,k\}$, we need to check whether $r(E_i\cup A \cup \ell)=r(E_i \cup A)$.
Doing this explicitly would require $O(nm)$ rank-equality tests.
To avoid this, we assign to each edge $e\in E$ a random $O(\log n)$-bit value, called its
\emph{pattern}, which serves as a fingerprint of the edge with respect to
the current pruned cover.

Roughly speaking, we extend the current matching $M$ to a spanning tree $M^*$ and assign random $O(\log n)$-bit weights to the edges of $M^*$.
For an edge $e=\{u,v\}\in E$, its pattern is the bitwise XOR of the weights on the
unique $u$-$v$ path in $M^*$.
We emphasize that the use of bitwise XOR is essential for handling indices of case~(iii) in Lemma~\ref{lem:structual_lemma}.
    Recall that every vector in $A$ has the form $\alpha_i=\sum_{\beta\in T_i}\beta$ with $T_i\subseteq M_i$, where $(A;E_0,\ldots,E_{k'})$ is the minimum cover computed in Line~\ref{line:sample_solve} and $M_i=E_i\cap M$.
    We write $\mathcal{T}_A$ for the family of the sets $T_i$ over all $\alpha_i\in A$.
    Since the sets $M_0,\ldots,M_{k'}$ are pairwise disjoint, the sets in $\mathcal{T}_A$ are pairwise disjoint subsets of $M$.
    We note that Line~\ref{line:removal} does not change $A$, so $\mathcal{T}_A$ may contain the set $T_i$ of a part $E_i$ removed in Line~\ref{line:removal}.
We now define patterns formally.

\begin{restatable}{definition}{pattern} \label{def:pattern}
    Let $M \subseteq E$ be a matroid matching, so that $M$ is a forest.
    We extend $M$ arbitrarily to a spanning tree $M^*$.\footnote{We may add auxiliary edges not present in $G$, since they are used only for defining $M^*$.}
    Let $K=100\log n$.
        We choose a weight function $w: M^{*} \to \{ 0, 1 \}^K $ uniformly at random among all assignments satisfying the following constraints: for every $T\in\mathcal{T}_A$, $\displaystyle \bigoplus_{\beta\in T} w(\beta)=0$. \footnote{This distribution can be sampled by choosing all but one value in each $T\in\mathcal{T}_A$ uniformly at random, and then setting the remaining value so that the XOR over $T$ is zero. Here, we regard $w(e)$ as a $K$-bit integer when implementing XOR operations.}
    Here, $\oplus$ denotes bitwise XOR.
    We call the resulting weighted tree $(M^*,w)$ the \emph{pattern tree}.
    
    For an edge $e=\{u,v\}\notin M^*$, let $C(e)$ denote its fundamental cycle with respect to $M^*$, i.e., the union of $e$ and the unique $u$-$v$ path in $M^*$.
    We define the \emph{pattern} of an edge $e$ by
    \begin{equation*}
    p(e)=
    \begin{cases}
    w(e), & \text{if } e\in M^* ,\\
    \displaystyle \bigoplus_{\beta\in C(e)\setminus \{e\}} w(\beta), & \text{if } e\notin M^* .
    \end{cases}
    \end{equation*}
\end{restatable}

\begin{restatable}{remark}{simpleobs} \label{rem:onservation}
    Viewing each edge as a vector in $\mathbb{F}_2^V$, for every edge $e\in E\setminus M^*$, we have
$\displaystyle e=\sum_{\beta\in C(e)\setminus\{e\}}\beta$.
    Since $M^*$ is a spanning tree, we view this identity as the representation of $e$ in the basis $M^*$.
    Thus, the definition of $p(e)$ is obtained by applying the weight function $w$ to this tree-basis representation and taking the bitwise XOR.

    We also note that the sets $T_0,\ldots,T_k$ are pairwise disjoint.
\end{restatable}

Once the weights on $M^*$ are fixed, we can easily build a simple data structure, called the \emph{pattern-tree data structure}, in $O(n)$ preprocessing time that returns the pattern $p(e)$ of any edge $e$ in $O(1)$ time.

As a warm-up, we present how to compute the value $r(A \cup \ell) - r(A)$ for any line $\ell \in L$ in $\tO(1)$ time using the pattern-tree data structure.
The key observation is that the pattern $p(e)$ is a randomized encoding of the fundamental cycle representation of edge vector $e$.
That is, we express edge vector $e$ as the sum of tree edges on its fundamental cycle over $\mathbb{F}_2$, i.e., $\displaystyle e = \sum_{\beta \in C(e) \setminus \{ e \}} \beta$, and replace this sum by the XOR of their random weights.

    Recall that every vector in $A$ has the form $\alpha_i=\sum_{\beta\in T_i}\beta$ with $T_i\in\mathcal{T}_A$, and that the sets in $\mathcal{T}_A$ are pairwise disjoint.
    The constraint $\bigoplus_{\beta\in T}w(\beta)=0$ for every $T\in\mathcal{T}_A$ ensures that each $\alpha_i\in A$ has pattern value zero.
By linearity, every vector in $\Span(A)$ also has value zero.
Thus, if $e\in \Span(A)$, then $p(e)=0$.
Conversely, if $e\notin\Span(A)$, then the representation of $e$ in the $M^*$-basis, as in Remark~\ref{rem:onservation}, is not a sum of whole sets in $\mathcal{T}_A$, and the randomness of the weights implies $p(e)\neq 0$ with high probability.
Thus, $p(e)$ lets us test whether $e \in \Span (A)$.
The same idea applies to a line $\ell=\{e,\bar e\}\in L$.
From the two pattern values $p(e)$ and $p(\bar e)$, we can test whether $e$, $\bar e$, and $e+\bar e$ belong to $\Span(A)$ by checking $p(e)=0$, $p(\bar e)=0$, and $p(e)=p(\bar e)$, respectively.
These tests determine the value $r(A\cup \ell)-r(A)$.

\begin{restatable}{lemma}{warmup} \label{lem:pattern_warmup}
    Given the pattern-tree data structure, the following queries can be answered with high probability in $O(1)$ time:
    \begin{itemize}
        \item for any edge $e\in E$, determine whether $r(A \cup \{e\}) - r(A)$ is $0$ or $1$,
        \item for any line $\ell = \{e, \bar{e} \} \in L$,  determine whether $r(A \cup \ell) - r(A)$ is $0$, $1$ or $2$. 
    \end{itemize}
\end{restatable}

For each $i\in\{0,\ldots,k\}$, we write $c_i=|M_i|$.
We show that, among all edges $e$ satisfying $r(E_i\cup A\cup \{e\})=r(E_i\cup A)$, there are only $O(c_i^2)$ possible values of $p(e)$.
This bound is the key to our fast algorithm.
For an edge $e$, write its edge vector using the spanning tree $M^*$ as $\displaystyle e=\sum_{\beta\in C(e)\setminus \{e\}}\beta$ over $\mathbb{F}_2$.
Recall that the pattern $p(e)$ is obtained by replacing this sum of edge vectors with the XOR of the corresponding random weights.
By Lemma~\ref{lem:structual_lemma}, the space $\Span(A\cup E_i)$ is generated by $A$ together with $M_i$, except in case~(iii), where one additional edge $x\in T_i\setminus M_i$ is needed.
The contribution of every vector in $A$ to the pattern is zero by the constraints in the definition of the pattern tree.
Thus, in cases~(i) and~(ii), the value $p(e)$ is determined only by how the path $C(e) \setminus \{ e \}$ intersects $M_i$.
After contracting every connected component of $M^*\setminus M_i$, the set $(C(e)\setminus\{e\})\cap M_i$ becomes a path in a tree with $c_i$ edges.
Since such a tree has only $O(c_i^2)$ paths, there are only $O(c_i^2)$ possible intersections.
In case~(iii), the use of XOR is crucial.
The additional generator $x$ contributes only a fixed XOR offset to the pattern
determined by $(C(e)\setminus\{e\})\cap M_i$.
Since this intersection has only $O(c_i^2)$ possibilities, the number of
possible patterns remains $O(c_i^2)$.

\begin{restatable}{lemma}{patternbound} \label{lem:pattern_bound}
    Fix $i\in\{0,\ldots,k\}$, and let $c_i=|M_i|$.
    Suppose that $r(A\cup E_i)>r(A)+1$.
    There is an algorithm that computes a set $\cP_i$ of patterns of size $O(c_i^2)$ such that the following hold:
    \begin{itemize}
        \item for every edge $e \in E$, if $r(E_i\cup A\cup \{e\})=r(E_i\cup A)$, then $p(e) \in \cP_i$, and
        \item with high probability, for every edge $e \in E$, if $r(E_i\cup A\cup \{e\})>r(E_i\cup A)$, then $p(e) \notin \cP_i$.
    \end{itemize}
    We call $\cP_i$ the \emph{pattern set} for $i\in \{0,\ldots,k\}$.
    The algorithm runs in $\tO(n + c_i^2)$ time. 
\end{restatable}

We now explain how the pattern sets $\cP_i$ are used to implement Line~\ref{line:nm_barrier} efficiently.
For a line $\ell=\{e,\bar{e}\}$, we call the pair $(p(e),p(\bar e))$ its \emph{line pattern}.
For a fixed  $i\in\{0,\ldots,k\}$, the condition $r(E_i\cup A\cup \ell)=r(E_i\cup A)$ holds if and only if both $r(E_i\cup A\cup \{e\})=r(E_i\cup A)$ and $r(E_i\cup A\cup {\bar e})=r(E_i\cup A)$ hold.
Thus, this condition can be tested by checking whether $p(e)\in \cP_i \text{ and } p(\bar e)\in \cP_i$.

Let $C$ be a threshold.
We first handle all $i\in\{0,\ldots,k\}$ with $c_i\leq C$ together.
We construct the pattern sets $\cP_i$ for all indices $i$ with $c_i\leq C$.
Using $\sum_i c_i=O(n)$ and $c_i \geq 1$, the total time for constructing these pattern sets is
\begin{equation*}
\sum_{i:\, c_i\leq C} \tO\left(
n + c_i^2
\right)
\leq \tO(n^2) + C  \sum_{i:\, c_i\leq C} \tO(c_i) =
\tO(n^2+Cn).
\end{equation*}
For each such $i\in\{0,\ldots,k\}$, define the line-pattern set $\cQ_i = \{(a,b) \mid a\in\cP_i,\ b\in\cP_i \}$.
Since $|\cP_i|=O(c_i^2)$, we have $|\cQ_i|=O(c_i^4)$.
We insert all elements of $\cQ_i$ into a global dictionary.
Since $\sum_i c_i=O(n)$, the number of inserted pairs is
\begin{equation*}
\sum_{i:\, c_i\leq C} O(c_i^4)
\leq
O(C^3)\sum_i c_i
=
O(C^3n).
\end{equation*}
After this preprocessing, for any line $\ell = \{e, \bar{e} \}$, we can check whether $(p(e), p(\bar{e})) \in \mathcal{Q}_i$ for some index $i$ with $c_i\leq C$ using a single dictionary lookup per line.
Hence, all $m$ lines can be processed in $\tO(m)$ time.
Thus, all indices with $c_i \leq C$ are handled in $\tO(n^2 + m+C^3n)$ time.

It remains to handle all $i$ with $c_i>C$.
There are at most $O(n/C)$ such indices, again because $\sum_i c_i=O(n)$.
For each fixed such $i\in\{0,\ldots,k\}$, we can build a data structure in $\tO(n)$ time that tests, for any line $\ell\in L$, whether $r(E_i\cup A\cup \ell)=r(E_i\cup A)$ in $\tO(1)$ time.
See Lemma~\ref{lem:large_part_query}.
Therefore, by scanning all $m$ lines for each such index $i$, the total time for handling all indices with $c_i>C$ is $\tO(mn/C)$ per iteration.

Choosing $C=m^{1/4}$ gives $\tO(n^2 + m+C^3n+mn/C) = \tO(n^2 + nm^{3/4})$ time per iteration.

\subparagraph*{Further improvement for the span-membership tests}

We now explain how to implement the inner for loop (Lines~\ref{line:forloop}--\ref{line:forloop_end}) in $\tO(n^2 + m)$ time per iteration of the outer for loop.
Recall that, for each line $\ell\in L$, we need to check whether there exists an index $i\in\{0,\ldots,k\}$ with $r(E_i\cup A\cup\ell)=r(E_i\cup A)$.

The overall plan is the same as in the previous paragraph, where all indices $i$ with $c_i\leq C$ were handled together in $\tO(n^2+m+C^3n)$ time per iteration.
We now improve the bound to $\tO(n^2+m+Cn)$.
Since $c_i\leq n$ for every $i$, we can take $C=n$ and handle all indices together in $\tO(n^2+m)$ time, so there is no need to treat indices $i$ with $c_i>C$ separately.

We now sketch the key additional idea for handling all indices $i$ with
$c_i\leq C$ together in $\tO(n^2+m+Cn)$ time per iteration.
See Section~\ref{sec:small_handle} for the full case analysis.

A direct use of the pattern sets would enumerate all line patterns in $\cQ_i = \{(a,b) \mid a\in\cP_i,\ b\in\cP_i \}$, which costs $O(c_i^4)$ for a single index.
Instead, we exploit Lemma~\ref{lem:structual_lemma} more carefully and combine it with data structures such as heavy-light decomposition. 
As a result, we can perform the span-membership tests for all lines by storing only $O(c_i^2)$ patterns in dictionaries for each index $i$, instead of the $O(c_i^4)$ line patterns in $\cQ_i$.
Since $\sum_i c_i\leq n$, the total size of the dictionaries over all indices $i$ with $c_i\leq C$ is $\displaystyle \sum_{i:\,c_i\leq C}O(c_i^2)\leq C\sum_i O(c_i)=O(Cn)$.

The key additional observation is that, for every edge $e\notin\Span(A)$, there is at most one index $i$ such that $e\in\Span(A\cup M_i)$.
Thus, by storing the patterns corresponding to $\Span(A\cup M_i)$ in a dictionary, the pattern of an edge identifies the only possible index.
In cases~(i) and~(ii) of Lemma~\ref{lem:structual_lemma}, we have $\Span(A\cup E_i)=\Span(A\cup M_i)$.
Hence, to find an index $i$ in case~(i) or~(ii) with $\ell\subseteq\Span(A\cup E_i)$ for a line $\ell=\{e,\bar e\}$ with $e,\bar e\notin\Span(A)$, it suffices to check the at most one index identified by $p(e)$ and $p(\bar e)$.

It remains to handle case~(iii) of Lemma~\ref{lem:structual_lemma}.
In this case, we can also show that each line needs to be checked against only $O(1)$ candidate indices.
Consequently, all indices with $c_i\leq C$ can be handled in $\tO(n^2+m+Cn)$ time.

\subparagraph*{Further improvement by modifying Assadi's framework}

In Algorithm~\ref{alg:streaming_graphic_matroid_parity}, a maximum matroid matching of the sampled instance is computed from scratch in every iteration.
Since the sampled instance has $\tO(n/\eps)$ lines, the computation takes $\tO(n^2/\eps)$ time per iteration by Gabow--Stallmann's algorithm, and $\tO(n^2/\eps^2)$ time over all $T=O(\log m/\eps)$ iterations.
To reduce the total time, we slightly modify the framework itself.

In iteration $t$, we add the lines of the matroid matching $M^{(t-1)}$ of the previous iteration to the sampled instance, and, starting from $M^{(t-1)}$, we repeatedly augment the current matroid matching by the augmenting-path algorithm of Gabow--Stallmann until it becomes maximum (see Algorithm~\ref{alg:warm_start}).
Each augmentation takes $\tO(n/\eps)$ time, and the number of augmentations in iteration $t$ is $(|M^{(t)}|-|M^{(t-1)}|)/2$, so the total number of augmentations over all iterations is at most $|M^{(T)}|/2\leq n/2$.
In addition, in each iteration, the augmenting-path algorithm is applied once more to conclude that the current matroid matching is maximum, which also takes $\tO(n/\eps)$ time.
Hence, computing maximum matroid matchings in the sampled instances requires $\tO(n\cdot n/\eps+T\cdot n/\eps)=\tO(n^2/\eps+n/\eps^2)$ time in total over all iterations, which is $\tO(n^2/\eps)$ for $\eps\geq 1/n$.

The approximation guarantee is unaffected by the modification.
Intuitively, if we add some extra lines to a properly sampled set of lines and solve the problem on the enlarged instance, the outcome is never worse than that of solving the problem on the sampled lines alone.

Combining the modified framework with the span-membership tests in $\tO(n^2+m)$ time per iteration, we obtain a $(1-\eps)$-approximation algorithm running in $\tO((m+n^2)/\eps)$ time, and hence an exact algorithm running in $\tO(n^{3/2}\sqrt{m}+\sqrt{n}m)$ time.
\section{Fast exact algorithm for graphic matroid parity (Proof of Theorem~{\ref{thm:graphic_matroid_parity_exact_main}})} \label{sec:fast_exact_algo_gmp}

In this section, we derive the exact algorithm of Theorem~\ref{thm:graphic_matroid_parity_exact_main} from the fast $(1-\eps)$-approximation algorithm of Theorem~\ref{thm:graphic_matroid_parity_approx_main}.
In our exact algorithm, we first compute a $(1-\eps)$-approximate solution.
Since any matroid matching has size at most $n$, at most $O(\eps n)$ augmentations
are needed to obtain an exact solution.
We then repeatedly find augmenting paths using the algorithm of Gabow--Stallmann~\cite{gabow1985efficient} until no augmenting path remains, thereby obtaining an exact solution.
The proof of Theorem~\ref{thm:graphic_matroid_parity_approx_main} is deferred to Section~\ref{sec:fast_approx_algo}.
We restate Theorems~\ref{thm:graphic_matroid_parity_exact_main} and \ref{thm:graphic_matroid_parity_approx_main} here.

\mainexact*

\mainapprox*

Gabow--Stallmann~\cite{gabow1985efficient} gave an algorithm to find one augmenting path in $\tO(m)$ time.


\begin{restatable}{lemma}{augpath} [From {\cite{gabow1985efficient}}] \label{lem:from_gabow_stallmann}
    Given a graphic matroid matching $M \subseteq E$, there is a deterministic algorithm that  either finds a graphic matroid matching $M'$ such that $|M'| = |M| + 2$, or concludes that $M$ is maximum.
    The algorithm runs in $\tO(m)$ time.
\end{restatable}

\begin{remark} \label{rem:gsone}
    Lemma~\ref{lem:from_gabow_stallmann} is not explicitly stated as a standalone lemma in the paper of Gabow--Stallmann~\cite{gabow1985efficient}.
    Their paper gives a fast implementation of the augmenting-path algorithm from~\cite{gabow1986augmenting} using range-search data structures.
    The augmenting-path algorithm finds augmenting paths one at a time. Each search is restarted from scratch with respect to the current matroid matching.
\end{remark}

By applying the algorithm in Lemma~\ref{lem:from_gabow_stallmann} repeatedly, Gabow--Stallmann obtained an exact algorithm running in $\tO(nm)$ time.
In our $(1-\eps)$-approximation algorithm, we need to solve graphic matroid parity on sampled instances with $\tO(n/\eps)$ lines.
For this purpose, we use the following theorem as a black box.

\begin{theorem}[From {\cite{gabow1985efficient}}] \label{thm:gabow_stallmann}
    There is a deterministic algorithm that finds a maximum graphic matroid matching in $\tilde{O}(nm)$ time.
\end{theorem}

We now prove Theorem~\ref{thm:graphic_matroid_parity_exact_main} by combining Theorem~\ref{thm:graphic_matroid_parity_approx_main} with Lemma~\ref{lem:from_gabow_stallmann}.

\begin{proof}[Proof of Theorem~\ref{thm:graphic_matroid_parity_exact_main}]
We may assume $m \geq n$, since otherwise the algorithm of Theorem~\ref{thm:gabow_stallmann} already runs in $\tO(nm) = \tO(n^{3/2}\sqrt{m})$ time.

We first run the $(1-\eps)$-approximation algorithm of Theorem~\ref{thm:graphic_matroid_parity_approx_main} with parameter $\eps =\max\{\sqrt{n/m},\, n^{-1/2}\}$.
We then repeatedly augment the obtained matroid matching by the algorithm of Lemma~\ref{lem:from_gabow_stallmann} until it concludes that the matroid matching is maximum.
Since any matroid matching has size at most $n$, the number of remaining augmentations is at most $O(\eps n)$.
By Lemma~\ref{lem:from_gabow_stallmann}, each augmentation can be performed in $\tO(m)$ time.
Hence, the total running time is
$\displaystyle
    \tO\left(
        \frac{m + n^2}{\eps}
        +
        \eps n m
    \right).
$

First, suppose that $m\leq n^{2}$.
Then, $\eps=\sqrt{n/m}$, and the running time is
\begin{equation*}
\tO\left(
\frac{m^{3/2}}{\sqrt{n}}
+
n^{3/2}\sqrt{m}
\right)
=
\tO\left(n^{3/2}\sqrt{m}\right),
\end{equation*}
where the equality uses $m\leq n^{2}$.

Next, suppose that $m>n^{2}$.
Then, $\eps=n^{-1/2}$, and the running time is
\begin{equation*}
\tO\left(
\sqrt{n} m
+
n^{5/2}
\right)
=
\tO(\sqrt{n} m),
\end{equation*}
where the equality uses $m>n^{2}$.
Combining the two cases, the running time is
\begin{equation*}
\tO\left(n^{3/2}\sqrt{m}+\sqrt{n}m\right).
\end{equation*}
\end{proof}



\section{Minimum Covers: Definitions and Structural Properties} \label{sec:structures}

In this section, we introduce minimum covers and the structural properties that will be used later.
As preparation for defining minimum covers, we first describe the relationship between linear matroid parity and graphic matroid parity.
We then define the notion of a minimum cover for linear matroid parity and present structural properties of the minimum cover produced by Gabow--Stallmann's algorithm for graphic matroid parity~\cite{gabow1986augmenting,gabow1985efficient}.
These properties are used both in our adaptation of Assadi's framework (Section~\ref{sec:adaptation_of_assadi}) and in our algorithm for breaking the $O(nm)$-time complexity barrier (Section~\ref{sec:fast_approx_algo}).

Graphic matroid parity can be viewed as a special case of linear matroid parity via the representation over $\mathbb{F}_2$ (see Section~\ref{sec:preliminaries}).
Whenever we use terminology from linear matroid parity for a graphic matroid parity instance, we mean the following associated linear matroid parity instance.

\begin{restatable}{definition}{associatedlinear} \label{def:associated_linear_matroid_parity}
    Let $(G=(V,E),L)$ be a graphic matroid parity instance, where $G=(V,E)$ is a multigraph and $L$ is a set of lines.
    Let $B_G\in \mathbb{F}_2^{V\times E}$ be the vertex-edge incidence matrix of $G$ over $\mathbb{F}_2$.
    For an edge $e=\{u,v\}\in E$, the column of $B_G$ indexed by $e$ is the vector with $1$ in the coordinates $u$ and $v$, and $0$ elsewhere.
    The \emph{associated linear matroid parity instance} is the linear matroid parity instance represented by $B_G$, together with the same partition $L$ of $E$ into lines.
\end{restatable}

\begin{remark}
    In the associated linear matroid parity instance of
    Definition~\ref{def:associated_linear_matroid_parity}, each edge $e\in E$ is regarded as its corresponding column vector in $\mathbb{F}_2^V$.
    Throughout this paper, we use the same symbol $e$ for both the edge and this vector.
\end{remark}

Lov{\'a}sz~\cite{lovasz1980selecting-independent-lines} proved the following min--max theorem for linear matroid parity.
Recall that a parity partition of $E$ is a partition $(E_0,\ldots,E_k)$ of $E$ into unions of lines.
Equivalently, for every $e\in E$ and every $i\in \{0,\ldots,k\}$, we have $e\in E_i$ if and only if $\bar e\in E_i$.

\minmax*

The following notion of a cover naturally arises from the min--max theorem (Theorem~\ref{thm:min_max_linear_matroid_parity}).

\cover*

Gabow--Stallmann~\cite{gabow1986augmenting} gave an augmenting-path algorithm for linear matroid parity.
They used the min--max theorem (Theorem~\ref{thm:min_max_linear_matroid_parity}) in their linear matroid parity algorithm to certify the optimality of the matroid matching returned by the algorithm.

For graphic matroid parity, Gabow--Stallmann~\cite{gabow1986augmenting} applied this augmenting-path framework to the associated linear matroid parity instance of Definition~\ref{def:associated_linear_matroid_parity}, obtaining an $O(n^2m)$-time algorithm.
Using range-searching data structures, Gabow--Stallmann~\cite{gabow1985efficient} later gave a faster implementation of the required operations, reducing the time per augmentation to $\tO(m)$ and the total running time to $\tO(nm)$.



Gabow--Stallmann's algorithm for graphic matroid parity~\cite{gabow1986augmenting,gabow1985efficient}
also can find a minimum cover of the associated linear matroid parity instance, as stated below.

\begin{restatable}{lemma}{gsmincover} [From {\cite{gabow1986augmenting, gabow1985efficient}}] \label{lem:one_can_find_cover}
    Given a multigraph $G=(V, E)$, a set $L$ of lines, and a maximum matroid matching $M \subseteq E$, there is a deterministic algorithm that, in $\tO(m)$ time, computes a minimum cover $(A;E_0,\ldots,E_k)$ of the associated linear matroid parity instance.

    More precisely, the algorithm computes a parity partition $(E_0,\ldots,E_k)$ and sets $T_i\subseteq E_i$ for $i\in\{0,\ldots,k\}$, with $T_0=\emptyset$ and $T_i\neq\emptyset$ for $i\in\{1,\ldots,k\}$.
    Let $M_i=E_i\cap M$.
    For each $i\in\{1,\ldots,k\}$ such that $T_i\subseteq M_i$, define
    $\displaystyle \alpha_i=\sum_{\beta\in T_i}\beta$, where the sum is taken over the linear representation of the graphic matroid over $\mathbb{F}_2$.
    Then, the set $A=\{\alpha_i\mid i\in\{1,\ldots,k\},\, T_i\subseteq M_i\}$ together with the partition $(E_0,\ldots,E_k)$ forms a minimum cover.

    Moreover, the values $r(A\cup E_i)$ for all $i\in\{0,\ldots,k\}$ are determined by the construction.
    The construction also guarantees that the set $A$ is linearly independent, and hence $r(A)=|A|$.
\end{restatable}

For the minimum cover found in Lemma~\ref{lem:one_can_find_cover}, the values
$r(A\cup E_i)$ admit the following structural property.

\structures*

\begin{remark}
    We store each vector in $A$ only by its support, i.e., the set of coordinates whose entries are equal to $1$.

    Although Lemmas~\ref{lem:one_can_find_cover} and~\ref{lem:structual_lemma} are implicit in Gabow--Stallmann's papers~\cite{gabow1986augmenting,gabow1985efficient}, they are not stated there in this form.
    In particular, Gabow--Stallmann do not explicitly present an algorithm for outputting a minimum cover. 
    However, when an augmenting-path search terminates without finding an augmenting path, the labels, blossoms, buds, and tips maintained by the algorithm contain enough information to construct such a cover.
    For completeness, we provide the details in Appendix~\ref{sec:gabow_stallmann_min_cover_appendix}.
\end{remark}

\section{Adapting Assadi's framework to graphic matroid parity} \label{sec:adaptation_of_assadi}

In this section, we adapt Assadi's~\cite{assadi2025simple} framework to graphic matroid parity, obtain Algorithm~\ref{alg:streaming_graphic_matroid_parity}, and prove that it outputs a $(1-\eps)$-approximate matroid matching (Lemma~\ref{lem:approx_correctness}).
For convenience, Algorithm~\ref{alg:streaming_graphic_matroid_parity} is restated as Algorithm~\ref{alg:restate_streaming_graphic_matroid_parity}.

In Section~\ref{sec:warm_start}, we modify Assadi's framework itself to obtain Algorithm~\ref{alg:warm_start}, an algorithm for graphic matroid parity that admits a faster implementation.
Since the modification to the algorithm is simple, we recommend reading this section first.

\begin{lemma} \label{lem:approx_correctness}
    Algorithm~\ref{alg:streaming_graphic_matroid_parity} outputs a $(1-\eps)$-approximate graphic matroid matching with exponentially high probability\footnote{We write ``with exponentially high probability'' if the failure probability is at most $\Exp(-\Theta(n))$.}.
\end{lemma}

Before describing our graphic matroid parity algorithm (Algorithm~\ref{alg:streaming_graphic_matroid_parity}) and proving Lemma~\ref{lem:approx_correctness}, we first introduce the following notion of a \emph{pruned cover}, which is used in our algorithm.
We then prove a property of pruned covers (Lemma~\ref{lem:new_min_max}) that will be needed in the proof of Lemma~\ref{lem:approx_correctness}.

\goodcover*

The following lemma shows that restricting attention to pruned covers does not change the min--max characterization.

\begin{lemma} \label{lem:new_min_max}
    Let $\nu(L)$ denote the maximum number of lines in a matroid matching.
    The minimum value of
    \begin{equation*}
    r(A)+\sum_{i=0}^{k}\left\lfloor \frac{r(A\cup E_i)-r(A)}{2}\right\rfloor
    \end{equation*}
    over all pruned covers $(A;E_0,\ldots,E_k)$ is equal to $\nu(L)$.
\end{lemma}

\begin{proof}
Every cover in Definition~\ref{def:matroid_parity_cover} is a pruned cover, since each line $\ell$ is contained in one part $E_i$ of the parity partition, and hence $r(E_i\cup A\cup \ell)=r(E_i\cup A)$.
Thus, covers in Definition~\ref{def:matroid_parity_cover} form a subclass of pruned covers.
Therefore, by Theorem~\ref{thm:min_max_linear_matroid_parity}, the minimum value over all pruned covers is at most $\nu(L)$.

Conversely, let $(A;E_0,\ldots,E_k)$ be a pruned cover.
We construct a cover $(A;\widehat{E}_0,\ldots,\widehat{E}_{k'})$ whose value is no larger.
For each line $\ell\in L$, if $r(A\cup \ell)-r(A)\leq 1$, we make $\ell$ a singleton part of the parity partition.
Otherwise, choose an index $i$ such that $r(E_i\cup A\cup \ell)=r(E_i\cup A)$ and assign $\ell$ to $i$.
Let $\widehat{E}_i$ be the union of the lines assigned to $i$.
Then, for every $i \in \{0, \ldots, k\}$, $\widehat{E}_i\subseteq \Span(A\cup E_i)$, and hence $r(A\cup \widehat{E}_i)\leq r(A\cup E_i)$.
Moreover, every singleton part $\ell$ with $r(A\cup \ell)-r(A)\leq 1$ contributes $\displaystyle \left\lfloor \frac{r(A\cup \ell)-r(A)}{2}\right\rfloor=0$ to the value.
Thus, we have 
$$
r(A)+\sum_{i=0}^{k}\left\lfloor \frac{r(A\cup E_i)-r(A)}{2}\right\rfloor \geq r(A)+\sum_{i=0}^{k'}\left\lfloor \frac{r(A\cup \widehat{E}_i)-r(A)}{2}\right\rfloor \geq \nu(L),
$$
where the second inequality follows from Theorem~\ref{thm:min_max_linear_matroid_parity}.
Therefore, every pruned cover has value at least $\nu(L)$.
This completes the proof.
\end{proof}

We now describe our graphic matroid parity algorithm (Algorithm~\ref{alg:streaming_graphic_matroid_parity}).
It follows the same high-level framework of Assadi~\cite{assadi2025simple}.
It maintains an importance value $q_{\ell}^{(t)}$ for each line $\ell\in L$, initially set to $1$.
In each iteration $t$, the algorithm samples lines with probabilities proportional to their current importance values, and computes a maximum matroid matching $M^{(t)}\subseteq E$ together with a minimum cover of the sampled instance using the algorithm of Gabow--Stallmann~\cite{gabow1986augmenting,gabow1985efficient}.
It then removes all parts $E_i$ with $r(A\cup E_i)-r(A)\leq 1$, obtaining a pruned cover.

The algorithm next scans the lines $\ell\in L$ one by one and updates their importance values based on whether they are covered by this pruned cover.
If a line $\ell$ is covered by the pruned cover, its importance value is kept unchanged.
Otherwise, its importance value is doubled.
After $T=4\log m/\eps$ iterations, the algorithm outputs the largest matroid matching among $M^{(1)},\ldots,M^{(T)}$.

\begin{algorithm}[t]
    \caption{Restatement of {Algorithm~\ref{alg:streaming_graphic_matroid_parity}}: A $(1-\eps)$-approximation algorithm for graphic matroid parity}
    \label{alg:restate_streaming_graphic_matroid_parity}
    Let $q_{\ell}^{(1)} \gets 1$, $Q^{(1)} \gets m$

    \For{$t = 1$ to $T := \frac{4}{\eps} \log m$ }{
        Sample each line $\ell \in L$ independently at random with probability $\displaystyle p_{\ell}^{(t)} = \min \left\{ 1,  \frac{100n \log(n)}{\eps} \cdot \frac{q_{\ell}^{(t)}}{Q^{(t)}} \right\}$. \label{line:sample2} \\
        Apply the algorithm of Gabow--Stallmann~\cite{gabow1986augmenting, gabow1985efficient} to compute a maximum matroid matching $M^{(t)} \subseteq E$ and a minimum cover $(A;E_0, \ldots,E_{k'})$ of the sampled instance. \label{line:sample_solve2} \\
        Remove every set $E_i$ satisfying $r(A \cup E_i)-r(A) \leq 1$ from the cover, and reindex the remaining sets as $(A;E_0,\ldots,E_k)$. \label{line:removal2} \\
        \For{$\ell \in L$ \label{line:forloop2}}{
            \If{($r(A \cup \ell ) - r(A) \leq 1$) or (there exists $i \in \{ 0, \ldots, k\}$ such that $r(E_i \cup A \cup \ell ) = r(E_i \cup A)$) \label{line:nm_barrier2}}{
                $q_{\ell}^{(t + 1)} \gets q_{\ell}^{(t)}$
            }
            \Else {
                $q_{\ell}^{(t + 1)} \gets q_{\ell}^{(t)} \cdot 2$
            }
        }
        $Q^{(t + 1)} \gets \sum_{\ell \in L} q_{\ell}^{(t + 1)}$
    }

    \Return{a largest matroid matching among $M^{(1)},\ldots,M^{(T)}$}
\end{algorithm}

We now prepare several lemmas needed for the proof of Lemma~\ref{lem:approx_correctness}, which gives the $(1-\eps)$-approximation guarantee of Algorithm~\ref{alg:streaming_graphic_matroid_parity}.
The proof follows the same overall strategy as Assadi's analysis~\cite{assadi2025simple}.
The only substantial new ingredient is Lemma~\ref{lem:candidate_bounds}, which provides the counting bound for pruned covers.
Once this counting lemma is in place, the proofs of Lemmas~\ref{lem:counterpate1} and~\ref{lem:counterpate2} are almost identical to the corresponding proofs in Assadi~\cite{assadi2025simple}.

A pruned cover $C=(A;E_0,\ldots,E_k)$ is called a \emph{candidate pruned cover} if it can be obtained in Line~\ref{line:removal} of Algorithm~\ref{alg:streaming_graphic_matroid_parity} for some choice of the sampled lines.
For a pruned cover $C=(A;E_0,\ldots,E_k)$, define its signature as
$$\sig(C)=\left(\Span(A);\Span(A\cup E_0),\ldots,\Span(A\cup E_k)\right).$$


The next lemma gives a family $\mathcal{S}$ of signatures that contains the signature of every candidate pruned cover.
The key point is that $|\mathcal{S}|\leq \exp(O(n\log n))$, which allows a union bound over a subset of $\mathcal{S}$ in the proof of Lemma~\ref{lem:counterpate1}.
The family $\mathcal{S}$ depends only on $G$ and $L$, and it contains the signature of the pruned cover obtained from the minimum cover of any subset $L'\subseteq L$ of lines, not only of the set of sampled lines.
The generality with respect to $L'$ will be used in Section~\ref{sec:warm_start}.
In the proof, we use the structural property of covers produced by Gabow--Stallmann from Section~\ref{sec:structures}.

\begin{lemma} \label{lem:candidate_bounds}
    Assume that $m=|L|\leq n^4$.
    There is a family $\mathcal{S}$ of signatures, depending only on $G$ and $L$, with $|\mathcal{S}|\leq n^{10n}$ and the following property.
    Let $L'\subseteq L$, let $M$ be a maximum matroid matching of the instance $L'$, and let $(A;E_0,\ldots,E_{k'})$ be the minimum cover of $L'$ computed from $M$ by Lemma~\ref{lem:one_can_find_cover}.
    Then, the signature of the pruned cover obtained from $(A;E_0,\ldots,E_{k'})$ by removing every part $E_i$ with $r(A\cup E_i)-r(A)\leq 1$ belongs to $\mathcal{S}$.
\end{lemma}

\begin{proof}
    We construct the family $\mathcal{S}$ as follows.
    Each member of $\mathcal{S}$ is specified by a sequence of four choices described below.
    We describe the choices one by one, bounding the number of possibilities at each step, which yields the bound on $|\mathcal{S}|$.
    We then show that, for every $L'\subseteq L$, the signature of the pruned cover obtained from the minimum cover of $L'$ computed by Lemma~\ref{lem:one_can_find_cover} arises from one of these sequences of choices, and hence belongs to $\mathcal{S}$.

    First, choose a matroid matching $M\subseteq E$ consisting of at most $n/2$ lines.
    The number of possible choices of $M$ is at most
    \begin{equation*}
        \sum_{j=0}^{\lfloor n/2\rfloor} \binom{m}{j}
        \leq
        \sum_{j=0}^{\lfloor n/2\rfloor} m^j
        \leq
        (n+1)m^{n/2}
        \leq
        n^{3n},
    \end{equation*}
    where we use $m\leq n^4$ and $n\geq 2$.

    Second, choose a partition of $M$ into at most $n$ labeled parts $M_0,\ldots,M_{k'}$.
    Since $|M|\leq n$, the number of possible such partitions is at most $n^n$, by assigning each element of $M$ to one of at most $n$ labeled parts.

    Third, for each part $M_i$, choose either nothing or a subset $T_i\subseteq M_i$, and let $A$ be the set of the vectors $\alpha_i=\sum_{\beta\in T_i}\beta$ over the parts for which a subset $T_i$ is chosen.
    For a fixed part $M_i$, there are at most $2^{|M_i|}+1$ choices, and hence the number of choices over all parts is at most $\prod_i(2^{|M_i|}+1)\leq 3^{|M|}\leq 3^n$.

    Fourth, for each $i\in\{0,\ldots,k'\}$, we choose the subspace $\Span(A\cup E_i)$ associated with the part $M_i$.
    Lemma~\ref{lem:structual_lemma} shows that, for a part $E_i$ with $r(A\cup E_i)-r(A)>1$, $\Span(A\cup E_i)$ is generated either by $A\cup M_i$, by $(A\setminus\{\alpha_i\})\cup M_i$, or by $A\cup M_i\cup\{x\}$ for some $x\in E$.
    Accordingly, for each part $M_i$, we choose one of the following $2m+3$ possibilities.
    The first possibility is that $r(A\cup E_i)-r(A)\leq 1$, in which case the part does not appear in the signature.
    The next two possibilities are that the part is assigned the subspace $\Span(A\cup M_i)$ or the subspace $\Span((A\setminus\{\alpha_i\})\cup M_i)$.
    The remaining $2m$ possibilities are that the part is assigned the subspace $\Span(A\cup M_i\cup\{x\})$ for an element $x\in E$, where we use $|E|=2m$.
    The signature determined by the four choices is $(\Span(A);U_{i_0},\ldots,U_{i_k})$, where $i_0<\cdots<i_k$ are the indices of the parts for which a subspace is assigned and $U_i$ is the subspace assigned to the part $M_i$.
    Therefore,
    \begin{equation*}
        |\mathcal{S}|
        \leq
        n^{3n}\cdot n^n \cdot 3^n \cdot (2m+3)^n
        \leq
        n^{3n}\cdot n^n \cdot 3^n \cdot (3n^4)^n
        =
        9^n n^{8n}
        \leq
        n^{10n},
    \end{equation*}
    where we use $m\leq n^4$ and $n\geq 3$.
    By construction, $\mathcal{S}$ depends only on $G$ and $L$.

    We now show that, for every $L'\subseteq L$, the signature of the pruned cover obtained from the minimum cover of $L'$ computed by Lemma~\ref{lem:one_can_find_cover} belongs to $\mathcal{S}$.
    Let $E'=\bigcup_{\ell\in L'}\ell$ be the edge set of the instance $L'$, let $M$ be a maximum matroid matching of $L'$, and let $(A;E_0,\ldots,E_{k'})$ be the minimum cover of $L'$ computed from $M$ by Lemma~\ref{lem:one_can_find_cover}.
    Let $T_i\subseteq E_i$, $M_i=E_i\cap M$, and $\alpha_i$ be as in Lemma~\ref{lem:one_can_find_cover}.
    Since $M$ is a matroid matching consisting of at most $n/2$ lines, it is one of the choices in the first step.

    Recall that the signature of the pruned cover consists of $\Span(A)$ and the subspaces $\Span(A\cup E_i)$ over the parts $E_i$ with $r(A\cup E_i)-r(A)>1$.
    For every such part, Lemma~\ref{lem:structual_lemma} gives $r(A\cup E_i)\leq r(A)+|M_i|+1$, and hence $M_i\neq\emptyset$.
    Since $(E_0,\ldots,E_{k'})$ is a partition of $E'$ and $M\subseteq E'$, the nonempty sets among $M_0,\ldots,M_{k'}$ form a partition of $M$ into at most $|M|\leq n$ parts.
    Labeling these parts in increasing order of $i$ gives one of the choices in the second step, and every part $E_i$ with $r(A\cup E_i)-r(A)>1$ corresponds to one of the labeled parts.

    By Lemma~\ref{lem:one_can_find_cover}, $A=\{\alpha_i\mid i\in\{1,\ldots,k'\},\,T_i\subseteq M_i\}$ with $\alpha_i=\sum_{\beta\in T_i}\beta$, which is one of the choices in the third step.

    Finally, consider an index $i$ corresponding to one of the labeled parts.
    If $r(A\cup E_i)-r(A)\leq 1$, then $E_i$ does not appear in the pruned cover, which corresponds to the first possibility in the fourth step.
    Otherwise, Lemma~\ref{lem:structual_lemma} shows that $\Span(A\cup E_i)$ is generated either by $A\cup M_i$, by $(A\setminus\{\alpha_i\})\cup M_i$, or by $A\cup M_i\cup\{x\}$ for some $x\in E$, which corresponds to one of the remaining possibilities in the fourth step.
    Therefore, the signature of the pruned cover arises from the four choices, and hence belongs to $\mathcal{S}$.
\end{proof}

We first prove the analogue of \cite[Lemma~3.3]{assadi2025simple}, which gives an upper bound on the total importance.

\begin{lemma}[Counterpart of {\cite[Lemma 3.3]{assadi2025simple}}] \label{lem:counterpate1}
    With exponentially high probability, $Q^{T + 1} \leq (1 + \eps/2)^T \cdot m$. 
\end{lemma}

\begin{proof}
    Fix an iteration $t$.
    Let $(A;E_0,\ldots,E_k)$ be the pruned cover obtained in this iteration, and let $F^{(t)} \subseteq L$ be the set of lines not covered by this pruned cover.
    That is, $\ell\in F^{(t)}$ if and only if $r(A\cup \ell)-r(A)>1$ and $r(E_i\cup A\cup \ell)>r(E_i\cup A)$ for every $i\in\{0,\ldots,k\}$.

    We first claim that, with exponentially high probability,
    \begin{equation*}
        \sum_{\ell\in F^{(t)}} q_{\ell}^{(t)}
        \leq
        \frac{\eps}{2} Q^{(t)} .
    \end{equation*}

    To prove the claim, condition on all randomness before iteration $t$.
    Then, the values $q_{\ell}^{(t)}$ and $Q^{(t)}$ are fixed, and the only remaining randomness is the sampling in iteration $t$.


    Let $\mathcal{S}$ be the family of signatures given by Lemma~\ref{lem:candidate_bounds}.
    Recall that $\mathcal{S}$ depends only on $G$ and $L$, and hence it does not depend on the randomness of the algorithm.
    For a signature $\sigma=(U;U_0,\ldots,U_k)\in\mathcal{S}$, let $F(\sigma)$ be the set of lines not covered by any candidate pruned cover with signature $\sigma$, that is, the set of lines $\ell\in L$ such that $r(U\cup\ell)-r(U)>1$ and $\ell\not\subseteq U_i$ for every $i\in\{0,\ldots,k\}$.
    We note that whether a line is covered depends only on the signature.
    Let $\sigma_t$ be the signature of the pruned cover obtained in iteration $t$.
    By Lemma~\ref{lem:candidate_bounds} applied to the set of sampled lines and $M^{(t)}$, we have $\sigma_t\in\mathcal{S}$, and $F^{(t)}=F(\sigma_t)$.

    Call a signature $\sigma$ bad if $\displaystyle \sum_{\ell\in F(\sigma)} q_{\ell}^{(t)} > \frac{\eps}{2}Q^{(t)}$.
    Fix a bad signature $\sigma \in \mathcal{S}$.
    If some line $\ell\in F(\sigma)$ has sampling probability $p_{\ell}^{(t)}=1$, then no pruned cover with signature $\sigma$ can be returned for the sampled instance.
    Thus, we may assume that $p_{\ell}^{(t)}<1$ for every $\ell\in F(\sigma)$.
    In this case, for every $\ell\in F(\sigma)$, we have
    $\displaystyle p_{\ell}^{(t)} = \frac{100n\log n}{\eps}\cdot \frac{q_{\ell}^{(t)}}{Q^{(t)}}$.
    Hence, the probability that no line in $F(\sigma)$ is sampled is 
    \begin{equation*}
        \prod_{\ell\in F(\sigma)}(1-p_{\ell}^{(t)})
        \leq
        \exp\left(-\sum_{\ell\in F(\sigma)}p_{\ell}^{(t)}\right)
        =
        \exp\left(
            - \frac{100n\log n}{\eps Q^{(t)}}
            \sum_{\ell\in F(\sigma)} q_{\ell}^{(t)}
        \right)
        <
        \exp(-50n\log n).
    \end{equation*}
    If some line of $F(\sigma)$ is sampled, then no tuple $(A;E_0, \ldots, E_k)$ with signature $\sigma$ can be a pruned cover for the sampled instance, since that sampled line is left uncovered.
    Therefore, for every bad signature $\sigma$,
    \begin{equation*}
        \Pr[\sigma_t=\sigma]
        \leq 
        \Pr[\text{no line in $F(\sigma)$ is sampled}]
        \leq
        \exp(-50n\log n).
    \end{equation*}

    The claim fails exactly when the returned signature $\sigma_t$ is bad.
    By a union bound over all bad signatures in $\mathcal{S}$, together with $|\mathcal{S}|\leq n^{10n}$ from Lemma~\ref{lem:candidate_bounds},
    \begin{equation*}
        \Pr\left[
            \sum_{\ell\in F^{(t)}} q_{\ell}^{(t)}
            >
            \frac{\eps}{2}Q^{(t)}
        \right]
        \leq
        |\mathcal{S}|\cdot \exp(-50n\log n)
        \leq
        n^{10n}\cdot \exp(-50n\log n)
        \leq
        \exp(-40n\log n),
    \end{equation*}
    which proves the claim for iteration $t$.
    
    Taking another union bound over all $T$ iterations, we may assume that the claim holds for every iteration $t\in\{1,\ldots,T\}$.

    On the event that the claim holds for every iteration $t\in\{1,\ldots,T\}$, by the update rule of the algorithm and the claim,
    \begin{equation*}
        Q^{(t+1)}
        =
        \sum_{\ell\in F^{(t)}} 2q_{\ell}^{(t)}
        +
        \sum_{\ell\in L\setminus F^{(t)}} q_{\ell}^{(t)}
        =
        Q^{(t)}
        +
        \sum_{\ell\in F^{(t)}} q_{\ell}^{(t)}
        \leq
        \left(1+\frac{\eps}{2}\right)Q^{(t)} .
    \end{equation*}
    Since $Q^{(1)}=m$, iterating the above inequality over $T$ iterations gives
    $Q^{(T+1)} \leq (1+\eps/2)^T m$.
\end{proof}

We next prove the analogue of \cite[Lemma~3.4]{assadi2025simple}, which gives a lower bound on the importance of some line when none of the sampled instances yields a $(1-\eps)$-approximate matroid matching.

\begin{lemma}[Counterpart of {\cite[Lemma 3.4]{assadi2025simple}}] \label{lem:counterpate2}
    Suppose in every iteration $t \in \{1, \ldots, T \}$, we have $\displaystyle \frac{|M^{(t)}|}{2} < (1 - \eps) \cdot \nu(L)$.
    Then, there exists at least one line $\ell \in L$ such that $q_{\ell}^{T + 1} \geq 2^{\eps \cdot T}$.
\end{lemma}

\begin{proof}
    Let $M^* \subseteq E$ be a maximum matroid matching, and let
    $L^* = \{\ell\in L \mid \ell\subseteq M^*\}$.
    Then, $|L^*|=\nu(L)$ and $|M^*|=2\nu(L)$.

    Fix an iteration $t \in \{ 1, \ldots, T \}$, and let $(A;E_0,\ldots,E_k)$ be the pruned cover obtained in this iteration.
    Let $L_{\mathrm{cov}}^{(t)}$ be the set of lines covered by this pruned cover, and define
    $M_{\mathrm{cov}}^* = \bigcup_{\ell\in L^*\cap L_{\mathrm{cov}}^{(t)}} \ell$.
    Since $M^*$ is a matroid matching, $M_{\mathrm{cov}}^*$ is also a matroid matching in the restricted instance $L_{\mathrm{cov}}^{(t)}$.

    The tuple $(A;E_0,\ldots,E_k)$ is a pruned cover for $L_{\mathrm{cov}}^{(t)}$.
    Hence, by Lemma~\ref{lem:new_min_max}, we have
    $$\frac{|M_{\mathrm{cov}}^*|}{2} \leq r(A)+\sum_{i=0}^{k}\left\lfloor \frac{r(A\cup E_i)-r(A)}{2}\right\rfloor.$$

    Let $(A;\widehat E_0,\ldots,\widehat E_{k'})$ be the minimum cover of the sampled instance before pruning, and let $(A;E_0,\ldots,E_k)$ be the pruned cover obtained from it.
    Here, we have
    \begin{equation*}
        r(A)+\sum_{i=0}^{k}
        \left\lfloor
        \frac{r(A\cup E_i)-r(A)}{2}
        \right\rfloor
        =
        r(A)+\sum_{i=0}^{k'}
        \left\lfloor
        \frac{r(A\cup \widehat{E}_i)-r(A)}{2}
        \right\rfloor
        =
        \frac{|M^{(t)}|}{2}.
    \end{equation*}
        The first equality follows because every set $\widehat{E}_i$ deleted in Line~\ref{line:removal} satisfies $r(A\cup\widehat{E}_i)-r(A)\leq 1$, and hence $\displaystyle \left\lfloor \frac{r(A\cup\widehat{E}_i)-r(A)}{2}\right\rfloor=0$.
    The second equality follows from the min--max theorem (Theorem~\ref{thm:min_max_linear_matroid_parity}) for the sampled instance.
    Therefore,
    \begin{equation*}
        \frac{|M_{\mathrm{cov}}^*|}{2}
        \leq
        \frac{|M^{(t)}|}{2}
        <
        (1-\eps) \cdot \nu(L).
    \end{equation*}
    Thus, in each iteration, at least $\eps\cdot\nu(L)$ lines of $L^*$ are not covered by the pruned cover.
    
    By an average argument, this implies that there exists at least one line $\ell \in L$ that is not covered in at least $\eps \cdot T$ iterations. Hence, by the update rule of the algorithm, the importance of this line increase to at least $2^{\eps \cdot T}$, which completes the proof.
\end{proof}

We now prove Lemma~\ref{lem:approx_correctness} by combining
Lemma~\ref{lem:counterpate1} and Lemma~\ref{lem:counterpate2}.
The argument is identical to \cite[Proof of Theorem~3.2]{assadi2025simple}.

\begin{proof}[Proof of Lemma~\ref{lem:approx_correctness}]
    Assume that the event in Lemma~\ref{lem:counterpate1} holds, i.e., $Q^{(T+1)} \leq (1+\eps/2)^T \cdot m$.
    Suppose, toward a contradiction, that Algorithm~\ref{alg:streaming_graphic_matroid_parity} does not find a $(1-\eps)$-approximate solution.
    Since the algorithm returns the largest matching among $M^{(1)},\ldots,M^{(T)}$, this means that, for every iteration $t\in\{1,\ldots,T\}$, we have
    $\displaystyle |M^{(t)}|/2 < (1-\eps) \cdot \nu(L)$.

    By Lemma~\ref{lem:counterpate2}, there exists a line $\ell\in L$ such that
    $q_{\ell}^{(T+1)} \geq 2^{\eps T}$.
    On the other hand, we have $q_{\ell}^{(T+1)} \leq Q^{(T+1)}$.
    Therefore,
    \begin{equation*}
        2^{\eps T}
        \leq
        q_{\ell}^{(T+1)}
        \leq
        Q^{(T+1)}
        \leq
        (1+\eps/2)^T \cdot m.
    \end{equation*}
    Using $(1+x)<2^{3x/2}$ for $x>0$, we have
    \begin{equation*}
        (1+\eps/2)^T \cdot m
        <
        2^{3\eps T/4+\log m}.
    \end{equation*}
    Since $T=4\log m/\eps$, the right-hand side is equal to $2^{\eps T}$.
    This contradicts the previous inequality.
    Hence, on the event of Lemma~\ref{lem:counterpate1}, the algorithm outputs a $(1-\eps)$-approximate graphic matroid matching.
    Since this event holds with exponentially high probability, the lemma follows.
\end{proof}

We also record the following easy sampling bound, which will be used in the complexity analysis.

\begin{lemma} \label{lem:small_size}
    With high probability, in every iteration, the total number of sampled lines in Line~\ref{line:sample} is $\displaystyle \tO\left(\frac{n}{\eps}\right)$. 
\end{lemma}

\begin{proof}
    Fix an iteration $t$.
    Here, the expected number of lines sampled in this iteration is at most $100n\log n/\eps$. 
    By a Chernoff bound, the number of sampled lines in this iteration is $O(n\log n/\eps)$ with high probability.
    Taking a union bound over all $T=O(\log m/\eps)$ iterations, the same bound holds for every iteration with high probability.
    Thus, in every iteration, the number of sampled lines is $\tO(n/\eps)$ with high probability.
\end{proof}

We now have all the ingredients needed to prove the streaming algorithm result, so we prove
Theorem~\ref{thm:graphic_matroid_parity_streaming_main}, which we restate here.
For simplicity, we state the space bound as $\tO(n/\eps^2)$, although a more careful implementation gives $\tO(n/\eps)$ space.

\mainstreaming*

\begin{proof}[Proof of Theorem~\ref{thm:graphic_matroid_parity_streaming_main}]
    We implement Algorithm~\ref{alg:streaming_graphic_matroid_parity} in the streaming model.
    At the beginning of iteration $t$, we store the pruned covers obtained in the previous iterations.
    Then, when a line $\ell\in L$ arrives in the stream, we can determine in which previous iterations the line $\ell$ was not covered, and hence compute its current importance value $q_{\ell}^{(t)}$.

    Each iteration can be implemented in $O(1)$ passes.
    In one pass, we compute $Q^{(t)}=\sum_{\ell\in L}q_{\ell}^{(t)}$.
    In another pass, we sample each line with probability $p_{\ell}^{(t)}$ and store the sampled instance.
    By Lemma~\ref{lem:small_size}, with high probability, the number of sampled lines is $\tO(n/\eps)$.
    On the stored sampled instance, we compute a maximum matroid matching $M^{(t)}$ and a minimum cover using the algorithm of Gabow--Stallmann~\cite{gabow1986augmenting,gabow1985efficient}, prune the cover, and store it for later iterations.

    Therefore, the space used in each iteration is $\tO(n/\eps)$, and storing the covers from all $T=O(\log n/\eps)$ iterations uses $\tO(n/\eps^2)$ space.
    The number of passes is $O(T)=O(\log n/\eps)$.
    Finally, by Lemma~\ref{lem:approx_correctness}, the algorithm outputs a $(1-\eps)$-approximate graphic matroid matching with high probability.
\end{proof}
\section{Modifying Assadi's framework for a faster implementation} \label{sec:warm_start}

In this section, we present Algorithm~\ref{alg:warm_start}, an algorithm for graphic matroid parity that admits a faster implementation than Algorithm~\ref{alg:streaming_graphic_matroid_parity}.
To this end, we slightly modify Assadi's~\cite{assadi2025simple} framework itself.

In Line~\ref{line:sample_solve} of Algorithm~\ref{alg:streaming_graphic_matroid_parity}, a maximum matroid matching of the sampled instance is computed from scratch in every iteration.
Since the number of sampled lines is $\tO(n/\eps)$ (Lemma~\ref{lem:small_size}), the computation takes $\tO(n^2/\eps)$ time per iteration by Theorem~\ref{thm:gabow_stallmann}.
Over all $T=O(\log m/\eps)$ iterations, the total time is $\tO(n^2/\eps^2)$.

Using a simple idea, we reduce the total time to $\tO(n^2/\eps + n/\eps^2)$.
Roughly speaking, instead of computing a maximum matroid matching from scratch in every iteration, we reuse the matroid matching computed in the previous iteration.
More precisely, let $L^{(t)}$ be the set of sampled lines together with the lines of the matroid matching $M^{(t-1)}$ of the previous iteration.
Starting from $M^{(t-1)}$, which is a matroid matching of the instance $L^{(t)}$, we repeatedly augment the current matroid matching by the algorithm of Lemma~\ref{lem:from_gabow_stallmann} until it becomes a maximum matroid matching of the instance $L^{(t)}$.
Then, in iteration $t$, the number of successful augmentations is $(|M^{(t)}|-|M^{(t-1)}|)/2$, and hence the total number of successful augmentations over all iterations is $(|M^{(T)}|-|M^{(0)}|)/2\leq n/2$.
The modified algorithm is given as Algorithm~\ref{alg:warm_start}.

\begin{algorithm}[t]
    \caption{A modified version of Algorithm~\ref{alg:streaming_graphic_matroid_parity} that reuses the matroid matching of the previous iteration}
    \label{alg:warm_start}
    Let $q_{\ell}^{(1)} \gets 1$, $Q^{(1)} \gets m$, $M^{(0)} \gets \emptyset$

    \For{$t = 1$ to $T := \frac{4}{\eps} \log m$ }{
        Sample each line $\ell \in L$ independently at random with probability $\displaystyle p_{\ell}^{(t)} = \min \left\{ 1,  \frac{100n \log(n)}{\eps} \cdot \frac{q_{\ell}^{(t)}}{Q^{(t)}} \right\}$, and let $S^{(t)} \subseteq L$ be the set of sampled lines. \label{line:ws_sample} \\
        Let $L^{(t)} \gets S^{(t)} \cup \{ \ell \in L \mid \ell \subseteq M^{(t-1)} \}$. \label{line:ws_instance} \\
        Starting from $M^{(t-1)}$, repeatedly apply the algorithm of Lemma~\ref{lem:from_gabow_stallmann} to the instance $L^{(t)}$ until it concludes that the current matroid matching is maximum. Let $M^{(t)} \subseteq E$ be the resulting maximum matroid matching of $L^{(t)}$, and compute a minimum cover $(A;E_0, \ldots,E_{k'})$ of $L^{(t)}$ by Lemma~\ref{lem:one_can_find_cover}. \label{line:ws_solve} \\
        Remove every set $E_i$ satisfying $r(A \cup E_i)-r(A) \leq 1$ from the cover, and reindex the remaining sets as $(A;E_0,\ldots,E_k)$. \label{line:ws_removal} \\
        \For{$\ell \in L$ \label{line:ws_forloop}}{
            \If{($r(A \cup \ell ) - r(A) \leq 1$) or (there exists $i \in \{ 0, \ldots, k\}$ such that $r(E_i \cup A \cup \ell ) = r(E_i \cup A)$) \label{line:ws_nm_barrier}}{
                $q_{\ell}^{(t + 1)} \gets q_{\ell}^{(t)}$
            }
            \Else {
                $q_{\ell}^{(t + 1)} \gets q_{\ell}^{(t)} \cdot 2$
            }
        }
        $Q^{(t + 1)} \gets \sum_{\ell \in L} q_{\ell}^{(t + 1)}$ \label{line:ws_forloop_end}
    }

    \Return{$M^{(T)}$}
\end{algorithm}

We first establish two simple properties of Algorithm~\ref{alg:warm_start}.

\begin{observation} \label{lem:warm_start_correctness}
    Algorithm~\ref{alg:warm_start} satisfies the following.
    \begin{enumerate}[(i)]
        \item For every $t\in\{1,\ldots,T\}$, $M^{(t-1)}$ is a matroid matching of the instance $L^{(t)}$.
        Consequently, $|M^{(1)}|\leq |M^{(2)}|\leq\cdots\leq |M^{(T)}|$, and the output $M^{(T)}$ is a largest matroid matching among $M^{(1)},\ldots,M^{(T)}$.
        \item With high probability, $|L^{(t)}|=\tO(n/\eps)$ for every $t\in\{1,\ldots,T\}$.
    \end{enumerate}
\end{observation}

\begin{proof}
    We first prove (i).
    Fix $t\in\{1,\ldots,T\}$.
    By Line~\ref{line:ws_instance}, every line of $M^{(t-1)}$ belongs to $L^{(t)}$.
    Moreover, $M^{(t-1)}$ is a forest, since it is empty when $t=1$ and is a matroid matching of $L^{(t-1)}$ when $t\geq 2$.
    Hence, $M^{(t-1)}$ is a matroid matching of the instance $L^{(t)}$.
    Since $M^{(t)}$ is a maximum matroid matching of $L^{(t)}$ by Line~\ref{line:ws_solve}, we have $|M^{(t)}|\geq |M^{(t-1)}|$.

    We next prove (ii).
    Line~\ref{line:ws_sample} is identical to Line~\ref{line:sample} of Algorithm~\ref{alg:streaming_graphic_matroid_parity}, so Lemma~\ref{lem:small_size} gives $|S^{(t)}|=\tO(n/\eps)$ for every $t$ with high probability.
    Since $M^{(t-1)}$ consists of at most $n/2$ lines and $\eps\leq 1$, we obtain $|L^{(t)}|\leq |S^{(t)}|+n/2=\tO(n/\eps)$.
\end{proof}

Despite the modification, Algorithm~\ref{alg:warm_start} still outputs a $(1-\eps)$-approximate graphic matroid matching.
For completeness, we give a full proof, but its essence is the same as the analysis of Algorithm~\ref{alg:streaming_graphic_matroid_parity} in Section~\ref{sec:adaptation_of_assadi}.
Intuitively, if we add some extra lines to a properly sampled set of lines and solve the problem on the enlarged instance, the outcome is never worse than solving the problem on the sampled lines alone.

\begin{lemma} \label{lem:ws_approx_correctness}
    Algorithm~\ref{alg:warm_start} outputs a $(1-\eps)$-approximate graphic matroid matching with exponentially high probability.
\end{lemma}

The proof is almost the same as the analysis of Section~\ref{sec:adaptation_of_assadi}.
We first prove the counterparts of Lemmas~\ref{lem:counterpate1} and~\ref{lem:counterpate2} for Algorithm~\ref{alg:warm_start}, and then prove Lemma~\ref{lem:ws_approx_correctness} by combining them.

The following lemma is the counterpart of Lemma~\ref{lem:counterpate1}.

\begin{lemma}[Counterpart of Lemma~\ref{lem:counterpate1}] \label{lem:ws_counterpate1}
    In Algorithm~\ref{alg:warm_start}, with exponentially high probability, $Q^{T + 1} \leq (1 + \eps/2)^T \cdot m$. 
\end{lemma}

The proof is the same as that of Lemma~\ref{lem:counterpate1}.
The only point to check is that every sampled line is covered by the pruned cover of iteration $t$, which was used in the proof of Lemma~\ref{lem:counterpate1}.
This still holds for Algorithm~\ref{alg:warm_start}, since the pruned cover of iteration $t$ covers every line of $L^{(t)}$ and $S^{(t)}\subseteq L^{(t)}$.

\begin{proof}[Proof of Lemma~\ref{lem:ws_counterpate1}]
    Fix an iteration $t$.
    Let $(A;E_0,\ldots,E_k)$ be the pruned cover obtained in this iteration, and let $F^{(t)} \subseteq L$ be the set of lines not covered by this pruned cover.
    That is, $\ell\in F^{(t)}$ if and only if $r(A\cup \ell)-r(A)>1$ and $r(E_i\cup A\cup \ell)>r(E_i\cup A)$ for every $i\in\{0,\ldots,k\}$.

    We first claim that, with exponentially high probability,
    \begin{equation*}
        \sum_{\ell\in F^{(t)}} q_{\ell}^{(t)}
        \leq
        \frac{\eps}{2} Q^{(t)} .
    \end{equation*}

    To prove the claim, condition on all randomness before iteration $t$.
    Then, the values $q_{\ell}^{(t)}$ and $Q^{(t)}$ are fixed.
        Moreover, the matroid matching $M^{(t-1)}$ is also fixed, and hence so is the set $\{\ell\in L\mid \ell\subseteq M^{(t-1)}\}$ added in Line~\ref{line:ws_instance}.
    The only remaining randomness is the sampling of $S^{(t)}$ in iteration $t$.

    Let $\mathcal{S}$ be the family of signatures given by Lemma~\ref{lem:candidate_bounds}.
    Recall that $\mathcal{S}$ depends only on $G$ and $L$, and hence it does not depend on the randomness of the algorithm.
    For a signature $\sigma=(U;U_0,\ldots,U_k)\in\mathcal{S}$, let $F(\sigma)$ be the set of lines not covered by any pruned cover with signature $\sigma$, that is, the set of lines $\ell\in L$ such that $r(U\cup\ell)-r(U)>1$ and $\ell\not\subseteq U_i$ for every $i\in\{0,\ldots,k\}$.
    We note that whether a line is covered depends only on the signature.
    Let $\sigma_t$ be the signature of the pruned cover obtained in iteration $t$.
    By Lemma~\ref{lem:candidate_bounds} applied to the instance $L^{(t)}$ and $M^{(t)}$, we have $\sigma_t\in\mathcal{S}$, and $F^{(t)}=F(\sigma_t)$.

    Call a signature $\sigma$ bad if $\displaystyle \sum_{\ell\in F(\sigma)} q_{\ell}^{(t)} > \frac{\eps}{2}Q^{(t)}$.
    Fix a bad signature $\sigma \in \mathcal{S}$.
    If some line $\ell\in F(\sigma)$ has sampling probability $p_{\ell}^{(t)}=1$, then no pruned cover with signature $\sigma$ can be returned for the instance $L^{(t)}$.
    Thus, we may assume that $p_{\ell}^{(t)}<1$ for every $\ell\in F(\sigma)$.
    In this case, for every $\ell\in F(\sigma)$, we have
    $\displaystyle p_{\ell}^{(t)} = \frac{100n\log n}{\eps}\cdot \frac{q_{\ell}^{(t)}}{Q^{(t)}}$.
    Hence, the probability that no line in $F(\sigma)$ is sampled is 
    \begin{equation*}
        \prod_{\ell\in F(\sigma)}(1-p_{\ell}^{(t)})
        \leq
        \exp\left(-\sum_{\ell\in F(\sigma)}p_{\ell}^{(t)}\right)
        =
        \exp\left(
            - \frac{100n\log n}{\eps Q^{(t)}}
            \sum_{\ell\in F(\sigma)} q_{\ell}^{(t)}
        \right)
        <
        \exp(-50n\log n).
    \end{equation*}
    If some line of $F(\sigma)$ is sampled, then it belongs to $L^{(t)}$ since $S^{(t)}\subseteq L^{(t)}$, and hence no tuple $(A;E_0, \ldots, E_k)$ with signature $\sigma$ can be a pruned cover for the instance $L^{(t)}$, since that sampled line is left uncovered.
    Therefore, for every bad signature $\sigma$,
    \begin{equation*}
        \Pr[\sigma_t=\sigma]
        \leq 
        \Pr[\text{no line in $F(\sigma)$ is sampled}]
        \leq
        \exp(-50n\log n).
    \end{equation*}

    The claim fails exactly when the returned signature $\sigma_t$ is bad.
    By a union bound over all bad signatures in $\mathcal{S}$, together with $|\mathcal{S}|\leq n^{10n}$ from Lemma~\ref{lem:candidate_bounds},
    \begin{equation*}
        \Pr\left[
            \sum_{\ell\in F^{(t)}} q_{\ell}^{(t)}
            >
            \frac{\eps}{2}Q^{(t)}
        \right]
        \leq
        |\mathcal{S}|\cdot \exp(-50n\log n)
        \leq
        n^{10n}\cdot \exp(-50n\log n)
        \leq
        \exp(-40n\log n),
    \end{equation*}
    which proves the claim for iteration $t$.
    
    Taking another union bound over all $T$ iterations, we may assume that the claim holds for every iteration $t\in\{1,\ldots,T\}$.

    On the event that the claim holds for every iteration $t\in\{1,\ldots,T\}$, by the update rule of the algorithm and the claim,
    \begin{equation*}
        Q^{(t+1)}
        =
        \sum_{\ell\in F^{(t)}} 2q_{\ell}^{(t)}
        +
        \sum_{\ell\in L\setminus F^{(t)}} q_{\ell}^{(t)}
        =
        Q^{(t)}
        +
        \sum_{\ell\in F^{(t)}} q_{\ell}^{(t)}
        \leq
        \left(1+\frac{\eps}{2}\right)Q^{(t)} .
    \end{equation*}
    Since $Q^{(1)}=m$, iterating the above inequality over $T$ iterations gives
    $Q^{(T+1)} \leq (1+\eps/2)^T m$.
\end{proof}

The following lemma is the counterpart of Lemma~\ref{lem:counterpate2}.

\begin{lemma}[Counterpart of Lemma~\ref{lem:counterpate2}] \label{lem:ws_counterpate2}
    Suppose that, in Algorithm~\ref{alg:warm_start}, $\displaystyle \frac{|M^{(t)}|}{2}<(1-\eps)\cdot\nu(L)$ for every $t\in\{1,\ldots,T\}$.
    Then, there exists a line $\ell\in L$ such that $q_{\ell}^{(T+1)}\geq 2^{\eps T}$.
\end{lemma}

The proof differs from that of Lemma~\ref{lem:counterpate2} only in that the min--max theorem is applied to the instance $L^{(t)}$ instead of the set of sampled lines.

\begin{proof}[Proof of Lemma~\ref{lem:ws_counterpate2}]
    Let $M^* \subseteq E$ be a maximum matroid matching, and let
    $L^* = \{\ell\in L \mid \ell\subseteq M^*\}$.
    Then, $|L^*|=\nu(L)$ and $|M^*|=2\nu(L)$.

    Fix an iteration $t \in \{ 1, \ldots, T \}$, and let $(A;E_0,\ldots,E_k)$ be the pruned cover obtained in this iteration.
    Let $L_{\mathrm{cov}}^{(t)}$ be the set of lines covered by this pruned cover, and define
    $M_{\mathrm{cov}}^* = \bigcup_{\ell\in L^*\cap L_{\mathrm{cov}}^{(t)}} \ell$.
    Since $M^*$ is a matroid matching, $M_{\mathrm{cov}}^*$ is also a matroid matching in the restricted instance $L_{\mathrm{cov}}^{(t)}$.

    The tuple $(A;E_0,\ldots,E_k)$ is a pruned cover for $L_{\mathrm{cov}}^{(t)}$.
    Hence, by Lemma~\ref{lem:new_min_max}, we have
    $$\frac{|M_{\mathrm{cov}}^*|}{2} \leq r(A)+\sum_{i=0}^{k}\left\lfloor \frac{r(A\cup E_i)-r(A)}{2}\right\rfloor.$$

    Let $(A;\widehat E_0,\ldots,\widehat E_{k'})$ be the minimum cover of the instance $L^{(t)}$ computed in Line~\ref{line:ws_solve}, and let $(A;E_0,\ldots,E_k)$ be the pruned cover obtained from it in Line~\ref{line:ws_removal}.
    Here, we have
    \begin{equation*}
        r(A)+\sum_{i=0}^{k}
        \left\lfloor
        \frac{r(A\cup E_i)-r(A)}{2}
        \right\rfloor
        =
        r(A)+\sum_{i=0}^{k'}
        \left\lfloor
        \frac{r(A\cup \widehat{E}_i)-r(A)}{2}
        \right\rfloor
        =
        \frac{|M^{(t)}|}{2}.
    \end{equation*}
        The first equality follows because every set $\widehat{E}_i$ deleted in Line~\ref{line:ws_removal} satisfies $r(A\cup\widehat{E}_i)-r(A)\leq 1$, and hence $\displaystyle \left\lfloor \frac{r(A\cup\widehat{E}_i)-r(A)}{2}\right\rfloor=0$.
    The second equality follows from the min--max theorem (Theorem~\ref{thm:min_max_linear_matroid_parity}) for the instance $L^{(t)}$, since $M^{(t)}$ is a maximum matroid matching of $L^{(t)}$.
    Therefore,
    \begin{equation*}
        \frac{|M_{\mathrm{cov}}^*|}{2}
        \leq
        \frac{|M^{(t)}|}{2}
        <
        (1-\eps) \cdot \nu(L).
    \end{equation*}
    Thus, in each iteration, at least $\eps\cdot\nu(L)$ lines of $L^*$ are not covered by the pruned cover.
    
    By an average argument, this implies that there exists at least one line $\ell \in L$ that is not covered in at least $\eps \cdot T$ iterations.
    Hence, by the update rule of the algorithm, the importance of this line increases to at least $2^{\eps \cdot T}$, which completes the proof.
\end{proof}

We now prove Lemma~\ref{lem:ws_approx_correctness} by combining Lemmas~\ref{lem:ws_counterpate1} and~\ref{lem:ws_counterpate2}.

\begin{proof}[Proof of Lemma~\ref{lem:ws_approx_correctness}]
    By Lemma~\ref{lem:ws_counterpate1}, with exponentially high probability, $Q^{(T+1)}\leq (1+\eps/2)^T\cdot m$.
    Assume that this event holds.
    Using $(1+x)<2^{3x/2}$ for $x>0$ and $T=4\log m/\eps$,
    \begin{equation*}
        Q^{(T+1)}
        \leq
        (1+\eps/2)^T\cdot m
        <
        2^{3\eps T/4+\log m}
        =
        2^{\eps T}.
    \end{equation*}
    Since $q_{\ell}^{(T+1)}\leq Q^{(T+1)}$ for every line $\ell\in L$, no line satisfies $q_{\ell}^{(T+1)}\geq 2^{\eps T}$.
    Hence, by Lemma~\ref{lem:ws_counterpate2}, there exists an iteration $t\in\{1,\ldots,T\}$ with $|M^{(t)}|/2\geq (1-\eps)\cdot\nu(L)$.
    By Observation~\ref{lem:warm_start_correctness}(i), $|M^{(T)}|\geq |M^{(t)}|$, and hence the output $M^{(T)}$ is a $(1-\eps)$-approximate graphic matroid matching.
\end{proof}

We next prove the lemma that is the key to the improvement in running time.
In Algorithm~\ref{alg:streaming_graphic_matroid_parity}, the corresponding part (Line~\ref{line:sample_solve}) takes $\tO(n^2/\eps^2)$ time in total.

\begin{lemma} \label{lem:warm_start}
    With high probability, the total running time of Lines~\ref{line:ws_instance}--\ref{line:ws_removal} of Algorithm~\ref{alg:warm_start} over all $T$ iterations is
    $\displaystyle \tO\left(\frac{n^2}{\eps}+\frac{n}{\eps^2}\right)$.
\end{lemma}

\begin{proof}
    Fix an iteration $t$, and let $a_t=(|M^{(t)}|-|M^{(t-1)}|)/2\geq 0$.
    In Line~\ref{line:ws_solve}, every application of the algorithm of Lemma~\ref{lem:from_gabow_stallmann} except the last one returns a matroid matching with exactly one more line than the current one, and the last one concludes that the current matroid matching is maximum.
    Hence, the algorithm of Lemma~\ref{lem:from_gabow_stallmann} is applied exactly $a_t+1$ times in iteration $t$.
    By Observation~\ref{lem:warm_start_correctness}(ii), with high probability, $|L^{(t)}|=\tO(n/\eps)$ for every iteration $t$.
    Applied to the instance $L^{(t)}$, each of the algorithms of Lemmas~\ref{lem:from_gabow_stallmann} and~\ref{lem:one_can_find_cover} takes $\tO(|L^{(t)}|)=\tO(n/\eps)$ time.
    By Lemma~\ref{lem:one_can_find_cover}, the cover comes with the values $r(A\cup E_i)$ for all $i\in\{0,\ldots,k'\}$ and $r(A)=|A|$, so Line~\ref{line:ws_removal} takes $O(|L^{(t)}|)$ time.
    Thus, Lines~\ref{line:ws_instance}--\ref{line:ws_removal} take $\tO((a_t+2)\,n/\eps)$ time in iteration $t$.

    Since $M^{(0)}=\emptyset$ and every matroid matching consists of at most $n/2$ lines, we have
    \begin{equation*}
        \sum_{t=1}^{T} a_t
        =
        \frac{|M^{(T)}|-|M^{(0)}|}{2}
        \leq
        \frac{n}{2}.
    \end{equation*}
    Summing over all $T=O(\log m/\eps)$ iterations, the total time is
    \begin{equation*}
        \sum_{t=1}^{T}\tO\left((a_t+2)\frac{n}{\eps}\right)
        =
        \tO\left((n+T)\frac{n}{\eps}\right)
        =
        \tO\left(\frac{n^2}{\eps}+\frac{n}{\eps^2}\right).
    \end{equation*}
\end{proof}

\section{Fast {$(1-\eps)$}-approximation algorithm for graphic matroid parity} \label{sec:fast_approx_algo}

In this section, we prove Theorem~\ref{thm:graphic_matroid_parity_approx_main}, which gives our fast $(1-\eps)$-approximation algorithm.
We restate the theorem here.

\mainapprox*

The algorithm is Algorithm~\ref{alg:warm_start}, and its approximation guarantee was proved in Section~\ref{sec:warm_start} (Lemma~\ref{lem:ws_approx_correctness}).
To prove our main approximation theorem (Theorem~\ref{thm:graphic_matroid_parity_approx_main}) it remains to perform efficiently the inner for loop (Lines~\ref{line:ws_forloop}--\ref{line:ws_forloop_end}), which updates the importance values of all lines.
In Section~\ref{sec:checking_span}, we prove the following lemma, which is the main technical contribution of the paper.
Once this lemma is established, Theorem~\ref{thm:graphic_matroid_parity_approx_main} follows by combining it with the results of the previous sections.

\begin{restatable}{lemma}{fastcheck} \label{lem:fast_span_checking}
    Fix an iteration $t$, and let $(A;E_0,\ldots,E_k)$ be the pruned cover obtained in this iteration.
    With high probability, the inner for loop (Lines~\ref{line:ws_forloop}--\ref{line:ws_forloop_end}) can be implemented in $\tO(n^2 + m)$ time.
    More precisely, with high probability, for all lines $\ell\in L$, we can decide whether
    \begin{equation*}
        r(A\cup \ell)-r(A)\leq 1
        \quad\text{or}\quad
        \text{there exists } i\in\{0,\ldots,k\}
        \text{ such that }
        r(E_i\cup A\cup \ell)=r(E_i\cup A)
    \end{equation*}
    in total $\tO(n^2 + m)$ time.
\end{restatable}

We have now assembled all the ingredients needed to obtain a fast $(1-\eps)$-approximation algorithm.
We first derive Theorem~\ref{thm:graphic_matroid_parity_approx_main} from Lemma~\ref{lem:fast_span_checking} and the results of the previous sections.

\begin{proof}[Proof of Theorem~\ref{thm:graphic_matroid_parity_approx_main}]
    Without loss of generality, we may assume that $\eps\geq 1/n$ and $m\leq n^4$.
    Indeed, if $\eps<1/n$, then a $(1-1/n)$-approximate matroid matching is a maximum matroid matching since $\nu(L)\leq n/2$, so we may replace $\eps$ by $1/n$.
    Also, we may remove duplicate lines, that is, lines represented by the same pair of vectors, since at most one of them can be used in a matroid matching, which leaves at most $O(n^4)$ lines.
    
    We now run Algorithm~\ref{alg:warm_start}.
    By Lemma~\ref{lem:ws_approx_correctness}, it outputs a $(1-\eps)$-approximate graphic matroid matching with exponentially high probability.
    It remains to bound the running time.

    Lines~\ref{line:ws_sample} and~\ref{line:ws_forloop_end} take $O(m)$ time per iteration.
    By Lemma~\ref{lem:warm_start}, with high probability, Lines~\ref{line:ws_instance}--\ref{line:ws_removal} take $\tO(n^2/\eps+n/\eps^2)=\tO(n^2/\eps)$ time in total over all iterations, where we use $\eps\geq 1/n$.
    By Lemma~\ref{lem:fast_span_checking}, with high probability, the inner for loop (Lines~\ref{line:ws_forloop}--\ref{line:ws_forloop_end}) takes $\tO(n^2+m)$ time in each iteration.
    Since $T=O(\log m/\eps)$, the total running time is
    \begin{equation*}
        \tO\left(\frac{n^2}{\eps}+T\cdot(n^2+m)\right)
        =
        \tO\left(\frac{m+n^2}{\eps}\right)
    \end{equation*}
    with high probability.
\end{proof}

\section{Fast span-membership tests via patterns (Proof of Lemma~\ref{lem:fast_span_checking})} \label{sec:checking_span}

In this section, we prove Lemma~\ref{lem:fast_span_checking}, which is the main technical contribution of the paper and is restated below.

\fastcheck*

We recommend first reading the technical overview in Section~\ref{sec:technical_overview}, where we sketch a simpler $\tO(n^2+nm^{3/4})$-time version of our approach.

At first glance, the inner for loop (Lines~\ref{line:ws_forloop}--\ref{line:ws_forloop_end}) of Algorithm~\ref{alg:warm_start} seems to require $\Omega(nm)$ time.
Since there are $m$ lines and $k+1$ parts, this gives $\Omega(km)$ rank-equality tests, which is $\Omega(nm)$ when $k=\Omega(n)$.
To avoid this bottleneck, we assign to each edge $e\in E$ a random $O(\log n)$-bit value, called its \emph{pattern}, which serves as a fingerprint of the edge with respect to the current pruned cover.
These patterns allow us to reinterpret the $\Omega(nm)$ rank-equality tests as a pattern-matching problem and handle many of them simultaneously, rather than checking them one by one.
To make this approach work, we crucially exploit the structural properties of the minimum cover produced by Gabow--Stallmann's algorithm~\cite{gabow1986augmenting,gabow1985efficient} presented in Section~\ref{sec:structures}.
The main technical novelty of this paper is the introduction of these patterns and their algorithmic use in updating the importance values efficiently.

We organize the proof as follows.
In Section~\ref{subsec:pattern}, we introduce the new notion of patterns.
In Section~\ref{subsec:ds}, we describe the data structures used to manipulate these patterns efficiently.
In Section~\ref{subsec:enumerate}, we show how to enumerate the relevant pattern set associated with each index $i$.
In Section~\ref{sec:small_handle}, we combine these ingredients to obtain the fast span-membership tests and prove Lemma~\ref{lem:fast_span_checking}.

\subparagraph*{Conventions.}
Throughout this section, we fix an iteration $t$ of Algorithm~\ref{alg:warm_start} and write $M=M^{(t)}$.
Let $(A;E_0,\ldots,E_{k'})$ be the minimum cover computed in Line~\ref{line:ws_solve}, and let $T_i\subseteq E_i$, $M_i=E_i\cap M$, and $\alpha_i$ be as in Lemma~\ref{lem:one_can_find_cover}.
Line~\ref{line:ws_removal} removes some parts and reindexes the remaining parts as $E_0,\ldots,E_k$, but it does not change $A$.
Each remaining part $E_i$ keeps its sets $T_i$ and $M_i$, and we use these sets with the new index $i$.
Since the three cases of Lemma~\ref{lem:structual_lemma} are distinguished by $T_i$ and $M_i$, each remaining part falls into exactly one of cases~(i), (ii), and~(iii), and the lemma applies to the pruned cover $(A;E_0,\ldots,E_k)$ with the new indices.
As in Section~\ref{sec:technical_overview}, let $\mathcal{T}_A$ be the family of the sets $T_i$ over all $\alpha_i\in A$, taken over all parts before Line~\ref{line:ws_removal}.
The sets in $\mathcal{T}_A$ are pairwise disjoint subsets of $M$, and a remaining part $E_i$ satisfies $T_i\in\mathcal{T}_A$ if and only if $E_i$ is in case~(ii).

\subsection{Patterns} \label{subsec:pattern}

In this subsection, we introduce the new notion of \emph{patterns}.
The purpose of patterns is to assign a random $O(\log n)$-bit value to each edge so that rank-equality tests can be performed using only these values.

\pattern*

In Definition~\ref{def:pattern}, we set $K=100\log n$ for concreteness.
To make the success probability in Lemma~\ref{lem:fast_span_checking} at least $1-1/n^{c}$ for any desired constant $c$, it suffices to set $K=C\log n$ for a sufficiently large constant $C$ depending on $c$.

    

We give the following remark, which will be useful for understanding the proofs of the subsequent lemmas.

\simpleobs*


The following lemma gives a simple data structure, which we call the \emph{pattern-tree data structure}, for querying pattern values.
Once the weights on $M^*$ are fixed, we root $M^*$ arbitrarily and store, for each vertex, the XOR of the weights on the path from the root to that vertex.
This preprocessing takes $O(n)$ time, and then the pattern $p(e)$ of any edge $e$ can be computed in $O(1)$ time.

\begin{lemma}[Pattern-tree data structure] \label{lem:pattern_computation_linear_time}
    After the weights on $M^*$ are fixed, there is a data structure that can be constructed in $O(n)$ time and returns $p(e)$ for any edge $e=\{u,v\}$ in $O(1)$ time.
\end{lemma}

\begin{proof}   
    Root the tree $M^*$ arbitrarily.
    For each vertex $v \in V$, store
    \begin{equation*}
        \pi(v)=\bigoplus_{\beta\in P(r,v)} w(\beta),
    \end{equation*}
    where $r$ is the root and $P(r,v)$ is the unique path from $r$ to $v$ in $M^*$.
    These values can be computed by a single BFS traversal of $M^*$ in $O(n)$ time.

    For any edge $e=\{u,v\}$, the edges common to the two root paths $P(r,u)$ and $P(r,v)$ cancel under XOR.
    Hence, $\pi(u)\oplus \pi(v)$ is exactly the XOR of the weights on the unique $u$-$v$ path in $M^*$.
    If $e\notin M^*$, this is $p(e)$ by definition.
    If $e\in M^*$, we simply return $p(e) = w(e)$.
    Therefore, $p(e)$ can be returned in $O(1)$ time.
\end{proof}

The first basic use of patterns is to test, for each edge $e\in E$, whether $e\in\Span(A)$, or equivalently, whether $r(A\cup\{e\})-r(A)=0$.
This can be done simply by checking whether $p(e)=0$.
By slightly extending this idea, we can also compute $r(A\cup \ell)-r(A)$ for each line $\ell=\{e,\bar e\}\in L$ using only the two pattern values $p(e)$ and $p(\bar e)$.
This is formalized in the following lemma.

\warmup*


\begin{proof}
    Let $\Phi:\Span(M^*)\to \mathbb{F}_2^K$ be the linear map obtained by setting
    $\Phi(\beta)=w(\beta)$ for every $\beta\in M^*$ and extending linearly over $\mathbb{F}_2$.
    For every edge $e\in E$, we have $\Phi(e)=p(e)$.
    Indeed, this is immediate if $e\in M^*$.
    If $e\notin M^*$, then
    $\displaystyle e=\sum_{\beta\in C(e)\setminus\{e\}}\beta$ over $\mathbb{F}_2$, where $C(e)$ denotes its fundamental cycle.
    Hence, $\displaystyle \Phi(e)=\bigoplus_{\beta\in C(e)\setminus\{e\}}w(\beta)=p(e)$.

        For every $T\in\mathcal{T}_A$, the constraint in Definition~\ref{def:pattern} gives $\Phi(\sum_{\beta\in T}\beta)=\bigoplus_{\beta\in T}w(\beta)=0$.
        Since every vector in $A$ is of the form $\sum_{\beta\in T}\beta$ with $T\in\mathcal{T}_A$, we have $\Phi(\alpha)=0$ for every $\alpha\in A$, and hence, by linearity, $\Phi(a)=0$ for every $a\in\Span(A)$.

    \begin{claim} \label{claim:middle}
        For every $x\in\Span(M^*)$, if $x \notin \Span(A)$, then we have
        $\Pr[\Phi(x)=0]=2^{-K}$.
    \end{claim}

    \begin{proof}[Proof of Claim~\ref{claim:middle}]
            Since $M^*$ is linearly independent, every vector $x\in\Span(M^*)$ has a unique representation
            $x=\sum_{\beta\in X}\beta$ for some $X\subseteq M^*$.
            For this representation, $\Phi(x)=\bigoplus_{\beta\in X}w(\beta)$.
    
            For each $T\in\mathcal{T}_A$, the constraint in Definition~\ref{def:pattern} is
            $\bigoplus_{\beta\in T}w(\beta)=0$.
            Thus, since the sets in $\mathcal{T}_A$ are pairwise disjoint, if $X$ is a union of some sets in $\mathcal{T}_A$, then
            $x$ is the sum of the corresponding vectors in $A$, and hence $x\in\Span(A)$.
            Therefore, if $x\notin\Span(A)$, then either $X$ contains an element not belonging to any set in $\mathcal{T}_A$, or there exists $T\in\mathcal{T}_A$ such that
            $X\cap T$ is a nonempty proper subset of $T$.
    
            Fix a coordinate $j\in\{1,\ldots,K\}$, and let $w_j(\beta)$ be the $j$-th bit of $w(\beta)$.
            Let $\Phi_j(x)$ be the $j$-th bit of $\Phi(x)$.
            Now, we have
            $\Phi_j(x)=\bigoplus_{\beta\in X} w_j(\beta)$.
            
            Thus, if $X$ contains an entire set $T\in\mathcal{T}_A$, then the contribution of this whole set to $\Phi_j(x)$ is zero.
            If $x\notin\Span(A)$, then $X$ is not a union of sets in $\mathcal{T}_A$.
            Hence, either $X$ contains an element not belonging to any set in $\mathcal{T}_A$, or there exists $T\in\mathcal{T}_A$ such that $X\cap T$ is a nonempty proper subset of $T$.
            
            In the first case, the contribution of such an element to $\Phi_j(x)$ is an independent uniformly random bit.
            In the second case, conditioned on the constraint
            $\bigoplus_{\beta\in T}w_j(\beta)=0$, the contribution
            $\bigoplus_{\beta\in X\cap T}w_j(\beta)$ is an independent uniformly random bit.
        Since the XOR of an independent uniformly random bit with any other independent value is uniformly random, $\Phi_j(x)$ is uniformly distributed over $\{0,1\}$.
        Therefore,
        \begin{equation*}
            \Pr[\Phi(x)=0]=2^{-K}.
        \end{equation*}
    \end{proof}

    We apply the claim to all vectors of the form $e$ for $e\in E$ and $e+\bar e$ for $\{e,\bar e\}\in L$.
    There are at most $3m$ such vectors.
    Since $K=100\log n$ and $m\leq n^4$, a union bound implies that, with high probability, for all such vectors $x$,
    \begin{equation*}
        \Phi(x)=0
        \quad\text{if and only if}\quad
        x\in\Span(A).
    \end{equation*}

    Condition on this event.
    For an edge $e\in E$, we have $r(A\cup\{e\})-r(A)=0$ if and only if $e\in\Span(A)$.
    By the event above and the identity $\Phi(e)=p(e)$, this is equivalent to $p(e)=0$.
    Otherwise, $r(A\cup\{e\})-r(A)=1$ .
    Thus, the first query is answered by checking whether $p(e)$ is zero.

    For a line $\ell=\{e,\bar e\}$, the value $r(A\cup \ell)-r(A)$ is determined by the following three membership tests:
    $e\in\Span(A)$, $\bar e\in\Span(A)$, and $e+\bar e\in\Span(A)$.
    Indeed, the rank increment is $0$ if and only if both $e$ and $\bar e$ belong to $\Span(A)$.
    It is at most $1$ if and only if some nonzero linear combination of $e$ and $\bar e$ belongs to $\Span(A)$.
    Over $\mathbb{F}_2$, this means that at least one of $e$, $\bar e$, and $e+\bar e$ belongs to $\Span(A)$.
    Therefore, the rank increment is $1$ if it is not $0$ and at least one of these three vectors belongs to $\Span(A)$, and it is $2$ otherwise.

    By the event above, these three membership tests are equivalent to checking whether
    $p(e)=0$, $p(\bar e)=0$, and $p(e)=p(\bar e)$, respectively.
    Indeed, the third condition follows from
    $\Phi(e+\bar e)=\Phi(e)\oplus\Phi(\bar e)=p(e)\oplus p(\bar e)$.
    The pattern-tree data structure returns $p(e)$ in $O(1)$ time.
    Therefore, the first query uses one pattern query, and the second query uses the two pattern queries $p(e)$ and $p(\bar e)$.
    Thus, both queries are answered in $O(1)$ time.
\end{proof}

\subsection{Preliminaries on data structures} \label{subsec:ds}

In this subsection, we describe the standard data structures required for an efficient implementation: dictionaries implemented by balanced binary search trees, lowest common ancestor (LCA) data structures, heavy-light decomposition, and segment trees for semigroup range queries.

We use the following standard dictionary data structure implemented by balanced binary search trees.


\begin{lemma}[Dictionaries via balanced binary search trees; see, e.g.,~{\cite[Chapter~13]{cormen2022introduction}}] \label{lem:map}
    Given $n$ key-value pairs with distinct comparable keys, there is a data structure that can be constructed in $\tO(n)$ time and, given a query key $x$, returns the value associated with $x$ if it exists, and otherwise reports that $x$ is not present, in $\tO(1)$ time.
\end{lemma}

We also use the standard lowest-common-ancestor (LCA) data structure~\cite{bender2000lca}. 
For two vertices $u$ and $v$ in a rooted tree, their LCA is the common ancestor of $u$ and $v$ that is farthest from the root.

\begin{lemma}[LCA data structures from {\cite{bender2000lca}}] \label{lem:lca}
    Given a rooted tree with $n$ vertices, there is a data structure that can be constructed in $O(n)$ time and answers LCA queries for any two vertices $u$ and $v$ in $O(1)$ time.
\end{lemma}

Using the LCA data structures in Lemma~\ref{lem:lca}, we obtain the following lemma.

\begin{lemma} \label{lem:path_symatric_dif}
    Given a rooted tree with $n$ vertices, there is a data structure that can be constructed in $O(n)$ time and answers the following query in $O(1)$ time. 
    Given two paths $P_1=P(u_1,v_1)$ and $P_2=P(u_2,v_2)$, specified by their endpoints, the query determines whether the edge-set symmetric difference $P_1 \oplus P_2$ is empty, is a single path, or can be decomposed into two edge-disjoint paths. 
    In the latter two cases, it returns the endpoints of the resulting path or paths.
\end{lemma}

The proof is based on the observation that $P_1\oplus P_2$ is determined by the four endpoints $u_1,v_1,u_2,v_2$ alone, and a query is answered by computing a constant number of tree distances, each with one LCA query.

\begin{proof}
The tree is rooted, and we build the LCA data structure of Lemma~\ref{lem:lca} in $O(n)$ time.
For a vertex $u$, let $\mathrm{depth}(u)$ denote the number of edges of the path from the root to $u$, which can be computed for all vertices in $O(n)$ time.
For two vertices $u$ and $v$, let $d(u,v)$ denote the number of edges of $P(u,v)$.
Let $w$ be the LCA of $u$ and $v$.
Since $d(u,v)=\mathrm{depth}(u)+\mathrm{depth}(v)-2\,\mathrm{depth}(w)$ and $w$ can be found in $O(1)$ time by Lemma~\ref{lem:lca}, we can compute $d(u,v)$ in $O(1)$ time.

Let $F=P_1\oplus P_2$, and let $T$ be the set of vertices that appear an odd number of times in the sequence $u_1,v_1,u_2,v_2$.
Then, $T$ is the set of odd-degree vertices of $F$, and $|T|\in\{0,2,4\}$.
We now have $F=\emptyset$ if $T=\emptyset$, and $F=P(a,b)$ if $T=\{a,b\}$.

Suppose that $T=\{a,b,c,d\}$.
Partition $T$ into two pairs $\{p,q\}$ and $\{r,s\}$ so that $d(p,q)+d(r,s)$ is minimized.
Then, $P(p,q)$ and $P(r,s)$ are edge-disjoint.
Indeed, if they shared an edge $(x, y)$ with $x$ on the side of $p$ and $r$, then $d(p,r)+d(q,s)\leq d(p,x)+d(x,r)+d(q,y)+d(y,s)=d(p,q)+d(r,s)-2$, contradicting the minimality.
Hence, $F=P(p,q)\cup P(r,s)$, since $P(p,q)\oplus P(r,s)$ has $T$ as its set of odd-degree vertices.
The minimizing partition is found by computing $d(p,q)+d(r,s)$ for the three partitions of $T$, which takes $O(1)$ time.
\end{proof}

We use the following data structures that can be obtained by combining segment trees for semigroup range queries (see, e.g., {\cite[Section 10.3]{de2000computational}} and {\cite{v021a009}}) with the heavy-light decomposition from~\cite{sleator1983data}. 
Given a sequence $s_1,\ldots,s_N$ over a semigroup $(S,\circ)$, a segment tree supports queries that return $s_\ell \circ s_{\ell+1}\circ \cdots \circ s_r$ for any interval $[\ell,r]$ in $O(\log N)$ time.
Given a rooted tree $T$ with $n$ vertices, a heavy-light decomposition partitions the edge set of $T$ into a collection $\mathcal{H}$ of edge-disjoint paths.
Here, every root-to-leaf path in $T$ can be expressed as a concatenation of $O(\log n)$ subpaths of paths in $\mathcal{H}$.

\begin{lemma}[Heavy-light decomposition from {\cite{sleator1983data}} and segment tree for semigroup range queries] \label{lem:hld}
    Let $T$ be a tree with $n$ vertices, let $(S,\circ)$ be a semigroup, and suppose that each edge $e$ of $T$ is assigned a weight $w(e)\in S$.
    There is a data structure that can be constructed in $\tO(n)$ time and answers the following query in $\tO(1)$ time.
    Given two vertices $u$ and $v$, return $w(e_1)\circ w(e_2)\circ \cdots \circ w(e_K)$, where $e_1,e_2,\ldots,e_{K}$ are the edges on the unique $u$-$v$ path in $T$.
\end{lemma}

\subsection{Fast enumeration of all patterns} \label{subsec:enumerate}

In this subsection, we prove the following lemma, which is a key tool for fast span-membership tests.
For each index $i \in \{ 0, \ldots, k\}$, we can construct a pattern set $\cP_i$ of size $O(c_i^2)$, where $c_i=|M_i|$.
For every edge $e\in E$, checking whether $p(e) \in \cP_i$ allows us to determine, with high probability, whether $e \in \Span(E_i\cup A)$.

\patternbound*


Before giving the construction, we first prove that a pattern set $\cP_i$ with the desired properties exists.

\begin{claim} \label{clm:pattern_counting}
    Fix $i\in\{0,\ldots,k\}$, and let $c_i=|M_i|$.
    Suppose that $r(A\cup E_i)>r(A)+1$.
    Then, there exists a set $\cP_i$ of patterns with $|\cP_i|=O(c_i^2)$ such that the following hold:
    \begin{itemize}
        \item for every edge $e \in E$, if $r(E_i\cup A\cup \{e\})=r(E_i\cup A)$, then $p(e) \in \cP_i$, and
        \item with high probability, for every edge $e \in E$, if $r(E_i\cup A\cup \{e\})>r(E_i\cup A)$, then $p(e) \notin \cP_i$.
    \end{itemize}
\end{claim}

\begin{proof}
Let $\Phi:\Span(M^*)\to \mathbb{F}_2^K$ be the linear map obtained by setting $\Phi(\beta)=w(\beta)$ for every $\beta\in M^*$ and extending linearly over $\mathbb{F}_2$.
As in the proof of Lemma~\ref{lem:pattern_warmup}, we have $\Phi(e)=p(e)$ for every edge $e\in E$, and $\Phi(a)=0$ for every $a\in\Span(A)$.

For each edge $e \in E$, let $P(e)$ denote the path in $M^*$ associated with $e$, defined as follows. If $e\in M^*$, then we set $P(e)={e}$. Otherwise, let $C(e)$ be the fundamental cycle of $e$ with respect to $M^*$, and set $P(e)=C(e)\setminus\{e\}$.
Here, we have $\displaystyle e=\sum_{\beta\in P(e)}\beta$ and $\displaystyle p(e)=\Phi(e)=\bigoplus_{\beta\in P(e)}w(\beta)$.
For each $e\in E$, define $\displaystyle \rho_i(e)=\bigoplus_{\beta\in P(e)\cap M_i} w(\beta)$.

We first record a simple observation.
Contract every connected component of $M^*\setminus M_i$, and let $H_i$ be the contracted tree.
The contracted graph is a tree whose edge set is $M_i$.
For any edge $e$, the set $P(e)\cap M_i$ is the set of edges of a path in the contracted tree $H_i$.
Since $H_i$ has $c_i$ edges, it has $O(c_i^2)$ paths.
Let $\mathcal{Q}_i$ be the collection of edge sets of these paths, including the empty path.
For each $Q\in\mathcal{Q}_i$, let $\displaystyle y_Q=\sum_{\beta\in Q}\beta$.
Here, we have $y_Q \in\Span(M_i)$ for each $Q\in\mathcal{Q}_i$.
Then, $|\mathcal{Q}_i|=O(c_i^2)$, and for every edge $e\in E$, we have $\rho_i(e)=\Phi(y_{Q})$, where $Q = P(e)\cap M_i$.

We now define a set $Y_i\subseteq \Span(A\cup E_i)$.
By Lemma~\ref{lem:structual_lemma}, under the assumption $r(A\cup E_i)>r(A)+1$, one of the following cases holds.

In cases~(i) and~(ii), the space $\Span(A\cup E_i)$ is generated by $A\cup M_i$ or by $(A\setminus\{\alpha_i\})\cup M_i$.
In these cases, define $Y_i=\{y_Q\mid Q\in\mathcal{Q}_i\}$.

In case~(iii), fix an edge $x\in T_i\setminus M_i$ such that $\Span(A\cup E_i)$ is generated by $A\cup M_i\cup\{x\}$.
Let $Q_x=P(x)\cap M_i$ and $y_x=\sum_{\beta\in Q_x}\beta$.
In this case, define
\begin{equation*}
    Y_i =
    \{y_Q\mid Q\in\mathcal{Q}_i\}
    \cup
    \{x+y_x+y_Q\mid Q\in\mathcal{Q}_i\}.
\end{equation*}
In all cases, $Y_i\subseteq \Span(A\cup E_i)$ and $|Y_i|=O(c_i^2)$.
We define $\cP_i=\{\Phi(y)\mid y\in Y_i\}$.
Then, $|\cP_i|\leq |Y_i|=O(c_i^2)$.

We prove the first property.
Let $e\in E$ satisfy $r(E_i\cup A\cup\{e\})=r(E_i\cup A)$, or equivalently, $e\in\Span(A\cup E_i)$.
Put $Q_e=P(e)\cap M_i$.

First consider cases~(i) and~(ii).
Then, we can write $e=a+s$ for some $a\in\Span(A)$ and some $s\in\Span(M_i)$, where in case~(ii) we may take $a\in\Span(A\setminus\{\alpha_i\})$.
Since $\Phi(a)=0$, we have $p(e)=\Phi(e)=\Phi(s)$.
Write $s=\sum_{\beta\in S}\beta$ for some $S\subseteq M_i$.
Here, the set $P(e)\cap M_i$ is either $S$ or $S\oplus T_i$, where the latter can occur only when $\alpha_i$ is used in the representation of $a$.
In the latter case, the constraint in the definition of the pattern tree gives $\bigoplus_{\beta\in T_i}w(\beta)=0$.
Hence, in either case, $\Phi(s)=\Phi(y_{Q_e})=\rho_i(e)$.
Therefore, $p(e)=\Phi(y_{Q_e})\in \cP_i$.

It remains to consider case~(iii).
In this case, we can write $e=a+s+\lambda x$ for some $a\in\Span(A)$, some $s\in\Span(M_i)$, and some $\lambda\in\{0,1\}$.
If $\lambda=0$, the same argument as above gives $p(e)=\Phi(y_{Q_e})\in\cP_i$.

Now suppose that $\lambda=1$.
Write $s=\sum_{\beta\in S}\beta$ for some $S\subseteq M_i$.
Every vector in $\Span(A)$ is represented over the tree basis $M^*$ using no elements of $M_i$.
Here, the equality $e=a+s+x$ implies $Q_e=S\oplus Q_x$, where $Q_e=P(e)\cap M_i$ and $Q_x=P(x)\cap M_i$.
Therefore, $\Phi(s)=\Phi(y_{Q_e})\oplus \Phi(y_x)$.
Since $\Phi(a)=0$, we obtain
\begin{equation*}
p(e)
=
\Phi(e)
=
\Phi(x)\oplus \Phi(y_x)\oplus \Phi(y_{Q_e})
=
\Phi(x+y_x+y_{Q_e}).
\end{equation*}
By the definition of $Y_i$ in case~(iii), the vector $x+y_x+y_{Q_e}$ belongs to $Y_i$.
Hence $p(e)\in\cP_i$.

We now prove the second property.
We use Claim~\ref{claim:middle}: for every fixed vector $z\in\Span(M^*)$, if $z\notin\Span(A)$, then $\Pr[\Phi(z)=0]=2^{-K}$.

Fix an edge $e\in E$ such that $r(E_i\cup A\cup\{e\})>r(E_i\cup A)$, or equivalently, $e\notin\Span(A\cup E_i)$.
For any $y\in Y_i$, we have $y\in\Span(A\cup E_i)$.
Therefore, $e+y\notin\Span(A\cup E_i)$, and in particular $e+y\notin\Span(A)$.
Hence, by Claim~\ref{claim:middle}, we have
\begin{equation*}
    \Pr[p(e)=\Phi(y)]
    =
    \Pr[\Phi(e+y)=0]
    =
    2^{-K}.
\end{equation*}

Taking a union bound over all $e\in E$ and all $y\in Y_i$, the probability that there exists an edge $e\in E$ such that $r(E_i\cup A\cup\{e\})>r(E_i\cup A)$ and $p(e)\in\cP_i$ is at most $|E|\cdot |Y_i|\cdot 2^{-K}$.
Since $|E|=2m$, $|Y_i|=O(c_i^2)\le O(n^2)$, $m\le n^4$, and $K=100\log n$, this probability is $n^{-\Omega(1)}$.
Therefore, with high probability, for every edge $e\in E$, if $r(E_i\cup A\cup\{e\})>r(E_i\cup A)$, then $p(e)\notin\cP_i$.
\end{proof}

We now give a construction of the pattern set $\cP_i$ in $\tO(n+c_i^2)$ time, completing the proof of Lemma~\ref{lem:pattern_bound}.

\begin{proof}[Proof of Lemma~\ref{lem:pattern_bound}]
By Claim~\ref{clm:pattern_counting}, it remains to show that the pattern set $\cP_i$ used in its proof can be constructed in $\tO(n+c_i^2)$ time.

We first construct the set $\cR_i=\{\Phi(y_Q)\mid Q\in\mathcal{Q}_i\}$, where $\mathcal{Q}_i$ is the collection of edge sets of paths in the tree obtained by contracting every connected component of $M^*\setminus M_i$.
Assign to each edge $\beta\in M^*$ the label $\widetilde{w}_i(\beta)=w(\beta)$ if $\beta\in M_i$, and $\widetilde{w}_i(\beta)=0$ otherwise.
Using the data structure in Lemma~\ref{lem:hld} to the tree $M^*$ with these labels, we obtain in $\tO(n)$ time a data structure that, given two vertices $u$ and $v$, returns
$\displaystyle \bigoplus_{\beta\in P_{M^*}(u,v)\cap M_i} w(\beta)$ in $\tO(1)$ time, where $P_{M^*}(u,v)$ denotes the unique $u$-$v$ path in $M^*$.

Let $U_i$ be the set of endpoints of the edges in $M_i$.
Initialize $\cR_i=\{0^K\}$.
For every pair of vertices $u,v\in U_i$, query the data structure on $u$ and $v$, and insert the returned value into $\cR_i$.
This enumerates all values $\Phi(y_Q)$ for $Q\in\mathcal{Q}_i$. 
Indeed, each vertex of the contracted tree is represented by some endpoint of an edge in $M_i$, and the query on two endpoints returns the XOR of the weights on the corresponding path in the contracted tree.
Since $|U_i|\le 2c_i$, this step requires $\tO(c_i^2)$ time.

By the argument in the proof of Claim~\ref{clm:pattern_counting}, we define $\cP_i$ as follows.
In cases~(i) and~(ii), we set $\cP_i=\cR_i$.
In case~(iii), choose an arbitrary edge $x\in T_i\setminus M_i$.
We then compute $p(x)=\Phi(x)$ using the pattern-tree data structure (Lemma~\ref{lem:pattern_computation_linear_time}).
We also compute $\rho_i(x)=\bigoplus_{\beta\in P(x)\cap M_i}w(\beta)$ by one query to the data structure above.
We then set
\begin{equation*}
    \cP_i
    =
    \cR_i
    \cup
    \{p(x)\oplus \rho_i(x)\oplus \rho \mid \rho\in\cR_i\}.
\end{equation*}
This is exactly the set $\{\Phi(y)\mid y\in Y_i\}$ from the proof of Claim~\ref{clm:pattern_counting} in case~(iii), because
$\Phi(x+y_x+y_Q)=p(x)\oplus\rho_i(x)\oplus\Phi(y_Q)$.

In all cases, $|\cP_i|=O(c_i^2)$.
The construction time is $\tO(n)$ for the data structure in Lemma~\ref{lem:hld}, plus $\tO(c_i^2)$ for computing all patterns in $\cR_i$.
Hence, the total running time is $\tO(n+c_i^2)$.
The correctness of the resulting set $\cP_i$ follows from Claim~\ref{clm:pattern_counting}.
\end{proof}

\subsection{Fast span-membership tests} \label{sec:small_handle}

In this subsection, we use the results of the previous subsections to give a fast algorithm for the span-membership tests, and prove Lemma~\ref{lem:fast_span_checking}.
To this end, we prove the following lemma.
Its statement involves an integer parameter $C$, following the exposition in technical overview (Section~\ref{sec:technical_overview}), although we apply the lemma only with $C=n$.
Since $|M_i|\leq n$ always holds, the lemma with $C=n$ allows us to perform the span-membership tests for all indices $i\in\{0,\ldots,k\}$ in $\tO(n^2+m)$ total time.

\begin{lemma} \label{lem:small_handle}
     With high probability, for all lines $\ell\in L$, we can decide whether there exists an index $i \in \{ 0, \ldots, k\}$ satisfying $r(A\cup E_i)>r(A)+1$ and $|M_i| \leq C$ such that $r(E_i\cup A\cup \ell)=r(E_i\cup A)$ in total $\displaystyle \tO\left(n^2 + m + Cn \right)$ time.
\end{lemma}

The technical overview in Section~\ref{sec:technical_overview} gives a simpler $\tO(n^2+m+C^3n)$-time implementation.
A simpler $\tO(n^2+m+C^3n)$-time implementation can be derived rather easily from Lemma~\ref{lem:pattern_bound}.
Here, we refine it to obtain the $\tO(n^2+m+Cn)$ bound.
The key tool for this refinement is the following lemma.

\begin{restatable}{lemma}{onlyone} \label{lem:only_one}
    For every edge $e \in E$ with $e \notin \Span(A)$, there exists at most one index $i\in\{0,\ldots,k\}$ such that $e\in\Span(A\cup M_i)$.
\end{restatable}

\begin{proof}
It suffices to prove the following statement: if $i\neq j$ and $e\in\Span(A\cup M_i)\cap\Span(A\cup M_j)$, then $e\in\Span(A)$.

Suppose that $e\in\Span(A\cup M_i)\cap\Span(A\cup M_j)$ for two distinct indices $i$ and $j$.
Then, there exist $a_i,a_j\in\Span(A)$ and sets $S_i\subseteq M_i$ and $S_j\subseteq M_j$ such that
$\displaystyle e=a_i+\sum_{\beta\in S_i}\beta=a_j+\sum_{\beta\in S_j}\beta$.
Therefore,
$\displaystyle \sum_{\beta\in S_i}\beta+\sum_{\beta\in S_j}\beta=a_i+a_j\in\Span(A)$.

We use the representation over the tree basis $M^*$, as in Remark~\ref{rem:onservation}.
    Every vector in $\Span(A)$ is represented over the tree basis $M^*$ by the symmetric difference of some sets in $\mathcal{T}_A$.
    Each set in $\mathcal{T}_A$ is contained in a single part of the partition $(M_0,\ldots,M_{k'})$ before Line~\ref{line:ws_removal}, and these parts are pairwise disjoint.
    This implies the following: if a vector in $\Span(A)$ is represented only using elements of $M_i\cup M_j$, then its part in $M_i$ is either empty or $T_i$, and its part in $M_j$ is either empty or $T_j$.

Applying this to the vector $\sum_{\beta\in S_i}\beta+\sum_{\beta\in S_j}\beta\in\Span(A)$, we get that $\sum_{\beta\in S_i}\beta\in\Span(A)$ and $\sum_{\beta\in S_j}\beta\in\Span(A)$.
Hence, $e=a_i+\sum_{\beta\in S_i}\beta\in\Span(A)$.
This proves the equivalent statement, and therefore proves the lemma.
\end{proof}

We now prove Lemma~\ref{lem:small_handle} using Lemmas~\ref{lem:pattern_bound} and~\ref{lem:only_one}, together with data structures from Section~\ref{subsec:ds}.




\begin{proof}[Proof of Lemma~\ref{lem:small_handle}]
    Write $c_i=|M_i|$, and let
    \begin{equation*}
        \mathcal I=\{i\in\{0,\ldots,k\}\mid c_i\le C
        \text{ and } r(A\cup E_i)>r(A)+1\}.
    \end{equation*}
    For each $i\in\mathcal I$, we run the construction in Lemma~\ref{lem:pattern_bound}.
    This gives the pattern set $\cP_i$ in $\tO(n+c_i^2)$ time.
    We also keep the auxiliary set
    $\cR_i=\{\Phi(y_Q)\mid Q\in\mathcal Q_i\}$ constructed in the proof of Lemma~\ref{lem:pattern_bound}, which has size $O(c_i^2)$.
    Since $\sum_i c_i=O(n)$, the total time for this construction step is
    \[
        \sum_{i\in\mathcal I}\tO(n+c_i^2)
        \le \tO(n^2)+\tO\left(C\sum_i c_i\right)
        =
        \tO(n^2+Cn).
    \]

    We build two global dictionaries.
    The first dictionary $D_{\mathrm{span}}$ stores, for every $i\in\mathcal I$ and every $\rho\in\cP_i$, the key $\rho$ together with one such index $i$.
    The second dictionary $D_{\mathrm{base}}$ stores, for every $i\in\mathcal I$ and every $\rho\in\cR_i$, the key $\rho$ together with one such index $i$.
    The total number of stored keys is $\sum_{i\in\mathcal I}O(c_i^2)=O(Cn)$, so these dictionaries are constructed in $\tO(Cn)$ time by Lemma~\ref{lem:map}.
    We also store each $\cP_i$ itself as a dictionary, so that membership in $\cP_i$ can be tested in $\tO(1)$ time.

    We will use the following interpretation of these dictionaries.
    By Lemma~\ref{lem:pattern_bound}, with high probability, for every edge $f\in E$ and every $i\in\mathcal I$,
    \[
        f\in\Span(A\cup E_i)
        \quad\Longleftrightarrow\quad
        p(f)\in\cP_i .
    \]
    Similarly, by the same argument applied to the auxiliary sets $\cR_i$, with high probability, for every edge $f\in E$ and every $i\in\mathcal I$,
    \[
        f\in\Span(A\cup M_i)
        \quad\Longleftrightarrow\quad
        p(f)\in\cR_i .
    \]
    Hence, for an edge $f\notin\Span(A)$, a lookup of $p(f)$ in $D_{\mathrm{base}}$ returns the unique index $i\in\mathcal I$ such that $f\in\Span(A\cup M_i)$, if such an index exists.
    The uniqueness follows from Lemma~\ref{lem:only_one}.

    It remains to describe how to process each line.
    We first build one more deterministic data structure.
    Let $\mathcal I_3\subseteq\mathcal I$ be the set of indices that fall into case~(iii) of Lemma~\ref{lem:structual_lemma}.
    Assign to each edge $\beta\in M^*$ the label $i$ if $\beta\in M_i$ for some $i\in\mathcal I_3$, and assign no label otherwise.
    Using Lemmas~\ref{lem:path_symatric_dif} and~\ref{lem:hld} with the semigroup that computes the minimum and maximum label on a path, we can determine in $\tO(1)$ time whether the edge set $P(e)\oplus P(\bar e)$ contains labels from no index, from exactly one index, or from at least two distinct indices.
    This data structure is constructed in $\tO(n)$ time.

    Now, fix a line $\ell=\{e,\bar e\}$.
    By Lemma~\ref{lem:pattern_warmup}, we can test in $\tO(1)$ time whether $e\in\Span(A)$ and whether $\bar e\in\Span(A)$.
    We consider the following three cases.

    \begin{enumerate}[(A)]
        \item Suppose that $e\in\Span(A)$ and $\bar e\in\Span(A)$.
        Then $e,\bar e\in\Span(A\cup E_i)$ for every $i\in\mathcal I$.
        Thus, the answer is yes if and only if $\mathcal I$ is nonempty.

        \item Suppose that exactly one of $e$ and $\bar e$ belongs to $\Span(A)$.
        By symmetry, assume that $\bar e\in\Span(A)$ and $e\notin\Span(A)$.
        Then $\ell$ is covered by some $i\in\mathcal I$ if and only if $e\in\Span(A\cup E_i)$ for some $i\in\mathcal I$.
        We decide this by looking up $p(e)$ in $D_{\mathrm{span}}$.

        \item Suppose that $e\notin\Span(A)$ and $\bar e\notin\Span(A)$.
        Let $i_e$ be the index returned by $D_{\mathrm{base}}$ on key $p(e)$, if it exists, and define $i_{\bar e}$ analogously.

        If $i_e$ exists and $p(\bar e)\in\cP_{i_e}$, then both $e$ and $\bar e$ belong to $\Span(A\cup E_{i_e})$, so we answer yes.
        Similarly, if $i_{\bar e}$ exists and $p(e)\in\cP_{i_{\bar e}}$, then we answer yes.

                    It remains to consider the case where, for every index $i\in\mathcal I$ with $\ell\subseteq\Span(A\cup E_i)$, neither $e$ nor $\bar e$ belongs to $\Span(A\cup M_i)$.
        Thus, any remaining solution must come from case~(iii) of Lemma~\ref{lem:structual_lemma}. That is, for the corresponding edge $x\in T_i\setminus M_i$, both $e$ and $\bar e$ belong to $\Span(A\cup M_i\cup\{x\})$, but neither belongs to $\Span(A\cup M_i)$. We inspect the labels appearing on $P(e)\oplus P(\bar e)$ to identify such possible indices.

        Let $I_\ell$ be the set of labels appearing on $P(e)\oplus P(\bar e)$. We then consider the following three cases.

        \begin{enumerate}[(a)]
            \item If $|I_\ell|\ge 2$, then no remaining solution exists, and we answer no.
            Indeed, suppose that there is an index $i$ in case~(iii) such that $\ell\subseteq \Span(A\cup E_i) = \Span(A \cup M_i \cup \{ x \})$.
            Since we are in the remaining case, neither $e$ nor $\bar e$ belongs to $\Span(A\cup M_i)$.
            Therefore, in the expression of $e$ or $\bar e$ using vectors from $A\cup M_i\cup\{x\}$, the coefficient of $x$ must be $1$.
            Equivalently, there exist $a_e,a_{\bar e}\in\Span(A)$ and $s_e,s_{\bar e}\in\Span(M_i)$ such that
            \begin{equation*}
            e=a_e+s_e+x
            \qquad\text{and}\qquad
            \bar e=a_{\bar e}+s_{\bar e}+x .
            \end{equation*}
            Thus, the two copies of $x$ cancel in $e+\bar e$.
            Therefore, the representation of $e+\bar e$ over the tree-basis $M^*$, as in Remark~\ref{rem:onservation}, can contain labeled edges only from $M_i$.
            Equivalently, the labels on $P(e)\oplus P(\bar e)$ can only be $i$.
            This contradicts $|I_\ell|\ge 2$.
            
            \item If $I_\ell=\{i\}$, then, by the argument in (a), $i$ is the only candidate index that can satisfy $\ell\subseteq\Span(A\cup E_i)$ in case~(iii).
            Thus, it remains only to check whether $\ell\subseteq\Span(A\cup E_i)$.
            We therefore answer yes if and only if both $p(e)$ and $p(\bar e)$ belong to $\cP_i$.
            
            \item Finally, suppose that $I_\ell=\emptyset$.
            If there is an index $i$ in case~(iii) such that $\ell\subseteq \Span(A\cup E_i) = \Span(A \cup M_i \cup \{ x \})$, then, in the expression of $e + \bar{e}$ using vectors from $A\cup M_i\cup\{x\}$, the coefficient of each $\beta \in M_i$ must be $0$.
            Since the two copies of $x\in T_i\setminus M_i$ cancel, this implies $e+\bar e\in\Span(A)$.
            We therefore check whether $p(e)=p(\bar e)$.
            If not, we answer no.
            If yes, then $\bar e=e+(e+\bar e)$ differs from $e$ by a vector in $\Span(A)$.
            Hence, for any index $i$, if $e\in\Span(A\cup E_i)$, then also $\bar e\in\Span(A\cup E_i)$.
            Thus, we answer yes if and only if the lookup of $p(e)$ in $D_{\mathrm{span}}$ succeeds.
        \end{enumerate}
    \end{enumerate}

    Each line is processed using only a constant number of dictionary lookups and path queries, so all $m$ lines are processed in $\tO(m)$ time.
    The total running time is therefore $\tO(n^2+Cn+m)$.

    The correctness holds with high probability by Lemmas~\ref{lem:pattern_bound} and~\ref{lem:pattern_warmup}, by the analogous guarantee for the auxiliary sets $\cR_i$, and by a union bound over all indices in $\mathcal I$ and all queried lines.
\end{proof}

Using Lemma~\ref{lem:small_handle}, we now prove Lemma~\ref{lem:fast_span_checking}, which is the goal of this section and is restated here.

\fastcheck*

\begin{proof}[Proof of Lemma~\ref{lem:fast_span_checking}]
    First, by Lemma~\ref{lem:pattern_computation_linear_time}, we build the pattern-tree data structure in $O(n)$ time.
    Then, by Lemma~\ref{lem:pattern_warmup}, we can decide, for each line $\ell\in L$, whether $r(A\cup\ell)-r(A)\leq 1$ in $\tO(m)$ total time.

    Since $M_i\subseteq M$ and $|M|\leq n$, every index $i\in\{0,\ldots,k\}$ satisfies $|M_i|\leq n$.
    Hence, by Lemma~\ref{lem:small_handle} with $C=n$, we can decide, for all lines $\ell\in L$, whether there exists an index $i\in\{0,\ldots,k\}$ with $r(E_i\cup A\cup\ell)=r(E_i\cup A)$ in $\tO(n^2+m+n\cdot n)=\tO(n^2+m)$ total time.

    Therefore, the total running time is $\tO(n^2+m)$.
    Correctness holds with high probability by Lemmas~\ref{lem:pattern_warmup} and~\ref{lem:small_handle}, together with a union bound over the polynomially many tests performed above.
\end{proof}

\section*{Acknowledgments of the use of AI}

All key ideas for breaking the $O(nm)$-time complexity barrier for graphic matroid parity, which lead to an $\tO(n^{4/3}m^{2/3}+\sqrt{n}m)$-time exact algorithm, are due to the author.
The key ideas are presented in Sections~\ref{sec:adaptation_of_assadi} and~\ref{sec:checking_span}.
After the author had written up the paper with the $\tO(n^{4/3}m^{2/3}+\sqrt{n}m)$-time exact algorithm, GPT~6 Astra (OpenAI) found an idea to improve the running time to $\tO(n^{3/2}\sqrt{m}+\sqrt{n}m)$.
The idea is the modification of Assadi's framework in Section~\ref{sec:warm_start},  which reduces a part of the $(1-\eps)$-approximation algorithm that took $\tO(n^2/\eps^2)$ time to $\tO(n^2/\eps)$ time.

The author used GPT~5.5, GPT~6 Astra (OpenAI), and Claude Fable~5.1 (Anthropic) for assistance in writing and editing parts of this paper.
These tools were also used to draft proofs of some lemmas after the author provided the lemma statements and the corresponding proof ideas.
All AI-generated text and proofs were carefully verified and revised by the author, who takes full responsibility for the contents of this paper.

\bibliography{biblio}

@article{harvey2009algebraic,
  title = {Algebraic algorithms for matching and matroid problems},
  author = {Harvey, Nicholas JA},
  journal = {SIAM Journal on Computing},
  volume = {39},
  number = {2},
  pages = {679--702},
  year = {2009},
  publisher = {SIAM},
  note = {Announced at FOCS 2006.},
  timestamp = {Sat, 27 May 2017 01:00:00 +0200},
  biburl = {https://dblp.org/rec/journals/siamcomp/Harvey09.bib},
  bibsource = {dblp computer science bibliography, https://dblp.org},
  doi = {10.1137/070684008},
  _bib2doi_selected = {dblp:/rec/journals/siamcomp/Harvey09.bib},
  _bib2doi_confirmed = {true},
}

@inproceedings{harvey2007algebraic,
  title = {An algebraic algorithm for weighted linear matroid intersection},
  author = {Harvey, Nicholas JA},
  booktitle = {Proceedings of the 8th Annual ACM-SIAM Symposium on Discrete Algorithms (SODA 2007)},
  pages = {444--453},
  year = {2007},
  timestamp = {Fri, 07 Dec 2012 00:00:00 +0100},
  biburl = {https://dblp.org/rec/conf/soda/Harvey07.bib},
  bibsource = {dblp computer science bibliography, https://dblp.org},
  url = {http://dl.acm.org/citation.cfm?id=1283383.1283430},
  _bib2doi_selected = {dblp:/rec/conf/soda/Harvey07.bib},
  _bib2doi_confirmed = {true},
  _bib2doi_finished = {true},
}

@article{gabow1996efficient,
  title = {Efficient theoretic and practical algorithms for linear matroid intersection problems},
  author = {Gabow, Harold N and Xu, Ying},
  journal = {Journal of Computer and System Sciences},
  volume = {53},
  number = {1},
  pages = {129--147},
  year = {1996},
  publisher = {Elsevier},
  timestamp = {Tue, 16 Feb 2021 00:00:00 +0100},
  biburl = {https://dblp.org/rec/journals/jcss/GabowX96.bib},
  bibsource = {dblp computer science bibliography, https://dblp.org},
  doi = {10.1006/jcss.1996.0054},
  _bib2doi_selected = {dblp:/rec/journals/jcss/GabowX96.bib},
  _bib2doi_confirmed = {true},
}

@inproceedings{gabow1989efficient,
  title = {Efficient algorithms for independent assignment on graphic and linear matroids},
  author = {Gabow, Harold N and Xu, Ying},
  booktitle = {Proceedings of the 30th Annual Symposium on Foundations of Computer Science (FOCS 1989)},
  pages = {106--111},
  year = {1989},
  organization = {IEEE Computer Society},
  timestamp = {Thu, 23 Mar 2023 00:00:00 +0100},
  biburl = {https://dblp.org/rec/conf/focs/GabowX89.bib},
  bibsource = {dblp computer science bibliography, https://dblp.org},
  doi = {10.1109/SFCS.1989.63463},
  _bib2doi_selected = {dblp:/rec/conf/focs/GabowX89.bib},
  _bib2doi_confirmed = {true},
}

@inproceedings{blikstad2021breaking,
  title = {Breaking {$O(nr)$} for matroid intersection},
  author = {Blikstad, Joakim},
  booktitle = {Proceedings of the 48th International Colloquium on Automata, Languages, and Programming (ICALP 2022)},
  pages = {31:1-31:17},
  year = {2021},
  volume = {198},
  doi = {10.4230/LIPIcs.ICALP.2021.31},
  timestamp = {Tue, 06 Jul 2021 01:00:00 +0200},
  biburl = {https://dblp.org/rec/conf/icalp/Blikstad21.bib},
  bibsource = {dblp computer science bibliography, https://dblp.org},
  _bib2doi_selected = {dblp:/rec/conf/icalp/Blikstad21.bib},
  _bib2doi_confirmed = {true},
}

@inproceedings{blikstad2021breaking_STOC,
  title = {Breaking the quadratic barrier for matroid intersection},
  author = {Blikstad, Joakim and van den Brand, Jan and Mukhopadhyay, Sagnik and Nanongkai, Danupon},
  booktitle = {Proceedings of the 53rd Annual ACM SIGACT Symposium on Theory of Computing (STOC 2021)},
  pages = {421--432},
  year = {2021},
  doi = {10.1145/3406325.3451092},
  timestamp = {Sat, 08 Jan 2022 00:00:00 +0100},
  biburl = {https://dblp.org/rec/conf/stoc/BlikstadBMN21.bib},
  bibsource = {dblp computer science bibliography, https://dblp.org},
  _bib2doi_selected = {dblp:/rec/conf/stoc/BlikstadBMN21.bib},
  _bib2doi_confirmed = {true},
}

@inproceedings{blikstad2023fast,
  title = {Fast Algorithms via Dynamic-Oracle Matroids},
  author = {Blikstad, Joakim and Mukhopadhyay, Sagnik and Nanongkai, Danupon and Tu, Ta-Wei},
  booktitle = {Proceedings of the 55th Annual ACM Symposium on Theory of Computing (STOC 2023)},
  pages = {1229--1242},
  year = {2023},
  doi = {10.1145/3564246.3585219},
  timestamp = {Sun, 19 Jan 2025 00:00:00 +0100},
  biburl = {https://dblp.org/rec/conf/stoc/BlikstadMNT23.bib},
  bibsource = {dblp computer science bibliography, https://dblp.org},
  _bib2doi_selected = {dblp:/rec/conf/stoc/BlikstadMNT23.bib},
  _bib2doi_confirmed = {true},
}

@inproceedings{chakrabarty2019faster,
  title = {Faster matroid intersection},
  author = {Chakrabarty, Deeparnab and Lee, Yin Tat and Sidford, Aaron and Singla, Sahil and Wong, Sam Chiu-wai},
  booktitle = {Proceedings of the 60th Annual Symposium on Foundations of Computer Science (FOCS 2019)},
  pages = {1146--1168},
  year = {2019},
  organization = {IEEE},
  doi = {10.1109/FOCS.2019.00072},
  timestamp = {Tue, 07 Jan 2020 00:00:00 +0100},
  biburl = {https://dblp.org/rec/conf/focs/ChakrabartyLS0W19.bib},
  bibsource = {dblp computer science bibliography, https://dblp.org},
  _bib2doi_selected = {dblp:/rec/conf/focs/ChakrabartyLS0W19.bib},
  _bib2doi_confirmed = {true},
}

@inproceedings{chekuri2016fast,
  title = {A fast approximation for maximum weight matroid intersection},
  author = {Chekuri, Chandra and Quanrud, Kent},
  booktitle = {Proceedings of the 27th Annual ACM-SIAM Symposium on Discrete Algorithms (SODA 2016)},
  pages = {445--457},
  year = {2016},
  organization = {SIAM},
  doi = {10.1137/1.9781611974331.ch33},
  timestamp = {Tue, 02 Feb 2021 00:00:00 +0100},
  biburl = {https://dblp.org/rec/conf/soda/ChekuriQ16.bib},
  bibsource = {dblp computer science bibliography, https://dblp.org},
  _bib2doi_selected = {dblp:/rec/conf/soda/ChekuriQ16.bib},
  _bib2doi_confirmed = {true},
}

@article{cunningham1986improved,
  title = {Improved bounds for matroid partition and intersection algorithms},
  author = {Cunningham, William H},
  journal = {SIAM Journal on Computing},
  volume = {15},
  number = {4},
  pages = {948--957},
  year = {1986},
  publisher = {SIAM},
  doi = {10.1137/0215066},
  timestamp = {Sat, 27 May 2017 01:00:00 +0200},
  biburl = {https://dblp.org/rec/journals/siamcomp/Cunningham86.bib},
  bibsource = {dblp computer science bibliography, https://dblp.org},
  _bib2doi_selected = {dblp:/rec/journals/siamcomp/Cunningham86.bib},
  _bib2doi_confirmed = {true},
}

@incollection{edmonds1979matroid,
  title = {Matroid intersection},
  author = {Edmonds, Jack},
  booktitle = {Annals of Discrete Mathematics},
  volume = {4},
  pages = {39--49},
  year = {1979},
  publisher = {Elsevier},
  doi = {10.1016/S0167-5060(08)70817-3},
  _bib2doi_finished = {true},
}

@inproceedings{edmonds1970submodular,
  title = {Submodular functions, matroids, and certain polyhedra},
  author = {Edmonds, Jack},
  booktitle = {Combinatorial Structures and Their Applications},
  pages = {69--87},
  year = {1970},
  _bib2doi_finished = {true},
}

@inproceedings{gabow1988forests,
  title = {Forests, frames, and games: algorithms for matroid sums and applications},
  author = {Gabow, Harold and Westermann, Herbert},
  booktitle = {Proceedings of the 20th Annual ACM Symposium on Theory of Computing (STOC 1988)},
  pages = {407--421},
  year = {1988},
  doi = {10.1145/62212.62252},
  timestamp = {Tue, 06 Nov 2018 00:00:00 +0100},
  biburl = {https://dblp.org/rec/conf/stoc/GabowW88.bib},
  bibsource = {dblp computer science bibliography, https://dblp.org},
  _bib2doi_selected = {dblp:/rec/conf/stoc/GabowW88.bib},
  _bib2doi_confirmed = {true},
}

@article{hopcroft1973n,
  title = {An {$n^{5/2}$} algorithm for maximum matchings in bipartite graphs},
  author = {Hopcroft, John E and Karp, Richard M},
  journal = {SIAM Journal on Computing},
  volume = {2},
  number = {4},
  pages = {225--231},
  year = {1973},
  publisher = {SIAM},
  doi = {10.1137/0202019},
  timestamp = {Wed, 14 Nov 2018 00:00:00 +0100},
  biburl = {https://dblp.org/rec/journals/siamcomp/HopcroftK73.bib},
  bibsource = {dblp computer science bibliography, https://dblp.org},
  _bib2doi_selected = {dblp:/rec/journals/siamcomp/HopcroftK73.bib},
  _bib2doi_confirmed = {true},
}

@article{huang2016exact,
  title = {Exact and approximation algorithms for weighted matroid intersection},
  author = {Huang, Chien-Chung and Kakimura, Naonori and Kamiyama, Naoyuki},
  journal = {Mathematical Programming},
  volume = {177},
  number = {1-2},
  pages = {85--112},
  year = {2019},
  publisher = {Springer},
  note = {Announced at SODA 2016.},
  doi = {10.1007/s10107-018-1260-x},
  timestamp = {Sun, 19 Jan 2025 00:00:00 +0100},
  biburl = {https://dblp.org/rec/journals/mp/HuangKK19.bib},
  bibsource = {dblp computer science bibliography, https://dblp.org},
  _bib2doi_selected = {dblp:/rec/journals/mp/HuangKK19.bib},
  _bib2doi_confirmed = {true},
}

@inproceedings{lee2015faster,
  title = {A faster cutting plane method and its implications for combinatorial and convex optimization},
  author = {Lee, Yin Tat and Sidford, Aaron and Wong, Sam Chiu-wai},
  booktitle = {Proceedings of the 56th Annual Symposium on Foundations of Computer Science (FOCS 2015)},
  pages = {1049--1065},
  year = {2015},
  organization = {IEEE},
  doi = {10.1109/FOCS.2015.68},
  timestamp = {Thu, 23 Mar 2023 00:00:00 +0100},
  biburl = {https://dblp.org/rec/conf/focs/LeeSW15.bib},
  bibsource = {dblp computer science bibliography, https://dblp.org},
  _bib2doi_selected = {dblp:/rec/conf/focs/LeeSW15.bib},
  _bib2doi_confirmed = {true},
}

@misc{nguyen2019note,
  title = {A note on {C}unningham's algorithm for matroid intersection},
  author = {Nguyen, Huy L},
  howpublished = {arXiv preprint arXiv:1904.04129},
  year = {2019},
  doi = {10.48550/arXiv.1904.04129},
  timestamp = {Thu, 25 Apr 2019 01:00:00 +0200},
  biburl = {https://dblp.org/rec/journals/corr/abs-1904-04129.bib},
  bibsource = {dblp computer science bibliography, https://dblp.org},
  _bib2doi_selected = {dblp:/rec/journals/corr/abs-1904-04129.bib},
  _bib2doi_confirmed = {true},
}

@book{schrijver2003combinatorial,
  title = {Combinatorial optimization: polyhedra and efficiency},
  author = {Schrijver, Alexander},
  volume = {24},
  year = {2003},
  publisher = {Springer},
  _bib2doi_finished = {true},
}

@inproceedings{u2022subquadratic,
  title = {Subquadratic weighted matroid intersection under rank oracles},
  author = {Tu, Ta-Wei},
  booktitle = {Proceedings of the 33rd International Symposium on Algorithms and Computation (ISAAC 2022)},
  pages = {63:1-63:14},
  year = {2022},
  volume = {248},
  doi = {10.4230/LIPIcs.ISAAC.2022.63},
  timestamp = {Sat, 30 Sep 2023 01:00:00 +0200},
  biburl = {https://dblp.org/rec/conf/isaac/Tu22.bib},
  bibsource = {dblp computer science bibliography, https://dblp.org},
  _bib2doi_selected = {dblp:/rec/conf/isaac/Tu22.bib},
  _bib2doi_confirmed = {true},
}

@article{shigeno1995dual,
  title = {A dual approximation approach to weighted matroid intersection},
  author = {Shigeno, Maiko and Iwata, Satoru},
  journal = {Operations {R}esearch {L}etters},
  volume = {18},
  number = {3},
  pages = {153--156},
  year = {1995},
  publisher = {Elsevier},
  timestamp = {Sun, 19 Jan 2025 00:00:00 +0100},
  biburl = {https://dblp.org/rec/journals/orl/ShigenoI95.bib},
  bibsource = {dblp computer science bibliography, https://dblp.org},
  doi = {10.1016/0167-6377(95)00047-X},
  _bib2doi_selected = {dblp:/rec/journals/orl/ShigenoI95.bib},
  _bib2doi_confirmed = {true},
}

@inproceedings{quanrud2024faster,
  title = {Faster exact and approximation algorithms for packing and covering matroids via push-relabel},
  author = {Quanrud, Kent},
  booktitle = {Proceedings of the 35th Annual ACM-SIAM Symposium on Discrete Algorithms (SODA 2024)},
  pages = {2305--2336},
  year = {2024},
  organization = {SIAM},
  timestamp = {Thu, 04 Apr 2024 01:00:00 +0200},
  biburl = {https://dblp.org/rec/conf/soda/Quanrud24.bib},
  bibsource = {dblp computer science bibliography, https://dblp.org},
  doi = {10.1137/1.9781611977912.82},
  _bib2doi_selected = {dblp:/rec/conf/soda/Quanrud24.bib},
  _bib2doi_confirmed = {true},
}

@inproceedings{huang2024robust,
  title = {Robust sparsification for matroid intersection with applications},
  author = {Huang, Chien-Chung and Sellier, Fran{\c{c}}ois},
  booktitle = {Proceedings of the 35th Annual ACM-SIAM Symposium on Discrete Algorithms (SODA 2024)},
  pages = {2916--2940},
  year = {2024},
  organization = {SIAM},
  timestamp = {Fri, 17 Jan 2025 00:00:00 +0100},
  biburl = {https://dblp.org/rec/conf/soda/HuangS24.bib},
  bibsource = {dblp computer science bibliography, https://dblp.org},
  doi = {10.1137/1.9781611977912.104},
  _bib2doi_selected = {dblp:/rec/conf/soda/HuangS24.bib},
  _bib2doi_confirmed = {true},
}

@article{feigenbaum2005graph,
  title = {On graph problems in a semi-streaming model},
  author = {Feigenbaum, Joan and Kannan, Sampath and McGregor, Andrew and Suri, Siddharth and Zhang, Jian},
  journal = {Theoretical Computer Science},
  volume = {348},
  number = {2-3},
  pages = {207--216},
  year = {2005},
  publisher = {Elsevier},
  note = {Announced at ICALP 2004.},
  timestamp = {Wed, 17 Feb 2021 00:00:00 +0100},
  biburl = {https://dblp.org/rec/journals/tcs/FeigenbaumKMSZ05.bib},
  bibsource = {dblp computer science bibliography, https://dblp.org},
  doi = {10.1016/j.tcs.2005.09.013},
  _bib2doi_selected = {dblp:/rec/journals/tcs/FeigenbaumKMSZ05.bib},
  _bib2doi_confirmed = {true},
}

@inproceedings{mcgregor2005finding,
  title = {Finding graph matchings in data streams},
  author = {McGregor, Andrew},
  booktitle = {Proceedings of the 8th International Workshop on Approximation Algorithms for Combinatorial Optimization Problems (APPROX 2005) and the 9th International Workshop on Randomized and Computation (RANDOM 2005)},
  pages = {170--181},
  year = {2005},
  organization = {Springer},
  timestamp = {Tue, 23 May 2017 01:00:00 +0200},
  biburl = {https://dblp.org/rec/conf/approx/McGregor05.bib},
  bibsource = {dblp computer science bibliography, https://dblp.org},
  doi = {10.1007/11538462_15},
  _bib2doi_selected = {dblp:/rec/conf/approx/McGregor05.bib},
  _bib2doi_confirmed = {true},
}

@inproceedings{eggert2009bipartite,
  title = {Bipartite graph matchings in the semi-streaming model},
  author = {Eggert, Sebastian and Kliemann, Lasse and Srivastav, Anand},
  booktitle = {Proceedings of the 17th Annual European Symposium on Algorithms (ESA 2009)},
  pages = {492--503},
  year = {2009},
  organization = {Springer},
  timestamp = {Fri, 26 May 2017 01:00:00 +0200},
  biburl = {https://dblp.org/rec/conf/esa/EggertKS09.bib},
  bibsource = {dblp computer science bibliography, https://dblp.org},
  doi = {10.1007/978-3-642-04128-0_44},
  _bib2doi_selected = {dblp:/rec/conf/esa/EggertKS09.bib},
  _bib2doi_confirmed = {true},
}

@article{ahn2013linear,
  title = {Linear programming in the semi-streaming model with application to the maximum matching problem},
  author = {Ahn, Kook Jin and Guha, Sudipto},
  journal = {Information and Computation},
  volume = {222},
  pages = {59--79},
  year = {2013},
  publisher = {Elsevier},
  note = {Announced at ICALP 2011.},
  timestamp = {Fri, 12 Feb 2021 00:00:00 +0100},
  biburl = {https://dblp.org/rec/journals/iandc/AhnG13.bib},
  bibsource = {dblp computer science bibliography, https://dblp.org},
  doi = {10.1016/j.ic.2012.10.006},
  _bib2doi_selected = {dblp:/rec/journals/iandc/AhnG13.bib},
  _bib2doi_confirmed = {true},
}

@article{eggert2012bipartite,
  title = {Bipartite matching in the semi-streaming model},
  author = {Eggert, Sebastian and Kliemann, Lasse and Munstermann, Peter and Srivastav, Anand},
  journal = {Algorithmica},
  volume = {63},
  number = {1},
  pages = {490--508},
  year = {2012},
  publisher = {Springer},
  timestamp = {Wed, 17 May 2017 01:00:00 +0200},
  biburl = {https://dblp.org/rec/journals/algorithmica/EggertKMS12.bib},
  bibsource = {dblp computer science bibliography, https://dblp.org},
  doi = {10.1007/s00453-011-9556-8},
  _bib2doi_selected = {dblp:/rec/journals/algorithmica/EggertKMS12.bib},
  _bib2doi_confirmed = {true},
}

@inproceedings{kapralov2013better,
  title = {Better bounds for matchings in the streaming model},
  author = {Kapralov, Michael},
  booktitle = {Proceedings of the 24th Annual ACM-SIAM Symposium on Discrete Algorithms (SODA 2013)},
  pages = {1679--1697},
  year = {2013},
  organization = {SIAM},
  timestamp = {Sun, 06 Oct 2024 01:00:00 +0200},
  biburl = {https://dblp.org/rec/conf/soda/Kapralov13.bib},
  bibsource = {dblp computer science bibliography, https://dblp.org},
  doi = {10.1137/1.9781611973105.121},
  _bib2doi_selected = {dblp:/rec/conf/soda/Kapralov13.bib},
  _bib2doi_confirmed = {true},
}

@inproceedings{tirodkar2018deterministic,
  title = {Deterministic algorithms for maximum matching on general graphs in the semi-streaming model},
  author = {Tirodkar, Sumedh},
  booktitle = {Proceedings of the 38th IARCS Annual Conference on Foundations of Software Technology and Theoretical Computer Science (FSTTCS 2018)},
  year = {2018},
  organization = {Schloss-Dagstuhl-Leibniz Zentrum f{\"u}r Informatik},
  doi = {10.4230/LIPIcs.FSTTCS.2018.39},
  timestamp = {Thu, 02 May 2019 01:00:00 +0200},
  biburl = {https://dblp.org/rec/conf/fsttcs/Tirodkar18.bib},
  bibsource = {dblp computer science bibliography, https://dblp.org},
  _bib2doi_selected = {dblp:/rec/conf/fsttcs/Tirodkar18.bib},
  _bib2doi_confirmed = {true},
}

@inproceedings{gamlath2019weighted,
  title = {Weighted matchings via unweighted augmentations},
  author = {Gamlath, Buddhima and Kale, Sagar and Mitrovic, Slobodan and Svensson, Ola},
  booktitle = {Proceedings of the 38th Symposium on Principles of Distributed Computing (PODC 2019)},
  pages = {491--500},
  year = {2019},
  timestamp = {Sun, 02 Oct 2022 01:00:00 +0200},
  biburl = {https://dblp.org/rec/conf/podc/GamlathKMS19.bib},
  bibsource = {dblp computer science bibliography, https://dblp.org},
  doi = {10.1145/3293611.3331603},
  _bib2doi_selected = {dblp:/rec/conf/podc/GamlathKMS19.bib},
  _bib2doi_confirmed = {true},
}

@inproceedings{assadi2021auction,
  title = {An auction algorithm for bipartite matching in streaming and massively parallel computation models},
  author = {Assadi, Sepehr and Liu, S Cliff and Tarjan, Robert E},
  booktitle = {Proceedings of the 4th Symposium on Simplicity in Algorithms (SOSA 2021)},
  pages = {165--171},
  year = {2021},
  organization = {SIAM},
  timestamp = {Wed, 17 Mar 2021 00:00:00 +0100},
  biburl = {https://dblp.org/rec/conf/sosa/AssadiLT21.bib},
  bibsource = {dblp computer science bibliography, https://dblp.org},
  doi = {10.1137/1.9781611976496.18},
  _bib2doi_selected = {dblp:/rec/conf/sosa/AssadiLT21.bib},
  _bib2doi_confirmed = {true},
}

@inproceedings{fischer2022deterministic,
  title = {Deterministic {$(1+ \varepsilon)$}-approximate maximum matching with poly {$(1/\varepsilon)$} passes in the semi-streaming model and beyond},
  author = {Fischer, Manuela and Mitrovi{\'c}, Slobodan and Uitto, Jara},
  booktitle = {Proceedings of the 54th Annual ACM SIGACT Symposium on Theory of Computing (STOC 2022)},
  pages = {248--260},
  year = {2022},
  timestamp = {Wed, 07 Dec 2022 00:00:00 +0100},
  biburl = {https://dblp.org/rec/conf/stoc/FischerMU22.bib},
  bibsource = {dblp computer science bibliography, https://dblp.org},
  doi = {10.1145/3519935.3520039},
  _bib2doi_selected = {dblp:/rec/conf/stoc/FischerMU22.bib},
  _bib2doi_confirmed = {true},
}

@inproceedings{huang20231,
  title = {{$(1- \varepsilon)$}-Approximate Maximum Weighted Matching in poly{$(1/\varepsilon, \log n)$} Time in the Distributed and Parallel Settings},
  author = {Huang, Shang-En and Su, Hsin-Hao},
  booktitle = {Proceedings of the 42nd Symposium on Principles of Distributed Computing (PODC 2023)},
  pages = {44--54},
  year = {2023},
  doi = {10.1145/3583668.3594570},
}

@article{ahn2011laminar,
  title = {Laminar families and metric embeddings: {Non-bipartite} maximum matching problem in the semi-streaming model},
  author = {Ahn, Kook Jin and Guha, Sudipto},
  journal = {arXiv preprint arXiv:1104.4058},
  year = {2011},
  doi = {10.48550/arXiv.1104.4058},
  timestamp = {Mon, 13 Aug 2018 01:00:00 +0200},
  biburl = {https://dblp.org/rec/journals/corr/abs-1104-4058.bib},
  bibsource = {dblp computer science bibliography, https://dblp.org},
  _bib2doi_selected = {dblp:/rec/journals/corr/abs-1104-4058.bib},
  _bib2doi_confirmed = {true},
}

@article{ahn2018access,
  title = {Access to data and number of iterations: {Dual} primal algorithms for maximum matching under resource constraints},
  author = {Ahn, Kook Jin and Guha, Sudipto},
  journal = {ACM Transactions on Parallel Computing (TOPC)},
  volume = {4},
  number = {4},
  pages = {1--40},
  year = {2018},
  publisher = {ACM},
  note = {Announced at SPAA 2015.},
  timestamp = {Sun, 19 Jan 2025 00:00:00 +0100},
  biburl = {https://dblp.org/rec/journals/topc/AhnG18.bib},
  bibsource = {dblp computer science bibliography, https://dblp.org},
  doi = {10.1145/3154855},
  _bib2doi_selected = {dblp:/rec/journals/topc/AhnG18.bib},
  _bib2doi_confirmed = {true},
}

@inproceedings{assadi2022semi,
  title = {Semi-Streaming Bipartite Matching in Fewer Passes and Optimal Space},
  author = {Assadi, Sepehr and Jambulapati, Arun and Jin, Yujia and Sidford, Aaron and Tian, Kevin},
  booktitle = {Proceedings of the 33rd Annual ACM-SIAM Symposium on Discrete Algorithms (SODA 2022)},
  pages = {627--669},
  year = {2022},
  organization = {SIAM},
  timestamp = {Tue, 12 Apr 2022 01:00:00 +0200},
  biburl = {https://dblp.org/rec/conf/soda/AssadiJJST22.bib},
  bibsource = {dblp computer science bibliography, https://dblp.org},
  doi = {10.1137/1.9781611977073.29},
  _bib2doi_selected = {dblp:/rec/conf/soda/AssadiJJST22.bib},
  _bib2doi_confirmed = {true},
}

@inproceedings{kale_et_al:LIPIcs.APPROX-RANDOM.2017.15,
  author = {Kale, Sagar and Tirodkar, Sumedh},
  title = {Maximum Matching in Two, Three, and a Few More Passes Over Graph Streams},
  booktitle = {Proceedings of the 20th International Workshop on Approximation Algorithms for Combinatorial Optimization Problems (APPROX 2017) and the 21st International Workshop on Randomized and Computation (RANDOM 2017)},
  pages = {15:1--15:21},
  year = {2017},
  volume = {81},
  doi = {10.4230/LIPIcs.APPROX-RANDOM.2017.15},
  timestamp = {Sun, 02 Oct 2022 01:00:00 +0200},
  biburl = {https://dblp.org/rec/conf/approx/KaleT17.bib},
  bibsource = {dblp computer science bibliography, https://dblp.org},
  _bib2doi_selected = {dblp:/rec/conf/approx/KaleT17.bib},
  _bib2doi_confirmed = {true},
}

@inproceedings{quanrud:LIPIcs.ICALP.2024.118,
  author = {Quanrud, Kent},
  title = {Adaptive Sparsification for Matroid Intersection},
  booktitle = {Proceedings of the 51st International Colloquium on Automata, Languages, and Programming (ICALP 2024)},
  pages = {118:1--118:20},
  year = {2024},
  volume = {297},
  doi = {10.4230/LIPIcs.ICALP.2024.118},
  timestamp = {Tue, 02 Jul 2024 01:00:00 +0200},
  biburl = {https://dblp.org/rec/conf/icalp/Quanrud24.bib},
  bibsource = {dblp computer science bibliography, https://dblp.org},
  _bib2doi_selected = {dblp:/rec/conf/icalp/Quanrud24.bib},
  _bib2doi_confirmed = {true},
}

@inproceedings{chekuri2015streaming,
  title = {Streaming algorithms for submodular function maximization},
  author = {Chekuri, Chandra and Gupta, Shalmoli and Quanrud, Kent},
  booktitle = {Proceedings of the 42nd International Colloquium on Automata, Languages, and Programming (ICALP 2015)},
  pages = {318--330},
  year = {2015},
  organization = {Springer},
  timestamp = {Tue, 23 May 2017 01:00:00 +0200},
  biburl = {https://dblp.org/rec/conf/icalp/ChekuriGQ15.bib},
  bibsource = {dblp computer science bibliography, https://dblp.org},
  doi = {10.1007/978-3-662-47672-7_26},
  _bib2doi_selected = {dblp:/rec/conf/icalp/ChekuriGQ15.bib},
  _bib2doi_confirmed = {true},
}

@article{feldman2018less,
  title = {Do less, get more: {Streaming} submodular maximization with subsampling},
  author = {Feldman, Moran and Karbasi, Amin and Kazemi, Ehsan},
  journal = {Advances in Neural Information Processing Systems 31: Proceedings of the 32th Annual Conference on Neural Information Processing Systems (Neurips 2018)},
  volume = {31},
  year = {2018},
}

@article{garg2023semi,
  title = {Semi-streaming algorithms for submodular matroid intersection},
  author = {Garg, Paritosh and Jordan, Linus and Svensson, Ola},
  journal = {Mathematical Programming},
  volume = {197},
  number = {2},
  pages = {967--990},
  year = {2023},
  publisher = {Springer},
  note = {Announced at IPCO 2021.},
  timestamp = {Mon, 28 Aug 2023 01:00:00 +0200},
  biburl = {https://dblp.org/rec/journals/mp/GargJS23.bib},
  bibsource = {dblp computer science bibliography, https://dblp.org},
  doi = {10.1007/s10107-022-01858-9},
  _bib2doi_selected = {dblp:/rec/journals/mp/GargJS23.bib},
  _bib2doi_confirmed = {true},
}

@inproceedings{blikstad2024efficient,
  title = {Efficient Matroid Intersection via a Batch-Update Auction Algorithm},
  author = {Blikstad, Joakim and Tu, Ta-Wei},
  booktitle = {Proceedings of the 8th Symposium on Simplicity in Algorithms (SOSA 2025)},
  pages = {226--237},
  year = {2025},
  organization = {SIAM},
  doi = {10.1137/1.9781611978315.18},
  timestamp = {Mon, 10 Mar 2025 00:00:00 +0100},
  biburl = {https://dblp.org/rec/conf/sosa/BlikstadT25.bib},
  bibsource = {dblp computer science bibliography, https://dblp.org},
  _bib2doi_selected = {dblp:/rec/conf/sosa/BlikstadT25.bib},
  _bib2doi_confirmed = {true},
}

@inproceedings{xu1994fast,
  title = {Fast algorithms for transversal matroid intersection problems},
  author = {Xu, Ying and Gabow, Harold N},
  booktitle = {Proceedings of the 5th International Symposium on Algorithms and Computation (ISAAC 1994)},
  pages = {625--633},
  year = {1994},
  organization = {Springer},
  timestamp = {Tue, 07 Nov 2017 00:00:00 +0100},
  biburl = {https://dblp.org/rec/conf/isaac/XuG94.bib},
  bibsource = {dblp computer science bibliography, https://dblp.org},
  doi = {10.1007/3-540-58325-4_231},
  _bib2doi_selected = {dblp:/rec/conf/isaac/XuG94.bib},
  _bib2doi_confirmed = {true},
}

@inproceedings{gabow1985efficient,
  title = {Efficient algorithms for graphic matroid intersection and parity},
  author = {Gabow, Harold N and Stallmann, Matthias},
  booktitle = {Proceedings of the 12th Colloquium on Automata, Languages and Programming (ICALP 1985)},
  pages = {210--220},
  year = {1985},
  organization = {Springer},
  doi={10.1007/BFb0015746}
}

@article{fujishige1995efficient,
  title = {An efficient cost scaling algorithm for the independent assignment problem},
  author = {Fujishige, Satoru and Xiaodong, Zhang},
  journal = {Journal of the Operations Research Society of Japan},
  volume = {38},
  number = {1},
  pages = {124--136},
  year = {1995},
  publisher = {The Operations Research Society of Japan},
}

@article{cheung2014algebraic,
  title = {Algebraic algorithms for linear matroid parity problems},
  author = {Cheung, Ho Yee and Lau, Lap Chi and Leung, Kai Man},
  journal = {ACM Transactions on Algorithms (TALG)},
  volume = {10},
  number = {3},
  pages = {1--26},
  year = {2014},
  publisher = {ACM New York, NY, USA},
  note = {Announced at SODA 2011.},
  timestamp = {Sun, 19 Jan 2025 00:00:00 +0100},
  biburl = {https://dblp.org/rec/journals/talg/CheungLL14.bib},
  bibsource = {dblp computer science bibliography, https://dblp.org},
  doi = {10.1145/2601066},
  _bib2doi_selected = {dblp:/rec/journals/talg/CheungLL14.bib},
  _bib2doi_confirmed = {true},
}

@inproceedings{gabow1979efficient,
  title = {Efficient algorithms for simple matroid intersection problems},
  author = {Gabow, Harold N and Tarjan, Robert E},
  booktitle = {Proceedings of the 20th Annual Symposium on Foundations of Computer Science (FOCS 1979)},
  pages = {196--204},
  year = {1979},
  organization = {IEEE},
  doi={10.1109/SFCS.1979.14}
}

@book{oxley2006matroid,
  title = {Matroid theory},
  author = {Oxley, James G},
  volume = {3},
  year = {2006},
  publisher = {Oxford University Press, USA},
}

@book{cygan2015parameterized,
  title = {Parameterized algorithms},
  author = {Cygan, Marek and Fomin, Fedor V and Kowalik, {\L}ukasz and Lokshtanov, Daniel and Marx, D{\'a}niel and Pilipczuk, Marcin and Pilipczuk, Micha{\l} and Saurabh, Saket},
  year = {2015},
  publisher = {Springer},
  timestamp = {Sun, 25 Oct 2020 01:00:00 +0200},
  biburl = {https://dblp.org/rec/books/sp/CyganFKLMPPS15.bib},
  bibsource = {dblp computer science bibliography, https://dblp.org},
  doi = {10.1007/978-3-319-21275-3},
  _bib2doi_selected = {dblp:/rec/books/sp/CyganFKLMPPS15.bib},
  _bib2doi_confirmed = {true},
}

@article{gabow1986augmenting,
  title={An augmenting path algorithm for linear matroid parity},
  author={Gabow, Harold N and Stallmann, Matthias},
  journal={Combinatorica},
  volume={6},
  number={2},
  pages={123--150},
  year={1986},
  publisher={Springer},
  note={Announced at FOCS 1984.},
  doi={10.1007/BF02579169}
}

@article{iwata2021weighted,
  title={A Weighted Linear Matroid Parity Algorithm},
  author={Iwata, Satoru and Kobayashi, Yusuke},
  journal={SIAM Journal on Computing},
  volume={51},
  number={2},
  pages={STOC17--238},
  year={2021},
  note={Announced at STOC 2017.},
  doi={10.1137/17M1141709}
}

@inproceedings{micali1980v,
  title={An {$O(\sqrt{V} E)$} algoithm for finding maximum matching in general graphs},
  author={Micali, Silvio and Vazirani, Vijay V},
  booktitle={Proceedings of the 21st Annual Symposium on Foundations of Computer Science (FOCS 1980)},
  pages={17--27},
  year={1980},
  organization={IEEE}
}

@article{lee2013matroid,
  title={Matroid Matching: The Power of Local Search},
  author={Lee, Jon and Sviridenko, Maxim and Vondr\'ak, Jan},
  journal={SIAM Journal on Computing},
  volume={42},
  number={1},
  pages={357--379},
  year={2013},
  publisher={SIAM},
  note={Announced at STOC 2010.}
}

@article{izumi2025forgetting,
  title={Forgetting Alternation and Blossoms: A New Framework for Fast Matching Augmentation and Its Applications to Sequential/Distributed/Streaming Computation},
  author={Izumi, Taisuke and Kitamura, Naoki and Yamaguchi, Yutaro},
  journal={arXiv preprint arXiv:2511.08210},
  year={2025}
}

@inproceedings{chuzhoy2024faster,
  title={A faster combinatorial algorithm for maximum bipartite matching},
  author={Chuzhoy, Julia and Khanna, Sanjeev},
  booktitle={Proceedings of the 35th Annual ACM-SIAM Symposium on Discrete Algorithms (SODA 2024)},
  pages={2185--2235},
  year={2024},
  organization={SIAM}
}

@inproceedings{chuzhoy2024maximum,
  title={Maximum Bipartite Matching in ${n^{2 + o(1)}}$ Time via a Combinatorial Algorithm},
  author={Chuzhoy, Julia and Khanna, Sanjeev},
  booktitle={Proceedings of the 56th Annual ACM Symposium on Theory of Computing (STOC 2024)},
  pages={83--94},
  year={2024}
}

@inproceedings{bernstein2024maximum,
  title={Maximum flow by augmenting paths in ${n^{2 + o(1)}}$ time},
  author={Bernstein, Aaron and Blikstad, Joakim and Saranurak, Thatchaphol and Tu, Ta-Wei},
  booktitle={Proceedings of the 65th Annual Symposium on Foundations of Computer Science (FOCS 2024)},
  pages={2056--2077},
  year={2024},
  organization={IEEE}
}

@inproceedings{bernstein2025combinatorial,
  title={Combinatorial Maximum Flow via Weighted Push-Relabel on Shortcut Graphs},
  author={Bernstein, Aaron and Blikstad, Joakim and Li, Jason and Saranurak, Thatchaphol and Tu, Ta-Wei},
  booktitle={Proceedings of the 66th Annual Symposium on Foundations of Computer Science (FOCS 2025)},
  pages={464-485},
  year={2025},
  organization={IEEE}
}

@book{lawler1976combinatorial,
  author    = {Lawler, Eugene L.},
  title     = {Combinatorial Optimization: Networks and Matroids},
  publisher = {Holt, Rinehart and Winston},
  year      = {1976}
}

@book{garey2002computers,
  title={Computers and intractability: {A} guide to the theory of {NP}-completeness},
  author={Garey, Michael R and Johnson, David S},
  year={1979},
  publisher={W.H. Freeman and company}
}

@inproceedings{arkhipov2026faster,
  title={Faster algorithms for packing forests in graphs and related problems},
  author={Arkhipov, Pavel and Kolmogorov, Vladimir},
  booktitle={Proceedings of the 37th Annual ACM-SIAM Symposium on Discrete Algorithms (SODA 2026)},
  pages={4023--4042},
  year={2026},
  organization={SIAM}
}

@article{assadi2025simple,
  title={A Simple {$(1-\varepsilon)$}-Approximation Semi-Streaming Algorithm for Maximum (Weighted) Matching},
  author={Assadi, Sepehr},
  journal={TheoretiCS},
  volume={4},
  year={2025},
  publisher={Episciences. org},
  note={Announced at SOSA 2024.}
}

@inproceedings{assadi2026better,
  title={Better Bounds for Semi-Streaming Single-Source Shortest Paths},
  author={Assadi, Sepehr and Hoppenworth, Gary and Sundaresan, Janani},
  booktitle={Proceedings of the 37th Annual ACM-SIAM Symposium on Discrete Algorithms (SODA 2026)},
  pages={5093--5116},
  year={2026},
  organization={SIAM}
}

@inproceedings{even1975n2,
  title={An {$O(n^{2.5})$} algorithm for maximum matching in general graphs},
  author={Even, Shimon and Kariv, Oded},
  booktitle={Proceedings of the 16th Annual Symposium on Foundations of Computer Science (FOCS 1975)},
  pages={100--112},
  year={1975},
  organization={IEEE}
}

@article{chen2025maximum,
  title={Maximum flow and minimum-cost flow in almost-linear time},
  author={Chen, Li and Kyng, Rasmus and Liu, Yang and Peng, Richard and Probst Gutenberg, Maximilian and Sachdeva, Sushant},
  journal={Journal of the ACM},
  volume={72},
  number={3},
  pages={1--103},
  year={2025},
  publisher={ACM New York, NY},
  note={Announced at FOCS 2022.}
}

@inproceedings{van2023deterministic,
  title={A deterministic almost-linear time algorithm for minimum-cost flow},
  author={Van Den Brand, Jan and Chen, Li and Kyng, Rasmus and Liu, Yang P and Peng, Richard and Gutenberg, Maximilian Probst and Sachdeva, Sushant and Sidford, Aaron},
  booktitle={Proceedings of the 64th Annual Symposium on Foundations of Computer Science (FOCS 2023)},
  pages={503--514},
  year={2023},
  organization={IEEE}
}

@article{terao2025faster,
  title={Faster matroid partition algorithms},
  author={Terao, Tatsuya},
  journal={ACM Transactions on Algorithms (TALG)},
  volume={21},
  number={2},
  pages={1--26},
  year={2025},
  publisher={ACM New York, NY},
  note={Announced at ICALP 2023.}
}

@article{edmonds1965paths,
  title={Paths, trees, and flowers},
  author={Edmonds, Jack},
  journal={Canadian Journal of mathematics},
  volume={17},
  pages={449--467},
  year={1965},
  publisher={Cambridge University Press}
}

@article{goldberg2004maximum,
  title={Maximum skew-symmetric flows and matchings},
  author={Goldberg, Andrew V and Karzanov, Alexander V},
  journal={Mathematical Programming},
  volume={100},
  number={3},
  pages={537--568},
  year={2004},
  publisher={Springer}
}

@article{jensen1982complexity,
  title={Complexity of matroid property algorithms},
  author={Jensen, Per M and Korte, Bernhard},
  journal={SIAM Journal on Computing},
  volume={11},
  number={1},
  pages={184--190},
  year={1982},
  publisher={SIAM}
}

@article{lovasz1980matroid,
  title={Matroid matching and some applications},
  author={Lov{\'a}sz, L{\'a}szl{\'o}},
  journal={Journal of Combinatorial Theory, Series B},
  volume={28},
  number={2},
  pages={208--236},
  year={1980},
  publisher={Elsevier}
}

@incollection{lovasz1978matroidmatching,
  author    = {Lov{\'a}sz, L{\'a}szl{\'o}},
  title     = {The matroid matching problem},
  booktitle = {Algebraic Methods in Graph Theory},
  series    = {Colloq. Math. Soc. J{\'a}nos Bolyai},
  volume    = {25},
  pages     = {495--517},
  year      = {1978}
}

@article{lovasz1980selecting-independent-lines,
  author  = {Lov{\'a}sz, L{\'a}szl{\'o}},
  title   = {Selecting independent lines from a family of lines in a space},
  journal = {Acta Scientiarum Mathematicarum},
  volume  = {42},
  pages   = {121--131},
  year    = {1980}
}

@inproceedings{madry2013navigating,
  title={Navigating central path with electrical flows: From flows to matchings, and back},
  author={Madry, Aleksander},
  booktitle={Proceedings of the 54th Annual Symposium on Foundations of Computer Science (FOCS 2013)},
  pages={253--262},
  year={2013},
  organization={IEEE}
}

@article{lee2019solving,
  title={Solving linear programs with {$\tilde{O}(\sqrt{\mathrm{rank}})$} linear system solves},
  author={Lee, Yin Tat and Sidford, Aaron},
  journal={arXiv preprint arXiv:1910.08033},
  year={2019}
}

@inproceedings{cohen2017negative,
  title={Negative-weight shortest paths and unit capacity minimum cost flow in {$\tilde{O}(m^{10/7} \log W)$} time},
  author={Cohen, Michael B and M{\k{a}}dry, Aleksander and Sankowski, Piotr and Vladu, Adrian},
  booktitle={Proceedings of the 28th Annual ACM-SIAM Symposium on Discrete Algorithms (SODA 2017)},
  pages={752--771},
  year={2017},
  organization={SIAM}
}

@inproceedings{liu2020faster,
  title={Faster energy maximization for faster maximum flow},
  author={Liu, Yang P and Sidford, Aaron},
  booktitle={Proceedings of the 52nd Annual ACM SIGACT Symposium on Theory of Computing (STOC 2020)},
  pages={803--814},
  year={2020}
}

@inproceedings{axiotis2020circulation,
  title={Circulation control for faster minimum cost flow in unit-capacity graphs},
  author={Axiotis, Kyriakos and M{\k{a}}dry, Aleksander and Vladu, Adrian},
  booktitle={Proceedings of the 61st Annual Symposium on Foundations of Computer Science (FOCS 2020)},
  pages={93--104},
  year={2020},
  organization={IEEE}
}

@inproceedings{van2020bipartite,
  title={Bipartite matching in nearly-linear time on moderately dense graphs},
  author={van den Brand, Jan and Lee, Yin-Tat and Nanongkai, Danupon and Peng, Richard and Saranurak, Thatchaphol and Sidford, Aaron and Song, Zhao and Wang, Di},
  booktitle={Proceedings of the 61st Annual Symposium on Foundations of Computer Science (FOCS 2020)},
  pages={919--930},
  year={2020},
  organization={IEEE}
}

@article{ford1956maximal,
  title={Maximal flow through a network},
  author={Ford Jr, Lester R and Fulkerson, Delbert R},
  journal={Canadian journal of Mathematics},
  volume={8},
  pages={399--404},
  year={1956},
  publisher={Cambridge University Press}
}

@article{gabow1976efficient,
  title={An efficient implementation of {E}dmonds' algorithm for maximum matching on graphs},
  author={Gabow, Harold N},
  journal={Journal of the ACM (JACM)},
  volume={23},
  number={2},
  pages={221--234},
  year={1976},
  publisher={ACM New York, NY, USA}
}

@InProceedings{huang_et_al:LIPIcs.APPROX/RANDOM.2020.62,
  author =	{Huang, Chien-Chung and Thiery, Theophile and Ward, Justin},
  title =	{Improved Multi-Pass Streaming Algorithms for Submodular Maximization with Matroid Constraints},
  booktitle =	{Proceedings of the 23th International Workshop on Approximation Algorithms for Combinatorial Optimization Problems (APPROX 2020) and the 24th International Workshop on Randomized and Computation (RANDOM 2020)},
  pages =	{62:1--62:19},
  year =	{2020},
  volume =	{176},
  doi =		{10.4230/LIPIcs.APPROX/RANDOM.2020.62},
}

@article{huang2023fpt,
  title={{FPT}-Algorithms for the {$\ell$}-Matchoid Problem with a Coverage Objective},
  author={Huang, Chien-Chung and Ward, Justin},
  journal={SIAM Journal on Discrete Mathematics},
  volume={37},
  number={2},
  pages={1053--1078},
  year={2023},
  publisher={SIAM},
  doi={10.1137/21M1442267}
}

@article{chakrabarti2015submodular,
  title={Submodular maximization meets streaming: matchings, matroids, and more},
  author={Chakrabarti, Amit and Kale, Sagar},
  journal={Mathematical Programming},
  volume={154},
  pages={225--247},
  year={2015},
  publisher={Springer},
  note={Announced at IPCO 2014.},
  doi={10.1007/s10107-015-0900-7}
}

@article{dudeja2026weighted,
  title={A Weighted-to-Unweighted Reduction for Matroid Intersection},
  author={Dudeja, Aditi and Grilnberger, Mara},
  journal={arXiv preprint arXiv:2602.15702},
  year={2026},
  note={To appear in IPCO 2026.}
}

@InProceedings{terao:LIPIcs.WADS.2025.50,
  author =	{Terao, Tatsuya},
  title =	{Deterministic $(2/3 - \varepsilon)$-Approximation of Matroid Intersection Using Nearly-Linear Independence-Oracle Queries},
  booktitle =	{Proceedings of the 19th International Symposium on Algorithms and Data Structures (WADS 2025)},
  pages =	{50:1--50:18},
  year =	{2025},
  volume =	{349},
  doi =		{10.4230/LIPIcs.WADS.2025.50}
}

@inproceedings{goel2012communication,
  title = {On the communication and streaming complexity of maximum bipartite matching},
  author = {Goel, Ashish and Kapralov, Michael and Khanna, Sanjeev},
  booktitle = {Proceedings of the 23rd Annual ACM-SIAM Symposium on Discrete Algorithms (SODA 2012)},
  pages = {468--485},
  year = {2012},
  organization = {SIAM},
  timestamp = {Sun, 06 Oct 2024 01:00:00 +0200},
  biburl = {https://dblp.org/rec/conf/soda/GoelKK12.bib},
  bibsource = {dblp computer science bibliography, https://dblp.org},
  doi = {10.1137/1.9781611973099.41},
  _bib2doi_selected = {dblp:/rec/conf/soda/GoelKK12.bib},
  _bib2doi_confirmed = {true},
}

@inproceedings{konrad_et_al:LIPIcs.STACS.2023.41,
  author = {Konrad, Christian and Naidu, Kheeran K. and Steward, Arun},
  title = {Maximum Matching via Maximal Matching Queries},
  booktitle = {Proceedings of the 40th International Symposium on Theoretical Aspects of Computer Science (STACS 2023)},
  pages = {41:1--41:22},
  year = {2023},
  volume = {254},
  doi = {10.4230/LIPIcs.STACS.2023.41},
  timestamp = {Tue, 21 Mar 2023 00:00:00 +0100},
  biburl = {https://dblp.org/rec/conf/stacs/0001NS23.bib},
  bibsource = {dblp computer science bibliography, https://dblp.org},
  _bib2doi_selected = {dblp:/rec/conf/stacs/0001NS23.bib},
  _bib2doi_confirmed = {true},
}

@article{feldman2024maximum,
  title = {Maximum matching sans maximal matching: {A} new approach for finding maximum matchings in the data stream model},
  author = {Feldman, Moran and Szarf, Ariel},
  journal = {Algorithmica},
  volume = {86},
  number = {4},
  pages = {1173--1209},
  year = {2024},
  publisher = {Springer},
  note = {Announced at APPROX/RANDOM 2022.},
  timestamp = {Mon, 15 Apr 2024 01:00:00 +0200},
  biburl = {https://dblp.org/rec/journals/algorithmica/FeldmanS24.bib},
  bibsource = {dblp computer science bibliography, https://dblp.org},
  doi = {10.1007/s00453-023-01190-4},
  _bib2doi_selected = {dblp:/rec/journals/algorithmica/FeldmanS24.bib},
  _bib2doi_confirmed = {true},
}

@inproceedings{esfandiari2016finding,
  title = {Finding large matchings in semi-streaming},
  author = {Esfandiari, Hossein and Hajiaghayi, MohammadTaghi and Monemizadeh, Morteza},
  booktitle = {Proceedings of the 16th International Conference on Data Mining Workshops (ICDMW 2016)},
  pages = {608--614},
  year = {2016},
  organization = {IEEE},
  timestamp = {Fri, 24 Mar 2023 00:00:00 +0100},
  biburl = {https://dblp.org/rec/conf/icdm/EsfandiariHM16.bib},
  bibsource = {dblp computer science bibliography, https://dblp.org},
  doi = {10.1109/ICDMW.2016.0092},
  _bib2doi_selected = {dblp:/rec/conf/icdm/EsfandiariHM16.bib},
  _bib2doi_confirmed = {true},
}

@inproceedings{konrad2018simple,
  title = {A simple augmentation method for matchings with applications to streaming algorithms},
  author = {Konrad, Christian},
  booktitle = {Proceedings of the 43rd International Symposium on Mathematical Foundations of Computer Science (MFCS 2018)},
  pages = {74--1},
  year = {2018},
  doi={10.4230/LIPIcs.MFCS.2018.74}
}

@article{bernstein2023improved,
  title = {Improved bounds for matching in random-order streams},
  author = {Bernstein, Aaron},
  journal = {Theory of Computing Systems},
  pages = {1--15},
  year = {2023},
  publisher = {Springer},
  note = {Announced at ICALP 2020.},
  doi={10.1007/s00224-023-10155-7}
}

@inproceedings{assadi_et_al:LIPIcs.ICALP.2021.19,
  author = {Assadi, Sepehr and Behnezhad, Soheil},
  title = {Beating Two-Thirds For Random-Order Streaming Matching},
  booktitle = {Proceedings of the 48th International Colloquium on Automata, Languages, and Programming (ICALP 2021)},
  pages = {19:1--19:13},
  year = {2021},
  volume = {198},
  doi = {10.4230/LIPIcs.ICALP.2021.19},
  timestamp = {Tue, 06 Jul 2021 01:00:00 +0200},
  biburl = {https://dblp.org/rec/conf/icalp/AssadiB21.bib},
  bibsource = {dblp computer science bibliography, https://dblp.org},
  _bib2doi_selected = {dblp:/rec/conf/icalp/AssadiB21.bib},
  _bib2doi_confirmed = {true},
}

@inproceedings{konrad2012maximum,
  title = {Maximum matching in semi-streaming with few passes},
  author = {Konrad, Christian and Magniez, Fr{\'e}d{\'e}ric and Mathieu, Claire},
  booktitle = {Proceedings of the 15th International Workshop on Approximation Algorithms for Combinatorial Optimization Problems (APPROX 2012) and the 16th International Workshop on Randomized and Computation (RANDOM 2012)},
  pages = {231--242},
  year = {2012},
  organization = {Springer},
  timestamp = {Tue, 28 Feb 2023 00:00:00 +0100},
  biburl = {https://dblp.org/rec/conf/approx/KonradMM12.bib},
  bibsource = {dblp computer science bibliography, https://dblp.org},
  doi = {10.1007/978-3-642-32512-0_20},
  _bib2doi_selected = {dblp:/rec/conf/approx/KonradMM12.bib},
  _bib2doi_confirmed = {true},
}

@article{sleator1983data,
  title={A data structure for dynamic trees},
  author={Sleator, Daniel D and Tarjan, Robert Endre},
  journal={Journal of Computer and System Sciences},
  volume={26},
  number={3},
  pages={362--391},
  year={1983},
  note={Announced at STOC 1981.}
}

@article{huang2026polynomial,
  title={Polynomial Kernels with Reachability for Weighted $d$-Matroid Intersection},
  author={Huang, Chien-Chung and Kakimura, Naonori and Kobayashi, Yusuke and Terao, Tatsuya},
  journal={arXiv preprint arXiv:2603.17345},
  year={2026},
  note={To appear in IPCO 2026.}
}

@article{cualinescu1998better,
  title={A better approximation algorithm for finding planar subgraphs},
  author={C{\u{a}}linescu, Gruia and Fernandes, Cristina G and Finkler, Ulrich and Karloff, Howard},
  journal={Journal of Algorithms},
  volume={27},
  number={2},
  pages={269--302},
  year={1998},
  publisher={Elsevier},
  note={Announced at SODA 1996.}
}

@inproceedings{promel1997rnc,
  title={{RNC}-approximation algorithms for the {S}teiner problem},
  author={Pr{\"o}mel, Hans J{\"u}rgen and Steger, Angelika},
  booktitle={Proceedings of the 14th Annual Symposium on Theoretical Aspects of Computer Science (STACS 1997)},
  pages={559--570},
  year={1997},
  organization={Springer}
}

@inproceedings{berman2006applications,
  title={Applications of the linear matroid parity algorithm to approximating Steiner trees},
  author={Berman, Piotr and F{\"u}rer, Martin and Zelikovsky, Alexander},
  booktitle={Proceedings of the 1st International Computer Science Symposium in Russia (CSR 2006)},
  pages={70--79},
  year={2006},
  organization={Springer}
}

@article{szigeti2003graphic,
  title={On the graphic matroid parity problem},
  author={Szigeti, Zolt{\'a}n},
  journal={Journal of Combinatorial Theory, Series B},
  volume={88},
  number={2},
  pages={247--260},
  year={2003},
  publisher={Elsevier}
}

@article{cualinescu2024improved,
  title={An improved algorithm for finding maximum outerplanar subgraphs},
  author={C{\u{a}}linescu, Gruia and Kaul, Hemanshu and Kudarzi, Bahareh},
  journal={Discrete Applied Mathematics},
  volume={342},
  pages={207--217},
  year={2024},
  publisher={Elsevier}
}

@article{kociumaka2014faster,
  title={Faster deterministic feedback vertex set},
  author={Kociumaka, Tomasz and Pilipczuk, Marcin},
  journal={Information Processing Letters},
  volume={114},
  number={10},
  pages={556--560},
  year={2014},
  publisher={Elsevier}
}

@article{cao2015feedback,
  title={On feedback vertex set: New measure and new structures},
  author={Cao, Yixin and Chen, Jianer and Liu, Yang},
  journal={Algorithmica},
  volume={73},
  number={1},
  pages={63--86},
  year={2015},
  publisher={Springer},
  note={Announced at SWAT 2010.}
}

@article{iwata2021improved,
  title={Improved analysis of highest-degree branching for feedback vertex set},
  author={Iwata, Yoichi and Kobayashi, Yusuke},
  journal={Algorithmica},
  volume={83},
  number={8},
  pages={2503--2520},
  year={2021},
  publisher={Springer},
  note={Announced at IPEC 2019.}
}

@article{cualinescu20161,
  title={$1.61$-approximation for min-power strong connectivity with two power levels},
  author={C{\u{a}}linescu, Gruia},
  journal={Journal of Combinatorial Optimization},
  volume={31},
  number={1},
  pages={239--259},
  year={2016},
  publisher={Springer}
}

@inproceedings{calinescu2010min,
  title={Min-power strong connectivity},
  author={C{\u{a}}linescu, Gruia},
  booktitle={Proceedings of the International Workshop on Randomization and Approximation Techniques in Computer Science (APPROX/RANDOM 2010)},
  pages={67--80},
  year={2010},
  organization={Springer}
}

@InProceedings{chalermsook_et_al:LIPIcs.STACS.2019.19,
  author =	{Chalermsook, Parinya and Schmid, Andreas and Uniyal, Sumedha},
  title =	{A Tight Extremal Bound on the {L}ov\'{a}sz Cactus Number in Planar Graphs},
  booktitle =	{Proceedings of the 36th International Symposium on Theoretical Aspects of Computer Science (STACS 2019)},
  pages =	{19:1--19:14},
  year =	{2019},
  volume =	{126},
  doi =		{10.4230/LIPIcs.STACS.2019.19},
}

@InProceedings{calinescu_et_al:LIPIcs.ESA.2024.38,
  author =	{C\u{a}linescu, Gruia and Uniyal, Sumedha},
  title =	{Local Optimization Algorithms for Maximum Planar Subgraph},
  booktitle =	{Proceedings of the 32nd Annual European Symposium on Algorithms (ESA 2024)},
  pages =	{38:1--38:18},
  year =	{2024},
  volume =	{308},
  doi =		{10.4230/LIPIcs.ESA.2024.38},
}

@inproceedings{orlin2008fast,
  title={A fast, simpler algorithm for the matroid parity problem},
  author={Orlin, James B},
  booktitle={Proceedings of the 13th International Conference on Integer Programming and Combinatorial Optimization (IPCO 2008)},
  pages={240--258},
  year={2008},
  organization={Springer}
}

@article{orlin1990solving,
  title={Solving the linear matroid parity problem as a sequence of matroid intersection problems},
  author={Orlin, James B and Vande Vate, John H},
  journal={Mathematical Programming},
  volume={47},
  number={1},
  pages={81--106},
  year={1990},
  publisher={Springer}
}

@inproceedings{gabow1991matroid,
  title={A matroid approach to finding edge connectivity and packing arborescences},
  author={Gabow, Harold N},
  booktitle={Proceedings of the 23rd annual ACM symposium on Theory of Computing (STOC 1991)},
  pages={112--122},
  year={1991}
}

@article{gabow1998algorithms,
  title={Algorithms for graphic polymatroids and parametrics-sets},
  author={Gabow, Harold N},
  journal={Journal of Algorithms},
  volume={26},
  number={1},
  pages={48--86},
  year={1998},
  publisher={Elsevier},
  note={Announced at SODA 1995.}
}

@InProceedings{ene_et_al:LIPIcs.ICALP.2019.54,
  author =	{Ene, Alina and Nguyen, Huy L.},
  title =	{Towards Nearly-Linear Time Algorithms for Submodular Maximization with a Matroid Constraint},
  booktitle =	{Proceedings of the 46th International Colloquium on Automata, Languages, and Programming (ICALP 2019)},
  pages =	{54:1--54:14},
  year =	{2019},
  volume =	{132},
  doi =		{10.4230/LIPIcs.ICALP.2019.54}
}

@InProceedings{henzinger_et_al:LIPIcs.ICALP.2023.74,
  author =	{Henzinger, Monika and Liu, Paul and Vondr\'{a}k, Jan and Zheng, Da Wei},
  title =	{Faster Submodular Maximization for Several Classes of Matroids},
  booktitle =	{Proceedings of the 50th International Colloquium on Automata, Languages, and Programming (ICALP 2023)},
  pages =	{74:1--74:18},
  year =	{2023},
  volume =	{261},
  doi =		{10.4230/LIPIcs.ICALP.2023.74}
}

@InProceedings{khanna2025optimal,
    author =	{Khanna, Sanjeev and Putterman, Aaron and Song, Junkai},
  title =	{Optimal Parallel Basis Finding in Graphic and Related Matroids},
  booktitle =	{Proceedings of the 53rd International Colloquium on Automata, Languages, and Programming (ICALP 2026)},
  pages =	{125:1--125:22},
  year =	{2026},
  volume =	{374},
  doi =		{10.4230/LIPIcs.ICALP.2026.125},
}

@article{roskind1985note,
  title={A note on finding minimum-cost edge-disjoint spanning trees},
  author={Roskind, James and Tarjan, Robert E},
  journal={Mathematics of Operations Research},
  volume={10},
  number={4},
  pages={701--708},
  year={1985},
  publisher={INFORMS}
}

@InProceedings{terao2026faster,
  author =	{Terao, Tatsuya},
  title =	{Faster Approximate Linear Matroid Intersection},
  booktitle =	{Proceedings of the 20th Scandinavian Symposium on Algorithm Theory (SWAT 2026)},
  pages =	{39:1--39:19},
  year =	{2026},
  volume =	{370},
  doi =		{10.4230/LIPIcs.SWAT.2026.39},
}

@InProceedings{mitrovic_et_al:LIPIcs.ICALP.2025.119,
  author =	{Mitrovi\'{c}, Slobodan and Mukherjee, Anish and Sankowski, Piotr and Sheu, Wen-Horng},
  title =	{Faster Semi-Streaming Matchings via Alternating Trees},
  booktitle =	{Proceedings of the 52nd International Colloquium on Automata, Languages, and Programming (ICALP 2025)},
  pages =	{119:1--119:19},
  year =	{2025},
  volume =	{334},
  doi =		{10.4230/LIPIcs.ICALP.2025.119},
}

@inproceedings{gabow1983linear,
  title={A linear-time algorithm for a special case of disjoint set union},
  author={Gabow, Harold N and Tarjan, Robert Endre},
  booktitle={Proceedings of the 15th Annual ACM Symposium on Theory of Computing (STOC 1983)},
  pages={246--251},
  year={1983}
}

@article{objois2025adaptive,
  title={Adaptive Sparsification for Linear Programming},
  author={Objois, {\'E}tienne and Vladu, Adrian},
  journal={arXiv preprint arXiv:2510.08348},
  year={2025},
  note={To appear in ESA 2026.}
}

@inproceedings{bender2000lca,
  title={The LCA problem revisited},
  author={Bender, Michael A and Farach-Colton, Martin},
  booktitle={Proceedings of 4th Latin American Symposium on Theoretical Informatics (LATIN 2000)},
  pages={88--94},
  year={2000},
  organization={Springer}
}

@book{cormen2022introduction,
  title={Introduction to algorithms},
  author={Cormen, Thomas H and Leiserson, Charles E and Rivest, Ronald L and Stein, Clifford},
  year={2022},
  publisher={MIT press}
}

@book{de2000computational,
  title={Computational geometry: {A}lgorithms and applications},
  author={De Berg, Mark},
  year={2000},
  publisher={Springer Science \& Business Media}
}

@article{v021a009,
 author = {Kulikov, Alexander S. and Mikhailin, Ivan and Mokhov, Andrey and Podolskii, Vladimir V.},
 title = {Complexity of Linear Operators},
 year = {2025},
 pages = {1--32},
 doi = {10.4086/toc.2025.v021a009},
 publisher = {Theory of Computing},
 journal = {Theory of Computing},
 volume = {21},
 number = {9},
}

@article{furst1988finding,
  title={Finding a maximum-genus graph imbedding},
  author={Furst, Merrick L and Gross, Jonathan L and McGeoch, Lyle A},
  journal={Journal of the ACM (JACM)},
  volume={35},
  number={3},
  pages={523--534},
  year={1988},
  publisher={ACM New York, NY, USA}
}

@article{gabow1991faster,
  title={Faster scaling algorithms for general graph matching problems},
  author={Gabow, Harold N and Tarjan, Robert E},
  journal={Journal of the ACM (JACM)},
  volume={38},
  number={4},
  pages={815--853},
  year={1991},
  publisher={ACM, New York, NY, USA}
}

@misc{OAI:Almost-Linear-Time-Maximum-Cardinality-Matching-in-Sparse-General-Graphs-September-24-2026,
  author = {{OpenAI}},
  title = {Almost-Linear-Time Maximum-Cardinality Matching in General Graphs},
  howpublished = {OpenAI Math Release preprint
                  \href{https://github.com/openai/math/blob/main/preprints/Almost-Linear-Time-Maximum-Cardinality-Matching-in-Sparse-General-Graphs-September-24-2026/main.pdf}{OAI:Almost-Linear-Time-Maximum-Cardinality-Matching-in-Sparse-General-Graphs-September-24-2026}},
  year = {2026}
}

@article{gabow2017weighted,
  title={The weighted matching approach to maximum cardinality matching},
  author={Gabow, Harold N},
  journal={Fundamenta Informaticae},
  volume={154},
  number={1-4},
  pages={109--130},
  year={2017},
}

@article{vazirani2024theory,
  title={A theory of alternating paths and blossoms from the perspective of minimum length},
  author={Vazirani, Vijay V},
  journal={Mathematics of Operations Research},
  volume={49},
  number={3},
  pages={2009--2047},
  year={2024},
  publisher={Informs}
}

@article{huang2026auction,
  title={Auction-Based Algorithms for Matroid Intersection: {Near-Linear} Query Complexity and Constant-Pass Semi-Streaming},
  author={Huang, Chien-Chung and Kobayashi, Yusuke},
  journal={arXiv preprint arXiv:2609.38685},
  year={2026}
}

@article{hamada2026shortest,
  title={A Shortest Augmenting Path Algorithm for Linear Matroid Parity},
  author={Hamada, Kou and Iwata, Satoru},
  journal={arXiv preprint arXiv:2610.03030},
  year={2026},
  note={To appear in FOCS'26.}
}

@article{terao2026fasterb,
  title={A Faster Auction Algorithm for Weighted Matroid Intersection},
  author={Terao, Tatsuya},
  journal={arXiv preprint arXiv:2610.00920},
  year={2026}
}

@article{li2026maximum,
  title={Maximum Cardinality Matching in $n^{2+o(1)}$ Time},
  author={Li, George Z. and Li, Jason and Zhang, Junkai},
  journal={arXiv preprint arXiv:2610.40591},
  year={2026}
}

\appendix

\section{Matroids} \label{appendix:matroids}


For the basics of matroid theory, see, e.g., the excellent textbooks of Oxley \cite{oxley2006matroid} or Schrijver \cite{schrijver2003combinatorial}.

A pair $\M = (E, \mathcal{I})$ of a finite set $E$ and a non-empty set family $\mathcal{I} \subseteq 2^{E}$ is called a {\em matroid} if the following properties are satisfied.
    
\begin{itemize}
    \item If $I \in \mathcal{I}$ and $J \subseteq I$, then $J \in \mathcal{I}$.
    \item If $I, J \in \mathcal{I}$ and $|J| < |I|$, then there exists $e \in I \setminus J$ such that $J \cup \{e\} \in \mathcal{I}$.
\end{itemize}

A set $I \subseteq E$ is called {\em independent} if $I \in \mathcal{I}$ and {\em dependent} otherwise.



\section{Handling indices with large {$|M_i|$}} \label{sec:large:handle}

In the technical overview (Section~\ref{sec:technical_overview}), we described a simpler version of the span-membership tests, which fixes an integer parameter $C$ and handles the indices $i\in\{0,\ldots,k\}$ with $|M_i|\leq C$ and those with $|M_i|>C$ separately.
The simpler version takes $\tO(n^2+nm^{3/4})$ time, whereas the final version in Section~\ref{sec:checking_span} takes $\tO(n^2+m)$ time (Lemma~\ref{lem:fast_span_checking}).
The final version applies Lemma~\ref{lem:small_handle} with $C=n$, under which every index satisfies $|M_i|\leq C$, and hence the indices with $|M_i|>C$ need no separate treatment.
For completeness, we present here the lemma for the indices with $|M_i|>C$, which is used only in the simpler version described in the technical overview.

\begin{lemma} \label{lem:large_handle}
    With high probability, for all lines $\ell\in L$, we can decide whether there exists an index $i \in \{ 0, \ldots, k\}$ satisfying $r(A\cup E_i)>r(A)+1$ and $|M_i| > C$ such that $r(E_i\cup A\cup \ell)=r(E_i\cup A)$ in total $\displaystyle \tO\left(\frac{mn}{C}\right)$ time.
\end{lemma}

Write $c_i=|M_i|$.
Since the sets $M_i$ are disjoint subsets of the matroid matching $M$, we have $\sum_i c_i\leq |M|=O(n)$.
Hence, there are only $O(n/C)$ indices $i$ with $c_i>C$.
Therefore, it suffices to design, for a fixed index $i \in \{0, \ldots, k\}$, a data structure that answers each line query in $\tO(1)$ time.
Then, by scanning all $m$ lines for all such indices, we obtain the desired $\tO(mn/C)$ time.
The required data structure is given by the following lemma.

\begin{restatable}{lemma}{largepartquery} \label{lem:large_part_query}
    Fix an index $i\in\{0,\ldots,k\}$ such that $r(E_i\cup A)>r(A)+1$.
    There is a data structure that can be constructed in $\tO(n)$ time and answers the following query in $\tO(1)$ time.
    Given a line $\ell=\{e,\bar e\}\in L$, decide whether $r(E_i\cup A\cup \ell)=r(E_i\cup A)$.
    For each query, the answer is correct with high probability.
\end{restatable}

\begin{proof}
    Let $B_i=\Span(A\cup M_i)$.
    We build an auxiliary pattern tree for the fixed index $i$.
    Starting from the same spanning tree $M^*$, we choose the weights as in
    Definition~\ref{def:pattern}, with the additional constraints
    \begin{equation*}
        w(\beta)=0 \qquad \text{for every } \beta\in M_i .
    \end{equation*}
    We then build the pattern-tree data structure of
    Lemma~\ref{lem:pattern_computation_linear_time}.
    This takes $\tO(n)$ time.
    Let $p_i(e)$ denote the pattern of an edge $e$ with respect to this auxiliary pattern tree.

    Let $\Phi_i:\Span(M^*)\to\mathbb{F}_2^K$ be the linear map corresponding to this auxiliary pattern tree, so that $\Phi_i(e)=p_i(e)$ for every edge $e\in E$.
    The constraints in Definition~\ref{def:pattern} imply that $\Phi_i(a)=0$ for every $a\in\Span(A)$, and the additional constraints imply that $\Phi_i(\beta)=0$ for every $\beta\in M_i$.
    Hence, $\Phi_i(b)=0$ for every $b\in B_i$.
    Moreover, the same argument as Claim~\ref{claim:middle} in the proof of Lemma~\ref{lem:pattern_warmup} implies the following claim.

    \begin{claim} \label{claim:middle_special}
        For every $z\in\Span(M^*)$, if $z \notin B_i = \Span(A\cup M_i)$, then we have
        $\Pr[\Phi_i(z)=0]=2^{-K}$.
    \end{claim}


    We first consider cases~(i) and~(ii) of Lemma~\ref{lem:structual_lemma}.
    In both cases, we have $\Span(A\cup E_i)=B_i$.
    Therefore, if $e\in\Span(A\cup E_i)$, then $p_i(e)=0$.
    Conversely, if $e\notin\Span(A\cup E_i)$, then $e\notin B_i$, and hence $p_i(e)\neq 0$ with high probability.
    Thus, in cases~(i) and~(ii), membership of an edge $e$ in $\Span(A\cup E_i)$ is tested by checking whether $p_i(e)=0$.

    It remains to consider case~(iii).
    Fix any edge $x\in T_i\setminus M_i$.
    By Lemma~\ref{lem:structual_lemma}, we have
    $\Span(A\cup E_i)=\Span(B_i\cup\{x\})$.
    If $e\in\Span(A\cup E_i)$, then either $e\in B_i$ or $e+x\in B_i$.
    Hence, $p_i(e)\in\{0,p_i(x)\}$.
    Conversely, if $e\notin\Span(A\cup E_i)$, then neither $e$ nor $e+x$ belongs to $B_i$.
    Therefore, by Claim~\ref{claim:middle_special}, if $e\notin\Span(A\cup E_i)$, we have
    \begin{equation*}
        \Pr[p_i(e)\in\{0,p_i(x)\}]
        \leq
        \Pr[\Phi_i(e)=0]+\Pr[\Phi_i(e+x)=0]
        \leq
        2^{-K+1}.
    \end{equation*}
    Thus, in case~(iii), membership of an edge $e$ in $\Span(A\cup E_i)$ is tested by checking whether $p_i(e)\in\{0,p_i(x)\}$.

    Finally, for a line $\ell=\{e,\bar e\}$, we have
    $r(E_i\cup A\cup \ell)=r(E_i\cup A)$ if and only if both $e$ and $\bar e$ belong to $\Span(A\cup E_i)$.
    We therefore apply the above membership test to both $e$ and $\bar e$.
    Each pattern value is returned in $\tO(1)$ time by the pattern-tree data structure, so the query time is $\tO(1)$.
    The answer is always correct when the equality holds, and is correct with high probability otherwise.
\end{proof}

We prove Lemma~\ref{lem:large_handle} using Lemma~\ref{lem:large_part_query}.

\begin{proof}[Proof of Lemma~\ref{lem:large_handle}]
    Write $c_i = |M_i|$.
    Since the sets $M_i$ are disjoint subsets of $M$, we have $\sum_i c_i\leq |M|=O(n)$.
    Hence, there are only $O(n/C)$ indices $i$ with $c_i>C$.

    Fix such an index $i$.
    Since we are working with the pruned cover, we have
    $r(E_i\cup A)>r(A)+1$, and hence Lemma~\ref{lem:large_part_query} applies.
    Then, by scanning all lines $\ell\in L$, we can decide in $\tO(m)$ time whether
    $r(E_i\cup A\cup \ell)=r(E_i\cup A)$ for each line $\ell$.

    Therefore, for all lines $\ell\in L$, we can decide whether there exists an index $i\in\{0,\ldots,k\}$ with $|M_i|>C$ such that $r(E_i\cup A\cup \ell)=r(E_i\cup A)$ in total $\tO(mn/C)$ time.

    The correctness holds with high probability by Lemma~\ref{lem:large_part_query} and a union bound over the $\tO(mn/C)$ line queries performed above.
\end{proof}
\section{Gabow--Stallmann's algorithm} \label{sec:gabow_stallmann_min_cover_appendix}

Lemmas~\ref{lem:one_can_find_cover} and~\ref{lem:structual_lemma} in Section~\ref{sec:structures} are implicit in Gabow--Stallmann's papers~\cite{gabow1986augmenting,gabow1985efficient}, but they are not stated there in this form.
In this section, for completeness, we explain these lemmas, which we restate here.

\gsmincover*




\structures*


Gabow--Stallmann~\cite{gabow1986augmenting} gave an augmenting-path algorithm for linear matroid parity.
For graphic matroid parity, Gabow--Stallmann~\cite{gabow1985efficient} applied this algorithm to the associated linear matroid parity instance (Definition~\ref{def:associated_linear_matroid_parity}) and gave a faster implementation using range-searching data structures.
In particular, their implementation finds one augmenting path in $\tO(m)$ time. See also Remark~\ref{rem:gsone}.

We first briefly describe Gabow--Stallmann's augmenting-path algorithm.
The algorithm searches for an augmenting path in an auxiliary graph, called a dependence graph, as in the matroid intersection algorithm.
As in graph matching algorithms, the search creates blossoms.
Roughly, when a blossom is formed, the algorithm designates one element as its \emph{bud} and some elements of the blossom as its \emph{tips}.
An element that was once a tip of a blossom but ceases to be a tip when that blossom is absorbed into a larger blossom is called an \emph{old tip}.
See Gabow--Stallmann~\cite[Definition~3.2]{gabow1986augmenting} for the formal definition.
In addition, each element is assigned a label during the search. (An element may also be unlabeled.)

The final unsuccessful search yields a certificate of optimality.
This is analogous to the graph matching algorithm, where the absence of an augmenting path yields a structure attaining equality in the Tutte--Berge formula.
In the linear matroid parity algorithm of Gabow--Stallmann~\cite{gabow1986augmenting}, the corresponding certificate is a cover in the sense of Definition~\ref{def:matroid_parity_cover}.
More precisely, to certify the optimality of the final matroid matching $M$ using the min--max theorem (Theorem~\ref{thm:min_max_linear_matroid_parity}), they construct a cover whose value is equal to $\frac{|M|}{2}$, i.e., a minimum cover.
The labels, blossoms, buds, and tips maintained at the end of the final unsuccessful search define this cover as follows. (See \cite[p.~145]{gabow1986augmenting}.)
\begin{align*}
&E_0 = \{e \mid e \text{ and } \bar{e} \text{ are unlabeled and are not in a (degenerate) blossom} \}, \\
&E_i \text{ is either $B \cap E$ for a blossom $B$, or a pair $\{e, \bar{e}\}$ (in $E$) with exactly one of $e, \bar{e}$ labeled, $i = 1, \ldots, k$.}
\end{align*}
We note that a blossom $B$ may contain a transform that is not in $E$.
Let $M_i = M\cap E_i$, where $M$ is the final matroid matching.
They define $T_i\subseteq E_i$ as follows:
\begin{align*}
&T_0=\emptyset,\\
&T_i=\{e\in E_i \mid \ e \text{ is unlabeled and is not an old tip}\}, \text{ if $i > 0$}
\end{align*}
Here, when the block $E_i$ corresponds to a blossom, the set $T_i$ is the set of current tips of the blossom.\footnote{The discussion in~\cite[p.~136]{gabow1986augmenting} implies that every unlabeled element in a current blossom is either a current tip of that blossom or an old tip. They then explicitly state that, if $E_i=B\cap E$, then $T_i$ consists of the tips of $B$~in \cite[p.~145]{gabow1986augmenting}.}
Since every blossom has at least one current tip, $T_i\neq\emptyset$ in this case.
In the other case, where $E_i = \{e, \bar{e}\}$, the definition of $T_i$ makes it a singleton.
Hence, for $i \in \{1, \ldots, k\}$, we have $T_i \neq \emptyset$.
The vector $\alpha_i$ is then defined as follows.
\begin{align*}
\alpha_i =
\begin{cases}
e, &
\text{if } E_i=\{e,\bar e\}\subseteq M_i
\text{ and } e \text{ is unlabeled},\\[2mm]
\displaystyle\sum_{\beta\in T_i} c(b_i,\beta)\beta, &
\text{if } E_i=B\cap E \text{ and } T_i\subseteq M_i,\\[2mm]
\text{undefined}, & \text{otherwise}.
\end{cases}
\end{align*}
Here, $b_i$ is a bud of $B$, and $c(b_i,\beta)$ is the coefficient that determines adjacency in the dependence graph: $c(b_i,\beta)\neq 0$ exactly when $b_i$ and $\beta$ are adjacent in the dependence graph.
Let $A=\{\alpha_i \mid \alpha_i\text{ is defined}\}$.

In the graphic case, and more generally in the binary case, \cite[Corollary 4.3(i)]{gabow1986augmenting}\footnote{\cite[Corollary 4.3(i)]{gabow1986augmenting} states that the bud $b$ of a blossom $B$ is adjacent to all tips of $B$.} implies that $c(b_i,\beta)=1$ for any $\beta \in T_i$. 
We note, however, that this property is not essential in the proofs in Sections~\ref{sec:adaptation_of_assadi} and~\ref{sec:fast_approx_algo}. 
What is essential is that $\alpha_i$ can be written as $\alpha_i=\sum_{\beta\in X_i}\beta$ for some $X_i\subseteq M_i$.

To prove the constructed cover $(A;E_0, \ldots, E_k)$ is a minimum cover, Gabow--Stallmann need to determine the values $r(E_i\cup A)$. 
See \cite[p.~146]{gabow1986augmenting}.
For this purpose, they prove the following properties. 
Using \cite[Lemma 4.8]{gabow1986augmenting}, they show that $M_0\cup A$ is a base of $E_0\cup A$, and that $M_i\cup (A\setminus\{\alpha_i\})$ is a base of $E_i\cup A$ for every $i>0$ with $T_i\subseteq M_i$. 
Moreover, using \cite[Lemmas 4.8 and 4.9]{gabow1986augmenting}, they show that, for every $i>0$ with $T_i\not\subset M_i$ and every $x\in T_i\setminus M_i$, the set $M_i\cup A\cup\{x\}$ is a base of $E_i\cup A$. 
Since $T_0=\emptyset$ and $T_i\neq\emptyset$ for every $i>0$, these properties imply Lemma~\ref{lem:structual_lemma}.

Gabow--Stallmann then show that $r(E_i\cup A)=r(A)+|M_i|+\delta_i$, where
\begin{equation*}
\delta_i=
\begin{cases}
0, & i=0,\\
+1, & i>0 \text{ and } T_i\not\subset M_i,\\
-1, & i>0 \text{ and } T_i\subseteq M_i.
\end{cases}
\end{equation*}
Note that the sets $M_0,\ldots,M_k$ are pairwise disjoint subsets of the independent set $M$, and that $\alpha_i\neq 0$ whenever $\alpha_i$ is defined.
Hence, the vectors in $A$ are linearly independent, and thus $r(A)=|A|$.
Then, they show that $\displaystyle r(A) + \sum_{i = 0}^{k} \left\lfloor \frac{r(A \cup E_i) - r(A)}{2} \right\rfloor$ is equal to $\displaystyle \frac{|M|}{2}$.
By the min--max theorem (Theorem~\ref{thm:min_max_linear_matroid_parity}), they show that the final matroid matching $M$ is maximum.
Therefore, the constructed cover is minimum in the sense of Definition~\ref{def:matroid_parity_cover}.

The implementation of Gabow and Stallmann~\cite{gabow1985efficient} for graphic matroid parity provides enough information about the labels, blossoms, buds, and tips maintained during the search.
Using this information, we can construct a minimum cover $(A;E_0,\ldots,E_k)$ for the associated linear matroid parity instance. 
Therefore, given a maximum matroid matching $M$, we can obtain a minimum cover in $\tO(m)$ time.
Indeed, we run one augmenting-path search from $M$. 
Since $M$ is maximum, this search terminates without finding an augmenting path, and the labels, blossoms, and tips maintained at termination define the minimum cover described above.
The discussion here gives Lemma~\ref{lem:one_can_find_cover}.

\end{document}